\documentclass[%
reprint
aps,
nofootinbib,floatfix,
longbibliography,
notitle,
superscriptaddress
]{revtex4-2}
\usepackage{graphicx}
\usepackage{dcolumn}
\usepackage{bm}
\usepackage{amsfonts}
\usepackage[colorlinks=true,citecolor=blue,urlcolor=blue]{hyperref}
\usepackage{mathrsfs}
\usepackage[dvipsnames]{xcolor}
\usepackage{subcaption}
\usepackage{amsmath,amssymb}
\usepackage{cleveref}
\usepackage{physics}
\usepackage{multirow}
\usepackage{tensor}
\usepackage[normalem]{ulem}

\allowdisplaybreaks
\DeclareMathAlphabet\mathbfcal{OMS}{cmsy}{b}{n}

\newcommand{\dif}{\mathrm{d}}

\newcommand{\s}{\mathrm{s}}
\newcommand{\iu}{\mathrm{i}}
\newcommand{\m}{m_{\rm{P}}}

\newcommand{\n}{\mathcal{N}}
\newcommand{\hphi}{\tilde{\Phi}}
\newcommand{\bfnmc}{\bar{F}}
\usepackage[T1]{fontenc}
\usepackage[utf8]{inputenc}
\usepackage{booktabs}
\renewcommand{\d}{\mathrm{d}}
\newcommand{\N}{\mathcal{N}}
\newcommand{\const}{\mathrm{const.}}

\allowdisplaybreaks
\begin{document}
\title{Non-Minimally Coupled Warm Inflation in the Defining Frame}
\author{Adri\'an Casado-Turri\'on}
\affiliation{Center for Physical Sciences and Technology (FTMC), Saul\.{e}tekio av.~3, 10257 Vilnius, Lithuania.}%
\author{Paulo B.~Ferraz} 
\affiliation{%
Faculdade de Ci\^{e}ncias e Tecnologia and CFisUC, Departamento de F\'isica,\\ 
Universidade de Coimbra, Rua Larga, P-3004-516 Coimbra, Portugal.
}%
\author{Mindaugas Kar\v{c}iauskas}
\affiliation{Center for Physical Sciences and Technology (FTMC), Saul\.{e}tekio av.~3, 10257 Vilnius, Lithuania.}%
\author{Jos\'e Jaime Terente D\'iaz}
\affiliation{%
Faculdade de Ci\^{e}ncias e Tecnologia and CFisUC, Departamento de F\'isica,\\ 
Universidade de Coimbra, Rua Larga, P-3004-516 Coimbra, Portugal.
}%

\begin{abstract}
\noindent
Warm inflation in $F(\Phi)R$ scalar-tensor theories of gravity is investigated in the `defining' frame, where the theory and its parameter values are specified. Translating the resulting dynamics to the Einstein frame, we find that the dissipation ratio is suppressed by the modified-gravity effects. Thus, although the effective warm-inflation dynamics can be consistently analysed in either frame, the dissipative regimes need not coincide between them. In particular, we find that quantum perturbations can dominate over thermal fluctuations in the scalar power spectrum even in a high-temperature, strong-dissipation regime in the defining frame. Finally, we compute the scalar spectral index and tensor-to-scalar ratio for a non-minimal coupling function $F(\Phi) = 1+\xi (\Phi/m_{\rm P})^2$ with a quartic potential and both constant and quadratic field-dependent dissipation coefficients, and identify benchmark points compatible with current CMB constraints. 
\end{abstract}
\maketitle
\begingroup
\hypersetup{linkcolor=black}
\endgroup
\section{Introduction}
\noindent
Cosmic inflation \cite{Dimopoulos:2020pjx,Achucarro:2022qrl} has, for more than four decades, remained the leading framework for understanding the very early Universe \cite{Starobinsky:1979ty,Starobinsky:1980te,Sato:1981ds,Sato:1981qmu,Guth:1980zm,Linde:1981mu}. By postulating a primordial phase of accelerated expansion, the inflationary paradigm addresses the horizon, flatness, and relic problems of the Hot Big Bang model, while also generating the primordial inhomogeneities that are imprinted in the Cosmic Microwave Background (CMB) and that subsequently grow, through gravitational instability, into the large-scale structures (LSS) observed today \cite{Bardeen:1980kt,Brandenberger:1983vj,Kodama:1984ziu,Mukhanov:1990me,Ma:1995ey,Hwang:2001fb,Weinberg:2003sw,Malik:2008im,Lyth:2009zz,Kinney:2009vz,Baumann:2009ds}. Recent observations from Planck~\cite{Planck:2018jri}, BICEP/Keck~\cite{BICEP:2021xfz}, ACT~\cite{AtacamaCosmologyTelescope:2025blo,AtacamaCosmologyTelescope:2025nti}, and complementary baryon-acoustic-oscillation (BAO) and LSS surveys, place increasingly stringent constraints on the scalar spectral index $n_\s$ and the tensor-to-scalar ratio $r$ \cite{eBOSS:2020yzd,DES:2021wwk,SPT-3G:2025bzu,Ferreira:2025lrd,McDonough:2025lzo,Balkenhol:2025wms}, thereby narrowing the space of viable inflationary models \cite{Martin:2013tda,CosmoVerseNetwork:2025alb,Kallosh:2025ijd}.

The simplest single-field scenarios, in which a slowly rolling `inflaton' field traverses a nearly flat potential \cite{Linde:1981mu,Albrecht:1982wi,Linde:1983gd,Liddle:1994dx,Leach:2002ar}, face well-known fine-tuning problems \cite{Baumann:2014nda}. If the flatness of the potential is not protected by any symmetry, radiative corrections and higher-dimensional operators can generate large contributions to the effective mass for the inflaton, spoiling the observational constraints on the scalar spectral index unless these effects are sufficiently suppressed or tuned away. Warm inflation \cite{Gleiser:1993ea,Berera:1995wh,Berera:1995ie,Berera:1996nv,Berera:1998gx,Yokoyama:1998ju,Berera:1998px,Berera:1999ws,Hall:2003zp,Moss:2008yb,Berera:2008ar,2009JCAP...07..013G,Bastero-Gil:2011rva,Ramos:2013nsa,Bastero-Gil:2014jsa,Bastero-Gil:2016qru,Bastero-Gil:2019gao,Bastero-Gil:2019rsp,Berghaus:2019whh,Ferraz:2023qia,Kamali:2023lzq,Berera:2023liv,Ballesteros:2023dno,Montefalcone:2023pvh,Rodrigues:2025neh,Berera:2025vsu} offers a different approach, in which the inflaton continuously dissipates energy into a concurrent quasi-thermal radiation bath during inflation \cite{Berera:2001gs,Berera:2006xq,Berera:2008ar,Bartrum:2014fla}. The associated friction relaxes the slow-roll requirements, alleviating the UV sensitivity of inflation \cite{Berera:2003yyp,Bartrum:2013fia,Ferraz:2023qia}, while also easing compatibility with the de Sitter swampland conjectures \cite{Motaharfar:2018zyb,Das:2018rpg,Kamali:2023lzq}. Moreover, the production of radiation eliminates the need for a separate reheating phase, allowing inflation to transition smoothly into a radiation-dominated era through a `graceful' exit \cite{Berera:2008ar,Allahverdi:2010xz,Bastero-Gil:2015nja}.

Provided that the coupling between the inflaton field and a sector containing light degrees of freedom is sufficiently strong, the bath can be continually replenished, maintaining a quasi-stationary radiation density throughout inflation. A concrete realisation of this mechanism is furnished by a two-stage process in which the inflaton interacts with heavy mediator fields that then decay into light radiation \cite{Bastero-Gil:2009sdq}. Microphysical, first-principles derivations of the dissipation coefficient within the Closed-Time-Path formalism have made this picture quantitatively precise \cite{Gleiser:1993ea,Berges:2004yj,Calzetta:2008iqa,Berera:2008ar,Bastero-Gil:2010dgy,Bartrum:2014fla}, strengthening the case for warm inflation as a self-consistent and observationally viable framework \cite{Yokoyama:1998ju,Berera:1998px,Berera:1999ws,Bartrum:2014fla,Bastero-Gil:2016qru,Bastero-Gil:2019gao,Levy:2020zfo,Ballesteros:2023dno,Ferraz:2023qia,Berera:2025vsu,Rodrigues:2025neh}.    

Nevertheless, the rich phenomenology of warm inflation and its improved agreement with observational data may be further explored by embedding the framework within a more general gravitational setting. Modified theories of gravity, in fact, predate the inflationary paradigm itself \cite{Clifton:2011jh}. Dirac's Large-Numbers Hypothesis \cite{Dirac:1937ti}, which suggested a gravitational constant varying with cosmological time \cite{Dirac:1938mt}, together with the Machian intuition that inertia is determined by the distribution of matter in the Universe \cite{Sciama:1953zz}, laid the foundations for theories in which the effective gravitational coupling is itself dynamical. These ideas were later incorporated into the scalar-tensor framework \cite{Fierz:1956zz,Jordan:1959eg,Brans:1961sx,Dicke:1961gz,Brans:1962zz,Bergmann:1968ve,Nordtvedt:1970uv,Wagoner:1970vr,Damour:1992we,Brans:1996tp,Goenner:2012cq,Will:2014kxa}, which now underlies a broad class of contemporary inflationary models \cite{Martin:2013tda,Odintsov:2023weg,CosmoVerseNetwork:2025alb}. In particular, non-minimal couplings of the form $F(\Phi)R$, with $F$ being a function of a scalar field $\Phi$ and $R$ denoting the Ricci scalar, are now regarded as essentially unavoidable from an effective-field-theory perspective \cite{Burgess:2007pt,Baumann:2014nda}. Even if absent at tree level, such couplings are permitted by the symmetries of the theory and are consequently generated by radiative corrections, appearing as necessary counterterms in quantum field theory on curved spacetime \cite{Callan:1970ze,Birrell:1982ix,Buchbinder:1992rb,Parker:2009uva}. Moreover, from the perspective of inflationary model-building, non-minimal couplings to gravity are of considerable interest, since they can relax constraints on dimensionless self-couplings and flatten the effective inflaton potential at large field values, rendering many otherwise excluded models compatible with observations \cite{Accetta:1985du,Lucchin:1985ip,Futamase:1987ua,McDonald:1990wy,Fakir:1990iu,Makino:1991sg,Kaiser:1993bq,Morris:2001ad,Bezrukov:2007ep,Park:2008hz,Chiba:2008ia,Boubekeur:2015xza,Rubio:2018ogq,Lerner:2009na,Tenkanen:2017jih,Ferreira:2018nav,Akin:2020mcr,Cheong:2021vdb,Kodama:2021yrm,Karciauskas:2022jzd,Pozdeeva:2025ied}.

These modifications of gravity, however, introduce an important conceptual subtlety. A given physical theory can be expressed in multiple field parametrisations related by conformal redefinitions of the metric tensor \cite{Shapiro:1995kt,Faraoni:1998qx,Flanagan:2004bz,Dabrowski:2008kx,Kaiser:2010ps,Kubota:2011re,Domenech:2016yxd}. Amongst these, the Jordan \cite{Fierz:1956zz,Jordan:1959eg} frame is conventionally identified as the one in which the scalar field $\Phi$ is non-minimally coupled to gravity, while the matter sector is not \cite{Boisseau:2000pr,Morris:2001ad,Fujii:2003pa,Faraoni:2004pi,Jarv:2021qpp,Jarv:2025qgo}. For the theories considered here, this identification is too restrictive, since the matter sector need not be minimally coupled. We therefore refer to the frame in which the model and its renormalisation conditions are specified as the `defining' frame, reserving the `Jordan frame' terminology for its original meaning, which is a particular frame where the direct scalar--matter couplings are absent (if such a frame exists). Through a suitable conformal transformation, the gravitational sector can instead be recast into the canonical Einstein-Hilbert form \cite{Maeda:1988ab}. The resulting `Einstein' frame features non-minimal couplings between $\Phi$ and the matter fields. Classically, the equivalence of the two frames is well established \cite{Magnano:1993bd,Kaiser:1994vs,Kaiser:1995nv,Faraoni:2006fx,Catena:2006bd,Catena:2006gk,Deruelle:2010ht,Gong:2011qe,White:2012ya,Chiba:2013mha,Prokopec:2013zya,Postma:2014vaa,Domenech:2015qoa,Jarv:2016sow,Karam:2017zno,Azri:2018gsz,Akin:2020mcr,Racioppi:2021jai,Karciauskas:2022jzd,Diaz:2023tma}. At the quantum level, by contrast, analyses of both the divergent and finite parts of the one-loop effective action have led to conflicting conclusions concerning frame equivalence \cite{Steinwachs:2011zs,Kamenshchik:2014waa}. Consequently, whether this equivalence extends to the quantum-gravitational regime remains considerably more debated than its classical counterpart \cite{Falls:2018olk}.          

Despite its conceptual appeal, the interplay between warm inflation and modified gravity has so far received comparatively limited attention, relative to the extensive literature on cold inflation \cite{Kamali:2018ylz,Amaek:2021cqs,Samart:2021eph,Samart:2021hgt,Payaka:2022jtb,Yeasmin:2022bqq,Eadkhong:2023ozb,Shiravand:2024ayw,Yeasmin:2024yel,Yuennan:2024nje,Yuennan:2025szw}. This combination is particularly relevant in light of the scalar--matter couplings that can arise, in the absence of symmetries forbidding them, even in the frame where the scalar field is non-minimally coupled to gravity \cite{tHooft:1979rat,Wetterich:1987fm,Damour:1994zq,Burgess:2007pt,Weinberg:2021exr,Copeland:2021qby,Peskin:2025lsg}. Several studies have investigated warm inflation with simple non-minimal gravitational couplings at the phenomenological level \cite{Nozari:2016nxc,Kamali:2018ylz,Samart:2021hgt,Amaek:2021cqs,Eadkhong:2023ozb,Cheng:2024uvn,Cheng:2025rmf,Cheng:2026tjl}. Others have examined Horndeski-type derivative couplings \cite{Nozari:2014seg,MohseniSadjadi:2014vsj,Goodarzi:2016iht,Motaharfar:2017dxh,Sebastiani:2017mkv,Motaharfar:2018mni,Arya:2023pod,Zhang:2024kcf,Zhang:2025suv,Zhao:2026jim,Zhang:2026zjv}, $f(R,\mathcal{T})$, and $f(\mathcal{G})$ modifications \cite{Sharif:2015vda}, where $\mathcal{T}$ denotes the trace of the matter energy-momentum tensor and $\mathcal{G}$ is the Gauss-Bonnet term, as well as the Higgs-Starobinsky scenario in both metric \cite{Samart:2021eph} and Palatini formulations \cite{Yuennan:2025szw}. In the case of non-minimally coupled models, which are most directly comparable to our present work, the inflationary observables are typically computed by first transforming the action conformally to the Einstein frame, and then using the standard flat-space expression for the dissipation coefficient in that frame. In effect, this procedure inherits the standard Markovian dissipation of General Relativity (GR)-based warm inflation, with the conformal transformation entering only through the rescaling of temperature \cite{Faraoni:2023gqg,Karolinski:2024ukr} and the friction term. Although formally simple, this approach overlooks an essential feature of non-minimally coupled scalar-tensor theories, which is the modification of the matter sector itself under the conformal rescaling of the metric, as mentioned above. 
Scalar--matter couplings that are constant in the original (defining) frame become field-dependent after the conformal transformation, and may contribute to the inflaton self-energy at one loop (and hence to the dissipation coefficient), in a way that a purely GR-based calculation cannot capture. Consequently, the dissipation coefficient and stochastic noise term, computed in the Einstein frame using flat-space techniques, need not, in general, coincide with those obtained by quantising the matter sector in the defining frame and subsequently translating the result to the Einstein frame. Since the model and its renormalisation conditions are specified in the defining frame, prior to any conformal or further field redefinition, we argue that the latter procedure provides the physically appropriate prescription. 

This approach should be understood within the treatment of gravity adopted here, which is that of a (semi-)classical background, so that classical frame equivalence holds at the level of the field equations. This equivalence does not, however, automatically carry over to the effective dynamics of the quantised matter sector, since the frame transformation acts non-trivially on the matter couplings. It is in this sense that the microphysical input of warm inflation (most notably, the dissipation coefficient and the associated stochastic noise) is sensitive to the underlying gravitational theory \cite{Wojnar:2022dvo}.

This work is organised as follows. In Sec.~\ref{sec:def-einstein-frames}, we derive an effective description of a generic interaction between the matter sector and the scalar degree of freedom of the gravitational sector in the defining frame. We then translate the resulting effective description into the Einstein frame by conformally transforming all the relevant quantities. Section~\ref{sec:nonminimally-coupled-warm-inflation} specialises to the case in which these interactions generate dissipative effects, thereby leading to a warm inflation scenario. We compute the slow-roll conditions in both the defining and Einstein frames, while imposing the necessary conditions for warm inflation to be realised. We also explain why, for consistency, one needs to first compute the quantum and thermal effects in the defining frame and then transform them into the Einstein frame. In Sec.~\ref{sec:cosmoperturb-einsteinframe}, the associated cosmological perturbations are computed, yielding the scalar spectral index and the tensor-to-scalar ratio separately in terms of defining- and Einstein-frame quantities. Finally, we assess the validity of two models with different temperature-independent dissipation coefficients, working with a quartic inflationary potential and a non-minimal coupling of the Higgs-inflation type model. A summary and discussion of our results are presented in Sec.~\ref{sec: summary}. Auxiliary details can be found in the appendices.

Throughout this work, natural units ($\hbar = c = k_{\textrm{B}} = 1$) are adopted. The reduced Planck mass is 
\begin{equation}
\m \equiv (8\pi G)^{-1/2} \simeq 2.44\times10^{18}\,\mathrm{GeV}~,
\end{equation}
and $G$ denotes Newton's gravitational constant. We employ the `mostly plus' metric signature, $(-,+,+,+)$. 

\section{\label{sec:def-einstein-frames}Defining and Einstein Frames}
\subsection{\label{sec:frame-transformations-def-Einstein}Frame Transformations}
\noindent
Consider the scalar-tensor theory of gravity defined by the following action \cite{Boisseau:2000pr,Domenech:2016yxd,Fujii:2003pa,Faraoni:2004pi}:
\begin{equation} \label{eq:defining frame action}
    S[g_{\mu\nu},\Phi,\Xi]=\int\dif^4 x\,\sqrt{-g} \left[\dfrac{\m^2}{2}F(\Phi)R-\dfrac{1}{2}g^{\mu\nu}\nabla_\mu\Phi\nabla_\nu\Phi-V(\Phi)+\mathcal{L}_{\textrm{m}}(g_{\mu\nu},\Xi)\right],
\end{equation}
where the matter action functional $S_{\textrm{m}}$ is
\begin{equation}
    S_{\textrm{m}}[g_{\mu\nu},\Xi] = \int \textrm{d}^4x\,\sqrt{-g}\,\mathcal{L}_{\textrm{m}}(g_{\mu\nu},\Xi)~.
\end{equation}
Here, $\Xi(x)$ collectively denotes the matter fields, $\Phi(x)$ is a scalar field with potential $V(\Phi)$, and $R$ is the Ricci scalar. The non-minimal coupling function $F(\Phi)$ is required to be strictly positive in order to ensure the correct sign of the kinetic term of the spin-2 mode \cite{Deser:1983rq,Batakis:1984hu,Bertolami:1987wj,Kaloper:1997sh,Esposito-Farese:2000pbo,Esposito-Farese:2004azw,Faraoni:2004qd,Bronnikov:2006pt,Gannouji:2006jm,Kase:2020qvz,Ballardini:2023mzm,Jarv:2024krk,Jarv:2025qgo}. Such a formulation of the action, in which the matter fields couple minimally to gravity while the gravitational sector is modified by one or more additional scalar fields, is customarily called the `Jordan frame' \cite{Fierz:1956zz,Jordan:1959eg,Goenner:2012cq,Magnano:1993bd,Faraoni:1998qx,Faraoni:2006fx,Kaiser:2010ps,Deruelle:2010ht,Chiba:2013mha,Postma:2014vaa,Azri:2018gsz,Racioppi:2021jai}. In this frame, the dimensionless couplings of the matter sector (such as the Yukawa and gauge couplings) are constant and independent of the scalar gravitational degree of freedom $\Phi$, at least at the classical level \cite{Shapiro:1995kt,Steinwachs:2011zs,Kamenshchik:2014waa}. The matter energy-momentum tensor (EMT) is defined as
\begin{equation} \label{eq:energy-momentum-tensor}
     \mathcal{T}_{\mu\nu}\equiv -\frac{2}{\sqrt{-g}}\frac{\delta S_{\textrm{m}}}{\delta g^{\mu\nu}}~.
\end{equation}
Because the matter fields are minimally coupled to gravity, the EMT is covariantly conserved, $\nabla_{\mu} \mathcal{T}^{\mu\nu}=0$ \cite{Koivisto:2005yk,Copeland:2021qby}. Consequently, an element of a pressureless perfect fluid (dust) moves along geodesics of the Jordan-frame metric $g_{\mu\nu}$ \cite{Bergmann:1968ve,Nordtvedt:1970uv,Faraoni:2004pi,Esposito-Farese:2004azw,Chiba:2013mha,Will:2014kxa,Joyce:2014kja,Jarv:2015kga}.

Nonetheless, unless forbidden by the symmetries of the theory, the scalar field $\Phi$ is generically expected to couple directly to (some of) the matter sector fields as well \cite{tHooft:1979rat,Wetterich:1987fm,Damour:1994zq}. Even if such couplings are taken to vanish at the tree level, the rules of effective field theory dictate that they will be generated by quantum effects \cite{Georgi:1993mps,Burgess:2007pt,Weinberg:2021exr,Peskin:2025lsg}. We therefore write
\begin{equation}
    \label{eq:the_action_nonminimal}S[g_{\mu\nu},\Phi,\Xi] = \int \textrm{d}^4x\,\sqrt{-g} \left[\frac{\m^2}{2}F(\Phi)R -\frac{1}{2}g^{\mu\nu}\nabla_{\mu} \Phi \nabla_{\nu} \Phi - V(\Phi)+\mathcal{L}_{\textrm{m}}(g_{\mu\nu},\Phi,\Xi)\right].
\end{equation}
Owing to the direct coupling of $\Phi$ to the matter fields, the covariant conservation of the EMT previously mentioned no longer holds, as shown in Sec.~\ref{sec: covariantfieldequations}. We reserve the term `Jordan frame' for the action in Eq.~\eqref{eq:defining frame action} \cite{Flanagan:2004bz,Jarv:2014hma,Jarv:2015kga}, and refrain from using it when matter fields couple directly to $\Phi$. Depending on the form of the coupling between $\Phi$ and $\Xi$, the action in Eq.~\eqref{eq:the_action_nonminimal} can, in some cases, be recast into the Jordan-frame representation of Eq.~\eqref{eq:defining frame action} via an appropriate field redefinition of the metric \cite{Bekenstein:1992pj}. We shall instead refer to Eq.~\eqref{eq:the_action_nonminimal} as the `defining-frame' action, meaning the frame in which the model is originally specified, irrespective of the structure of the matter or gravitational sectors. In particular cases, the defining frame may coincide with the Jordan frame, or with another one, contingent on the starting point of the analysis.

A convenient reformulation of the gravitational action can be achieved by means of a conformal rescaling of the metric \cite{Dicke:1961gz,Faraoni:1998qx,Catena:2006gk}:\footnote{
    This ensures that, in the general-relativistic case ($F=1$), the two metrics coincide. Note also that a conformal transformation is \textit{not} a coordinate transformation, but rather a redefinition of the metric tensor field.
    \label{footnote:clarification-conformalrescaling}
} 
\begin{equation}
    \label{eq:conformal-Einstein}\hat{g}_{\mu\nu}(x) \equiv F[\Phi(x)] g_{\mu\nu}(x)~.
\end{equation}
The condition $F>0$, already required on stability grounds \cite{Gannouji:2006jm,Ballardini:2023mzm,Jarv:2024krk,Jarv:2025qgo}, additionally guarantees that the transformed metric preserves the Lorentzian signature of the defining-frame metric, so that distances, volumes, and causal structure remain well defined in the new frame. Under this transformation, the square root of the determinant of the metric changes as $\sqrt{-g} \rightarrow F^2 \sqrt{-g}$. Upon integration by parts and neglecting a boundary term, the action in Eq.~\eqref{eq:the_action_nonminimal} becomes (see Ref.~\cite{Dabrowski:2008kx} for the conformal transformations of all relevant geometrical quantities, including the Ricci scalar)  
\begin{equation} \label{eq:einstein frame action}
    S[\hat{g}_{\mu\nu},\Phi,\Xi]=\int\dif^4 x\,\sqrt{-\hat{g}}\,\left\{\dfrac{\m^2}{2}\hat{R}-\dfrac{\mathcal{K}(\Phi)}{2}\hat{g}^{\mu\nu}\hat{\nabla}_\mu\Phi\hat{\nabla}_\nu\Phi-U(\Phi)+\hat{\mathcal{L}}_{\textrm{m}}\left[\dfrac{\hat{g}_{\mu\nu}}{F(\Phi)},\Phi,\Xi\right]\right\},
\end{equation}
where
\begin{eqnarray}
   \mathcal{K}(\Phi) &\equiv& \frac{1}{F(\Phi)}\left\{1+\dfrac{3}{2}\dfrac{\m^2\left[F_{,\Phi}(\Phi)\right]^2}{F(\Phi)}\right\},
    \label{eq:non-canonical-function}\\
    U(\Phi) &\equiv& \dfrac{V(\Phi)}{F^2(\Phi)}~,
   \label{eq:potentialU}
\end{eqnarray}
and 
\begin{equation}
    \label{eq:trans-mathcalL-m-hat}\hat{\mathcal{L}}_{\textrm{m}} \equiv F^{-2}(\Phi) \mathcal{L}_{\textrm{m}}~.
\end{equation}
A subscript comma denotes differentiation with respect to the argument; \textit{e.g.}, $F_{,\Phi}(\Phi)\equiv\dif F/\dif\Phi$. Also, $\nabla_{\mu} \Phi = \partial_{\mu} \Phi = \hat{\nabla}_{\mu} \Phi$ (and similarly for any scalar function). The conformal transformation of the metric tensor does not alter the scalar field, but it casts the gravitational sector into the canonical form given by the Einstein-Hilbert action \cite{Maeda:1988ab}. For this reason, the new frame is usually referred to as the `Einstein frame'. 

Irrespective of whether one starts from the action in Eq.~\eqref{eq:the_action_nonminimal} or Eq.~\eqref{eq:defining frame action}, after the conformal transformation of the metric, the scalar field becomes coupled to all fields in the matter sector through the diffeomorphism-invariant volume measure $\textrm{d}^4x\sqrt{-g}$. Additionally, in the hatted frame, the scalar field $\Phi$ appears with a non-canonical kinetic term. This kinetic function encodes part of the effective self-interaction of the scalar gravitational degree of freedom, which does not propagate as a free field even in the absence of a potential term, due to its coupling to $R$ in the defining frame \cite{Mimoso:1994wn,Brans:1996tp}. We can introduce an additional scalar field $\tilde{\Phi}$, which is related to $\Phi$ by \cite{Esposito-Farese:2000pbo}\footnote{
    We deliberately use a tilde to denote the scalar field with a canonical kinetic term in the Einstein frame, distinguishing the scalar-field redefinition from the conformal rescaling of the metric.
} 
\begin{equation}
    \label{eq:sf-redefinition}\left(\frac{\textrm{d}\hphi}{\textrm{d}\Phi}\right)^2\equiv \mathcal{K}(\Phi)~.
\end{equation}
The benefit of introducing such a field is that its kinetic term is canonical, although closed-form solutions to Eq.~\eqref{eq:sf-redefinition} are generally difficult to obtain \cite{Mimoso:1994wn, Garcia-Bellido:2008ycs,Domcke:2017rzu,Tang:2021lcn}. The full action in terms of this field can then be written as
\begin{equation} \label{eq:einstein frame action-canonical}
    S[\hat{g}_{\mu\nu},\hphi,\Xi]=\int\dif^4 x\,\sqrt{-\hat{g}}\,\left\{\dfrac{\m^2}{2}\hat{R}-\dfrac{1}{2}\hat{g}^{\mu\nu}\hat{\nabla}_\mu\hphi\hat{\nabla}_\nu\hphi-U[\Phi(\hphi)]+\hat{\mathcal{L}}_{\textrm{m}}\left[\dfrac{\hat{g}_{\mu\nu}}{F[\Phi(\hphi)]},\Phi(\hphi),\Xi\right]\right\}.
\end{equation}
The first two terms enclosed by the curly brackets take the form of the Einstein-Hilbert action with a canonical scalar field, as mentioned above. However, the modification of gravity is not removed by this field redefinition: it manifests itself via the universal coupling of matter fields to $\tilde{\Phi}$.

\subsection{\label{sec: covariantfieldequations}Covariant Field Equations and the Energy-Momentum Conservation}
\noindent
The classical field equations in the defining frame are obtained by applying the principle of stationary action to $S[g_{\mu\nu},\Phi,\Xi] = S_{\textrm{grav}}[g_{\mu\nu},\Phi]+S_{\textrm{m}}[g_{\mu\nu},\Phi,\Xi]$ given in Eq.~\eqref{eq:the_action_nonminimal}. Varying with respect to the metric, the scalar field $\Phi$, and the matter fields $\Xi$, and discarding boundary terms, the variation can be written as
\begin{equation}
    \delta S = \int \dif^4x\,\sqrt{-g} \left[\frac{1}{2}\left(E^{(g)}_{\mu\nu}-\mathcal{T}_{\mu\nu}\right)\delta g^{\mu\nu}+\left(E^{(\Phi)}-\mathcal{J}\right)\delta \Phi +\frac{1}{\sqrt{-g}}\frac{\delta S_{\textrm{m}}}{\delta \Xi}\delta \Xi\right], 
\end{equation}
where the Euler-Lagrange expressions on the gravitational side, and the matter sources, are defined by
\begin{align}
    E^{(g)}_{\mu\nu} &\equiv \frac{2}{\sqrt{-g}}\dfrac{\delta S_\mathrm{grav}}{\delta g^{\mu\nu}}
    =\m^2\left[F(\Phi) G_{\mu\nu}-(\nabla_\mu\nabla_\nu-g_{\mu\nu}\Box)F(\Phi)\right]-\nabla_\mu\Phi\nabla_\nu\Phi+\left[\dfrac{1}{2}g^{\rho\sigma}\nabla_\rho\Phi\nabla_\sigma\Phi+V(\Phi)\right]g_{\mu\nu}~,
    \label{eq:Eofg}\\
    E^{(\Phi)} &\equiv \frac{1}{\sqrt{-g}}\dfrac{\delta S_\mathrm{grav}}{\delta\Phi}=\Box\Phi-V_{,\Phi}(\Phi)+\m^2\dfrac{F_{,\Phi}(\Phi)}{2}R~,
    \label{eq:Eofg-DEFINITION}\\
    \mathcal{J} &\equiv -\frac{1}{\sqrt{-g}}\frac{\delta S_{\textrm{m}}}{\delta \Phi}~.
    \label{eq:source-term}
\end{align}
Here
\begin{equation}
\Box \equiv g^{\mu\nu}\nabla_{\mu} \nabla_{\nu}
\end{equation}
is the d'Alembertian, and
\begin{equation}
G_{\mu\nu} \equiv R_{\mu\nu}-\frac{1}{2} g_{\mu\nu}R
\end{equation}
is the Einstein tensor, which satisfies the contracted Bianchi identity $\nabla_{\nu} G^{\mu\nu}=0$ \cite{Carroll:1997ar}, where $\nabla_{\mu}$ is the covariant derivative associated with the Levi-Civita connection of the defining-frame metric $g_{\mu\nu}$. The matter EMT $\mathcal{T}_{\mu\nu}$ is again given by Eq.~\eqref{eq:energy-momentum-tensor}, but now taking into account that the matter action \emph{does} depend on the scalar field $\Phi$, since interactions between $\Phi$ and the matter fields are being considered. Stationarity of $S$ under arbitrary $\delta g^{\mu\nu}$, $\delta \Phi$, and $\delta \Xi$, yields the classical field equations
\begin{eqnarray}
    E^{(g)}_{\mu\nu}&=& \mathcal{T}_{\mu\nu}~,
    \label{eq:Eg-equation-classical}\\
    E^{(\Phi)} &=& \mathcal{J}~,
    \label{eq:Ephi-equation-classical}\\
    \frac{\delta S_{\textrm{m}}}{\delta \Xi}&=&0~.
    \label{eq:MATTER-equation-classical}
\end{eqnarray}

From Eqs.~\eqref{eq:Eofg-DEFINITION} and \eqref{eq:Eg-equation-classical} one readily verifies that
\begin{equation}
    \label{eq:stress-energy conservation}\nabla_{\mu}\tensor{\mathcal{T}}{^{\mu}_{\nu}} = -E^{(\Phi)}\nabla_{\nu}\Phi~,
\end{equation}
so that the EMT $\mathcal{T}_{\mu\nu}$ is conserved when $E^{(\Phi)} = 0$.\footnote{From Eq.~\eqref{eq:stress-energy conservation}, the EMT is also conserved when $\Phi=\Phi_0 \equiv \mathrm{const.}$, in which case the theory reduces to standard General Relativity (GR) with an effective cosmological constant, $\Lambda_\mathrm{eff}\equiv V(\Phi_0)$ \cite{Clifton:2011jh}.} However, the explicit dependence of the action functional $S_{\textrm{m}}$ on $\Phi$ results in an exchange of energy and momentum between the matter sector and the scalar field, as seen in Eq.~\eqref{eq:Ephi-equation-classical}. Hence,
\begin{equation}
    \label{eq:exchange-matter-phi-dissipation-explicit}\nabla_{\mu}\tensor{\mathcal{T}}{^{\mu}_{\nu}} = -\mathcal{J} \nabla_{\nu}\Phi~.
\end{equation}
The metric and scalar-field equations \eqref{eq:Eg-equation-classical} and \eqref{eq:Ephi-equation-classical} can then be written as  
\begin{eqnarray}
    FG_{\mu\nu}-\left(\nabla_{\mu}\nabla_{\nu}-g_{\mu\nu}\Box\right)F&=&\m^{-2}\left(\mathcal{T}_{\mu\nu}+T_{\mu\nu}^{\Phi}\right),
    \label{eq:metric-fieldeqs-defframe}\\
    \Box \Phi -V_{,\Phi}+\frac{\m^2}{2}F_{,\Phi}R &=& \mathcal{J}~,
    \label{eq:sf-equation-covariant}
\end{eqnarray}
where $T_{\mu\nu}^{\Phi}$ is the scalar field EMT:
\begin{equation}
    \label{eq:Defining-frame-field}T_{\mu\nu}^{\Phi} \equiv \nabla_{\mu}\Phi \nabla_{\nu} \Phi -g_{\mu\nu}\left(\frac{1}{2} g^{\rho\sigma} \nabla_{\rho} \Phi \nabla_{\sigma} \Phi+V\right).
\end{equation}
Equivalently, Eq.~\eqref{eq:metric-fieldeqs-defframe} can be recast as the Einstein field equations by moving the geometric terms involving $F$ to the right-hand side. The resulting effective source admits a description as an `imperfect' fluid \cite{Faraoni:2018qdr}.

On the other hand, the field equations for the matter fields are those given in Eq.~\eqref{eq:MATTER-equation-classical}. For a coarse-grained description of the matter sector under the assumption of local isotropy and homogeneity, we may adopt a perfect-fluid form for the EMT \cite{Maartens:1996vi}:
\begin{equation}
  \label{eq:perfect-fluid-form-emtensor}\mathcal{T}_{\mu\nu}\equiv \rho_{\textrm{m}} u_{\mu} u_{\nu}+ P_{\textrm{m}}h_{\mu\nu}~, 
\end{equation}
where $u_{\mu}$ is the normalised four-velocity field of the matter fluid, satisfying $u_{\mu} u^{\mu} = -1$, and $h_{\mu\nu}\equiv g_{\mu\nu}+u_{\mu}u_{\nu}$ are the covariant components of the projection tensor onto spatial hypersurfaces orthogonal to $u^{\mu}$. The normalisation condition implies that $u^{\mu}$ is metric- (and hence frame-) dependent.\footnote{\label{footnote:fourveloc}The coordinate components of the four-velocity (unit time-like vector) are 
\begin{equation}
    \label{eq:four-velocity-generaldef}u^{\mu} \equiv \frac{1}{\sqrt{-g_{\rho\sigma}\frac{\textrm{d}x^{\rho}}{\textrm{d}\lambda}\frac{\textrm{d}x^{\sigma}}{\textrm{d}\lambda}}}\frac{\textrm{d}x^{\mu}}{\textrm{d}\lambda}~,
\end{equation}
where $\lambda$ is an affine parameter. While the coordinates $x^{\mu}$ and the parameter $\lambda$ are unchanged under a conformal transformation, the metric components enter explicitly through the infinitesimal proper interval $\sqrt{ -g_{\mu\nu}\textrm{d}x^{\mu}\textrm{d}x^{\nu}}$, so that $u^{\mu}$ does transform non-trivially under conformal rescalings \cite{Chiba:2013mha}.} Using the covariant conservation equation~\eqref{eq:exchange-matter-phi-dissipation-explicit}, we obtain
\begin{equation}
    \label{eq:conservation-eq-mattersector}u^{\mu} u^{\nu}\nabla_{\nu}\rho_{\textrm{m}}+h^{\mu\nu}\nabla_{\nu} P_{\textrm{m}}+\left(\Theta u^{\mu}+a^{\mu}\right)\left(\rho_{\textrm{m}}+P_{\textrm{m}}\right) = -\mathcal{J}\nabla^{\mu} \Phi~,
\end{equation}
with $\rho_{\textrm{m}}\equiv u^{\mu}u^{\nu}\mathcal{T}_{\mu\nu}$ being the energy density, $P_{\textrm{m}}\equiv \frac{1}{3}h^{\mu\nu}\mathcal{T}_{\mu\nu}$ the (locally) isotropic pressure, $\Theta \equiv \nabla_{\mu} u^{\mu}$ the scalar expansion rate, and $a^{\mu}\equiv u^{\nu}\nabla_{\nu} u^{\mu}$ the four-acceleration. Projecting Eq.~\eqref{eq:conservation-eq-mattersector} along $u^{\mu}$ and onto the space-like hypersurfaces orthogonal to $u^{\mu}$, and exploiting the normalisation condition $u_{\mu}u^{\mu}=-1$, together with $u_{\mu}a^{\mu}=0$ and $u^{\mu} h_{\mu\nu}=0$, we arrive at the continuity and Euler equations:
\begin{eqnarray}
    u^{\mu}\nabla_{\mu} \rho_{\textrm{m}}+\Theta \left(\rho_{\textrm{m}}+P_{\textrm{m}}\right) &=& \mathcal{J}u^{\mu}\nabla_{\mu} \Phi~,
    \label{eq:continuity-matter}\\
    h^{\mu\nu}\nabla_{\nu}P_{\textrm{m}}+a^{\mu}\left(\rho_{\textrm{m}}+P_{\textrm{m}}\right)&=&-\mathcal{J}h^{\mu\nu}\nabla_{\nu}\Phi~,
    \label{eq:Euler-matter}
\end{eqnarray}
respectively.

Equations~\eqref{eq:metric-fieldeqs-defframe}, \eqref{eq:sf-equation-covariant}, \eqref{eq:continuity-matter}, and \eqref{eq:Euler-matter} hold in the defining frame; their Einstein-frame counterparts are obtained via the conformal transformation of Eq.~\eqref{eq:conformal-Einstein}, yielding
\begin{eqnarray}
    \label{eq:metric-rescaled-Einstein}\m^2\hat{G}_{\mu\nu} &=& \hat{\mathcal{T}}_{\mu\nu}+\hat{T}^{\Phi}_{\mu\nu}~,\\
    \label{eq:scalar-rescaled-Einstein}\hat{\Box} \Phi -\hat{\nabla}_{\mu}\ln F \hat{\nabla}^{\hat{\mu}}\Phi -\frac{V_{,\Phi}}{F}+\frac{\m^2}{2}F_{,\Phi} \left(\hat{R}+3\hat{\Box}\ln F -\frac{3}{2}\hat{\nabla}_{\mu}\ln F \hat{\nabla}^{\hat{\mu}}\ln F \right)&=&\frac{\mathcal{J}}{F}~,\\
    \label{eq:continuity-rescaled-Einstein}\hat{u}^{\mu}\hat{\nabla}_{\mu}\hat{\rho}_{\textrm{m}}+\hat{\Theta}\left(\hat{\rho}_{\textrm{m}}+\hat{P}_{\textrm{m}}\right)+\frac{1}{2}\left(\hat{\rho}_{\textrm{m}}-3\hat{P}_{\textrm{m}}\right)\hat{u}^{\mu}\hat{\nabla}_{\mu}\ln F &=& \frac{\mathcal{J}}{F^2}\hat{u}^{\mu}\hat{\nabla}_{\mu}\Phi~,\\
    \label{eq:Euler-rescaled-Einstein}\hat{h}^{\mu\nu}\hat{\nabla}_{\nu}\hat{P}_{\textrm{m}} +\hat{a}^{\mu}\left(\hat{\rho}_{\textrm{m}}+\hat{P}_{\textrm{m}}\right)-\frac{1}{2}\left(\hat{\rho}_{\textrm{m}}-3\hat{P}_{\textrm{m}}\right)\hat{h}^{\mu\nu}\hat{\nabla}_{\nu}\ln F&=& -\frac{\mathcal{J}}{F^2}\hat{h}^{\mu\nu}\hat{\nabla}_{\nu}\Phi~,
\end{eqnarray}
where 
\begin{eqnarray}
    \hat{\mathcal{T}}_{\mu\nu}&\equiv& \frac{1}{F}\mathcal{T}_{\mu\nu}~,
    \label{eq:rescaling-energymomentumtensor}\\
    \hat{T}^{\Phi}_{\mu\nu} &\equiv& \frac{1}{F}T^{\Phi}_{\mu\nu}+\frac{3\m^2}{2}\left(\hat{\nabla}_{\mu}\ln F\hat{\nabla}_{\nu}\ln F-\frac{1}{2}\hat{g}_{\mu\nu}\hat{g}^{\rho\sigma}\hat{\nabla}_{\rho}\ln F \hat{\nabla}_{\sigma}\ln F\right),
\end{eqnarray}
with 
\begin{equation}
    \hat{\Box} \equiv \hat{\nabla}_{\mu}\hat{\nabla}^{\hat{\mu}} = \hat{g}^{\mu\nu}\hat{\nabla}_{\mu}\hat{\nabla}_{\nu}
\end{equation}
being the d'Alembertian associated with the hatted metric $\hat{g}_{\mu\nu}$, and\footnote{\label{footnote:clarification-trans-mattersector}These transformations may alternatively be derived from Eqs.~\eqref{eq:energy-momentum-tensor} and \eqref{eq:trans-mathcalL-m-hat} (\textit{i.e.} by accompanying the metric rescaling with field redefinitions of the matter fields such that the matter action $S_{\textrm{m}}$ retains its functional form in the new variables), such that
\begin{equation}
    \hat{\mathcal{T}}_{\mu\nu} = -\frac{2}{\sqrt{-\hat{g}}}\frac{\delta \hat{S}_{\textrm{m}}}{\delta \hat{g}^{\mu\nu}} = \frac{1}{F} \mathcal{T}_{\mu\nu}~.
\end{equation}
For $u^{\mu}$, the definition in Eq.~\eqref{eq:four-velocity-generaldef} gives $u^{\mu} \rightarrow F^{-1/2} u^{\mu}$, as explained in the footnote. The transformations of $\rho_{\textrm{m}}$ and $P_{\textrm{m}}$ in Eqs.~\eqref{eq:rescale-rhom} and \eqref{eq:rescale-pm} follow straightforwardly from $\rho_{\textrm{m}} \equiv u^{\mu} u^{\nu} \mathcal{T}_{\mu\nu}$ and $P_{\textrm{m}}\equiv \frac{1}{3}h^{\mu\nu}\mathcal{T}_{\mu\nu}$, introduced below Eq.~\eqref{eq:conservation-eq-mattersector}.}
\begin{eqnarray}
    \hat{\rho}_{\textrm{m}}&=&F^{-2}\rho_{\textrm{m}}~,
    \label{eq:rescale-rhom}\\
    \hat{P}_{\textrm{m}}&=&F^{-2}P_{\textrm{m}}~,
    \label{eq:rescale-pm}\\
    \hat{u}^{\mu} &=& F^{-1/2}u^{\mu}~,
    \label{eq:rescale-velocity}\\
    \hat{a}^{\mu} &=& F^{-1}\left(a^{\mu}+\frac{1}{2}h^{\mu\nu}\nabla_{\nu}\ln F\right),
    \label{eq:rescale-acceleration}
\end{eqnarray}
together with $\hat{h}^{\mu\nu} = F^{-1}h^{\mu\nu}$. The hatted Ricci scalar and Einstein tensors read \cite{Dabrowski:2008kx}
\begin{eqnarray}
    \hat{R} &=& F^{-1}\left[R-3\left(\Box \ln F +\frac{1}{2}g^{\mu\nu}\nabla_{\mu}\ln F\nabla_{\nu}\ln F\right)\right],\\
    \hat{G}_{\mu\nu} &=& G_{\mu\nu}-\nabla_{\mu}\nabla_{\nu}\ln F +\frac{1}{2}\nabla_{\mu}\ln F \nabla_{\nu}\ln F +\frac{1}{2}g_{\mu\nu}\left(2\Box \ln F +\frac{1}{2}g^{\sigma\rho}\nabla_{\sigma}\ln F \nabla_{\rho}\ln F\right),
\end{eqnarray}
respectively. The scalar expansion rate transforms accordingly as
\begin{equation}
    \hat{\Theta}\equiv \hat{\nabla}_{\mu}\hat{u}^{\mu} = \frac{1}{\sqrt{F}}\left(\Theta+\frac{3}{2}u^{\mu}\nabla_{\mu}\ln F\right).
\end{equation}
With the exception of the metric field equations \eqref{eq:metric-rescaled-Einstein}, the remaining ones are not simplified by the conformal transformation, despite the metric redefinition. This is to be expected: as already noted, the Einstein frame is the one in which the gravitational sector takes on the Einstein-Hilbert form, while the matter and scalar sectors generally acquire non-minimal couplings and a non-canonical kinetic term, respectively, and therefore do not simplify in tandem. Equations~\eqref{eq:continuity-rescaled-Einstein} and \eqref{eq:Euler-rescaled-Einstein} further show that radiation, for which $P_{\textrm{m}}/\rho_{\textrm{m}} = \hat{P}_{\textrm{m}}/\hat{\rho}_{\textrm{m}}=1/3$, is frame-insensitive at the level of the conservation equations.  

In addition to the conformal transformation of the metric, we may eliminate the hatted Ricci scalar from Eq.~\eqref{eq:scalar-rescaled-Einstein} by means of the trace of Eq.~\eqref{eq:metric-rescaled-Einstein}, obtaining
\begin{equation}
    \label{eq:intermediate-form-sfeqs}\hat{\Box}\Phi -F\left(\frac{V}{F^2}\right)_{,\Phi}+\frac{F_{,\Phi}}{2}\left(\hat{\rho}_{\textrm{m}}-3\hat{P}_{\textrm{m}}\right)+\frac{3\m^2}{2}F_{,\Phi}\hat{\Box}\ln F -\frac{F_{,\Phi}}{2F}\hat{\nabla}_{\mu}\Phi \hat{\nabla}^{\hat{\mu}}\Phi=\frac{\mathcal{J}}{F}~, 
\end{equation}
where $\hat{\mathcal{T}} \equiv \hat{g}^{\mu\nu}\hat{\mathcal{T}}_{\mu\nu} = -\hat{\rho}_{\textrm{m}}+3\hat{P}_{\textrm{m}}$ and $\hat{\nabla}^{\hat{\mu}} \equiv \hat{g}^{\mu\nu}\hat{\nabla}_{\nu}$.\footnote{\label{footnote:validity-of-some-expressions}This expression for the trace remains valid for an imperfect-fluid EMT, since both the energy flux (orthogonal to the four-velocity) and the anisotropic stress (traceless) contribute zero.} The scalar-field redefinition in Eq.~\eqref{eq:sf-redefinition}, along with Eqs.~\eqref{eq:non-canonical-function} and \eqref{eq:potentialU}, then yields
\begin{eqnarray}
    \m^2 \hat{G}_{\mu\nu} &=& \hat{\mathcal{T}}_{\mu\nu}+\hat{T}^{\hphi}_{\mu\nu}~,
    \label{eq:METRIC-eq-rescaledandredefined}\\
    \hat{\Box}\hphi -U_{,\hphi} &=&\frac{1}{F}\left[\frac{\mathcal{J}}{F\sqrt{\mathcal{K}}}-\frac{F_{,\hphi}}{2}\left(\hat{\rho}_{\textrm{m}}-3\hat{P}_{\textrm{m}}\right)\right],
    \label{eq:sf-eq-rescaledandredefined}\\
    \hat{u}^{\mu}\hat{\nabla}_{\mu}\hat{\rho}_{\textrm{m}}+\hat{\Theta}\left(\hat{\rho}_{\textrm{m}}+\hat{P}_{\textrm{m}}\right) &=& \frac{1}{F}\left[\frac{\mathcal{J}}{F\sqrt{\mathcal{K}}}-\frac{F_{,\hphi}}{2}\left(\hat{\rho}_{\textrm{m}}-3\hat{P}_{\textrm{m}}\right)\right]\hat{u}^{\mu}\hat{\nabla}_{\mu}\hphi~,
    \label{eq:continuity-eq-rescaledandredefined}\\
    \hat{h}^{\mu\nu}\hat{\nabla}_{\nu}\hat{P}_{\textrm{m}}+\hat{a}^{\mu}\left(\hat{\rho}_{\textrm{m}}+\hat{P}_{\textrm{m}}\right)&=& -\frac{1}{F}\left[\frac{\mathcal{J}}{F\sqrt{\mathcal{K}}}-\frac{F_{,\hphi}}{2}\left(\hat{\rho}_{\textrm{m}}-3\hat{P}_{\textrm{m}}\right)\right]\hat{h}^{\mu\nu}\hat{\nabla}_{\nu}\hphi~.
    \label{eq:Euler-eq-rescaledandredefined}
\end{eqnarray}
The potential $U$, the non-minimal coupling function $F$, and the kinetic function $\mathcal{K}$ are now regarded as functions of the redefined field $\hphi$ since $\Phi = \Phi(\hphi)$. The scalar-field EMT,
\begin{equation}
    \hat{T}^{\hphi}_{\mu\nu} \equiv \hat{\nabla}_{\mu}\hphi \hat{\nabla}_{\nu}\hphi -\hat{g}_{\mu\nu}\left(\frac{1}{2}\hat{g}^{\rho\sigma}\hat{\nabla}_{\rho}\hphi \hat{\nabla}_{\sigma}\hphi +U\right),
\end{equation}
takes its canonical form, in agreement with Eq.~\eqref{eq:Defining-frame-field} in the special case in which $\Phi$ is itself canonical. Using Eqs.~\eqref{eq:sf-eq-rescaledandredefined} and \eqref{eq:continuity-eq-rescaledandredefined}, the projection of the contracted Bianchi identity along $\hat{u}^{\mu}$, \textit{i.e.} $\hat{u}^{\nu}\hat{g}^{\mu\sigma}\hat{\nabla}_{\sigma}\hat{G}_{\mu\nu} = 0$, may be verified directly from $\hat{u}^{\nu}\hat{g}^{\mu\sigma}\hat{\nabla}_{\sigma}\hat{T}^{\hphi}_{\mu\nu} = \left(\hat{\Box}\hphi - U_{,\hphi}\right) \hat{u}^{\mu}\hat{\nabla}_{\mu}\hphi$, and $\hat{u}^{\nu}\hat{g}^{\mu\sigma}\hat{\nabla}_{\sigma}\hat{\mathcal{T}}_{\mu\nu} = -\hat{u}^{\mu}\hat{\nabla}_{\mu}\hat{\rho}_{\textrm{m}}-\hat{\Theta}\left(\hat{\rho}_{\textrm{m}}+\hat{P}_{\textrm{m}}\right)$.\footnote{The Einstein tensor is identically divergence-free under its associated Levi-Civita connection. Since any metric, including those conformally-related, admits its own Levi-Civita connection, the same property holds for $\hat{G}_{\mu\nu}$ with respect to $\hat{g}_{\mu\nu}$.}

\section{\label{sec:nonminimally-coupled-warm-inflation}Non-Minimally Coupled Warm Inflation}
\subsection{\label{sec:the_eqs_WI_effective}The Effective Equations of Warm Inflation}
\noindent
In the previous section, we derived general equations for the system in which the scalar field $\Phi$ couples non-minimally to the fields $\Xi$ of the matter sector. Henceforth, a spatially flat Friedmann-Lema\^{i}tre-Robertson-Walker (FLRW) spacetime is assumed, with line element
\begin{equation}
    \label{eq:line-element-flrw-definingframe-anyframereally}
    \d s^2 = -\N^2 \d t^2 + a^2(t)\delta_{ij}\d x^i \d x^j~,
\end{equation}
where $\mathcal{N}$ is the lapse function, $t$ and $x^i$ denote the temporal and spatial coordinates, and $a(t)$ is the homogeneous scale factor. The proper time $\d\tau$ of an observer comoving with the above slicing (the normal observer)\footnote{As defined in Ref.~\cite{gourgoulhon:2012book}, a normal observer is one whose four-velocity coincides with the vector field normal to the spatial slicing. Such an observer is also sometimes referred to as an Eulerian or comoving observer.} is given by
\begin{equation}
    \d\tau = \N\d t~.
\end{equation}
The Hubble parameter associated with the constant-time hypersurfaces can be written as \cite{Karciauskas:2022jzd}
\begin{equation}
    H\equiv\frac{1}{a}\frac{\d a}{\d\tau}=\frac{\dot{a}}{\N a}~.
\end{equation}
In this work, we use overdots to indicate derivatives with respect to coordinate time $t$. For the background cosmology, we can decompose the scalar field $\Phi$ as
\begin{equation}
    \label{phi-dec}
    \Phi\left(t,\mathbf{x}\right) = \phi(t) + \delta\phi\left(t,\mathbf{x}\right)~.
\end{equation}
$\phi(t)$ is the homogeneous background field and $\delta\phi(x)$ its perturbation. The latter sources cosmological perturbations, whereas only the homogeneous field $\phi$ is relevant for the background dynamics considered in this section.

Interactions between the $\Phi$ field and the matter sector give rise to a non-vanishing source term $\mathcal{J}$ (see Eq.~\eqref{eq:source-term}), whose form depends on the underlying microphysics \cite{Berera:2008ar,Bastero-Gil:2014jsa,Ballesteros:2022hjk}. Quite generically, $\mathcal{J}$ may be approximated as
\begin{equation}
    \mathcal{J} = \Upsilon u^{\mu}\nabla_{\mu}\Phi-\xi_T~,\label{eq: source1}
\end{equation}
where $\Upsilon$ is the local dissipation coefficient, and $\xi_T$ is a stochastic noise term with variance
\begin{align}
    \langle \xi_T(t,\mathbf{x})\xi_T(t',\mathbf{x}')\rangle =\frac{2\Upsilon T}{a^{3}\mathcal{N}}\delta(t-t')\delta^{(3)}(\mathbf{x}-\mathbf{x}')~,
\end{align}
which only sources the perturbations, and $T$ denotes the temperature of the thermal bath. Within the framework of GR, the effective equations of motion for the scalar field have been studied extensively in the literature \cite{Gleiser:1993ea, Berera:1998gx, Berera:2004kc, Moss:2006gt,Berera:2007qm,Berera:2008ar}; for a more thorough exposition we refer the reader to Ref.~\cite{Berera:2008ar} and the textbook \cite{Calzetta:2008iqa}. For convenience, Appendix~\ref{app:dissipative-Minkowski-dyn} provides a brief review of the relevant formalism based on Ref.~\cite{Berera:2008ar}. Therein, the conditions required for the consistent implementation of warm inflation in scalar-tensor theories with a non-minimally coupled inflaton are also discussed.  

Since the effective Planck mass is time dependent in the modified gravity theories considered here, one might expect additional requirements for the validity of the equations presented in Appendix~\ref{app:dissipative-Minkowski-dyn}, which were originally derived within GR. In practice, however, only one extra condition arises, while the rest remain unchanged. Thermal effects dominate over the curvature of the spatial hypersurfaces provided that
\begin{equation}
\label{eq:thermal-domination-cond}T \gg H~,
\end{equation}
which is identical to the corresponding condition in GR. The other requirements ensure the consistency of the adiabatic approximation, and are given by
\begin{equation}
\label{eq:some-conditions-keepinmind}
\tau_i^{-1} \gg H~, \qquad \tau_i^{-1} \gg \frac{1}{\phi}\frac{\d\phi}{\d\tau}~, \qquad \tau_i^{-1} \gg \frac{1}{2F(\phi)}\frac{\d F(\phi)}{\d\tau}~.
\end{equation}
In these expressions, $\tau_i$ denotes the relaxation proper time of the species that mediate dissipation, as seen by the normal observer. The first two conditions coincide with the analogous expressions in GR \cite{Berera:2023liv}, and guarantee, respectively, that the thermal bath remains close to equilibrium and that the local approximation $\Upsilon \dot\phi$ for the dissipative term is valid (see Appendix~\ref{app:dissipative-Minkowski-dyn}). The third condition is new and arises from the modification of gravity. Noting that $(\d F(\phi)/\d\tau)/(2F(\phi)) = (\d\m^{\mathrm{eff}}/\d\tau)/\m^{\mathrm{eff}}$, where $\m^{\mathrm{eff}}(\phi)\equiv \sqrt{F(\phi)}\m$ is the effective reduced Planck mass (in Planck units), the third condition requires the effective Planck mass to evolve on a timescale much longer than the relaxation one. Physically, this ensures that any Planck-suppressed contributions to the dynamics remain negligible compared with the thermal contributions sourced by the matter fields at all times. Moreover, the third condition also guarantees that approximate thermal equilibrium can be achieved sufficiently fast. As shown in the following subsections, this requirement is automatically satisfied throughout slow-roll inflation. 

Finally, the self-interactions of the scalar field can in principle contribute to the self-energy, with the magnitude of this contribution determined by the size of the self-coupling \cite{Berera:2004kc}. If the scalar field is identified with the inflaton, observational constraints on the scalar perturbation power spectrum tightly bound this coupling (for instance, $\lambda \sim 10^{-13}$ in a quartic potential \cite{Bezrukov:2007ep,Lyth:2009zz,Rubio:2018ogq}) so that, in standard inflation scenarios, these effects are negligible. In standard warm inflation, the bound depends on the form of the dissipation coefficient $\Upsilon \propto T^{c}$; for the physically motivated cases $c\geq 0$, it is in fact tighter than in cold inflation by at least three orders of magnitude \cite{Montefalcone:2022owy}, with thermal contributions further suppressing the required value of $\lambda$ \cite{Berera:2023liv}.

In warm inflation, the field $\phi$ plays the role of the inflaton. During the inflationary phase, $\phi$ continuously dissipates its energy into radiation. Consequently, the dominant matter component is radiation, $\rho_{\textrm{m}} \equiv \rho_r$, with equation of state $P_{\textrm{m}} \equiv P_r = \rho_r/3$ \cite{Bartrum:2013fia,Bastero-Gil:2019gao}. This dissipation modifies the field dynamics. Using the source term $\mathcal{J}$ in Eq.~\eqref{eq: source1}, we readily see that Eqs.~\eqref{eq:metric-fieldeqs-defframe}, \eqref{eq:sf-equation-covariant}, \eqref{eq:continuity-matter}, and \eqref{eq:Euler-matter} reduce to
\begin{align}
    \label{eq:f-eq-DF-Q}&3H^2\m^2 = \frac{1}{\bfnmc}\left(\frac{\dot \phi^2}{2\mathcal{N}^2}+\bar{V}-3H\m^2\frac{\dot\bfnmc}{\mathcal{N}}+\bar{\rho}_r\right),\\
    \label{eq:tH-eq-DF-Q}&2\frac{\dot H}{\mathcal{N}}\m^2=-\frac{1}{\bfnmc}\left[\frac{\dot \phi^2}{\mathcal{N}^2}-\m^2\left(H+\frac{\dot{\mathcal{N}}}{\mathcal{N}^2}\right)\frac{\dot \bfnmc}{\mathcal{N}}+\m^2\frac{\ddot \bfnmc}{\mathcal{N}^2}+\frac{4}{3}\bar{\rho}_r\right],\\
    \label{eq:sf-eq-DF-Q}&\frac{\ddot \phi}{\mathcal{N}^2}+\left(3H+\bar{\Upsilon}-\frac{\dot{\mathcal{N}}}{\mathcal{N}^2}\right)\frac{\dot \phi}{\mathcal{N}}+\bar{V}_{,\phi}-3\m^2\bfnmc_{,\phi}\left(\frac{\dot H}{\mathcal{N}}+2H^2\right) = 0~,\\
    \label{eq:rad-eq-DF-Q}&\frac{\dot{\bar{\rho}}_r}{\mathcal{N}} =-4H \bar{\rho}_r +\bar{\Upsilon} \frac{\dot \phi^2}{\mathcal{N}^2}~,
\end{align}
where overbars denote background quantities (adopting the convention introduced in Appendix~\ref{sec:app-linearperturbations}), and we recall that $\phi=\phi(t)$ corresponds to the homogeneous mode of the scalar field, which is insensitive to the stochastic noise term present in the source $\mathcal{J}$, as mentioned before. In addition, $\bar{F}=F(\phi)$, $\bar{V}_{,\phi} \equiv \left.V_{,\Phi}\right|_{\Phi=\phi}$, and similarly for related quantities. Combining Eqs.~\eqref{eq:f-eq-DF-Q}, \eqref{eq:tH-eq-DF-Q}, and \eqref{eq:sf-eq-DF-Q} yields
\begin{equation}
    \label{eq:sf-eq-rewritten}\frac{\ddot \phi}{\mathcal{N}^2}+\left(3H+\frac{\bar{\Upsilon}}{\bfnmc\bar{\mathcal{K}}}+\frac{\dot{\bar{\mathcal{K}}}}{2\mathcal{N}\bar{\mathcal{K}}}+\frac{\dot \bfnmc}{\mathcal{N}\bfnmc}-\frac{\dot{\mathcal{N}}}{\mathcal{N}^2}\right)\frac{\dot \phi}{\mathcal{N}} = -\frac{\bfnmc}{\bar{\mathcal{K}}}\left(\frac{\bar{V}}{\bfnmc^2}\right)_{,\phi}~,
\end{equation}
where we have divided by $\bfnmc\bar{\mathcal{K}}$. This product is non-vanishing (in fact, strictly positive), as follows from Eq.~\eqref{eq:non-canonical-function}:
\begin{equation}
    \label{eq:just-a-reminder}\bfnmc\bar{\mathcal{K}} = 1+\frac{3\m^2 \bfnmc^2_{,\phi}}{2\bfnmc}~,
\end{equation}
and from the condition $\bfnmc >0$, which is required to avoid a graviton ghost \cite{Gannouji:2006jm,Ballardini:2023mzm,Jarv:2024krk,Jarv:2025qgo}.

The equations above are valid in the defining frame. To express them in the Einstein frame, we perform the conformal rescaling given in Eq.~\eqref{eq:conformal-Einstein} which transforms the FLRW line element \eqref{eq:line-element-flrw-definingframe-anyframereally} into
\begin{equation}
    \dif \hat{s}^2 = -\hat{\mathcal{N}}^2\dif t^2+\hat{a}^2(t)\delta_{ij}\dif x^i\dif x^j~.
\end{equation}
The hatted and unhatted quantities are related by
\begin{eqnarray}
    \hat{\mathcal{N}}&=&\sqrt{\bar{F}}\mathcal{N}~,
    \label{eq:transformation-background-lapse-conformal} \\
    \hat{a} &=& \sqrt{\bar{F}}a~.
    \label{eq:transformation-background-scale-factor}
\end{eqnarray}
Note that we keep the spacetime slicing unchanged; that is, the coordinate time $t$ is not redefined in the hatted frame. The proper time of the normal observer, however, \emph{does} transform according to \cite{Karciauskas:2022jzd}
\begin{equation}
    \d\hat{\tau} = \hat{\n} \d t = \sqrt{\bar{F}}\d\tau~.
    \label{hattau-def}
\end{equation}
The Einstein-frame Hubble parameter is related to its defining-frame counterpart via the relation \cite{Chiba:2013mha, Kuusk:2016rso, Karciauskas:2022jzd}
\begin{equation}
    \label{eq:conformal-transformation-Hubbleexprate}\hat{H} = \frac{\dot{\hat{a}}}{\hat{\mathcal{N}}\hat{a}} = \frac{H}{\sqrt{\bar{F}}}\left(1+\frac{\dot{\bar{F}}}{2H\mathcal{N}\bar{F}}\right).
\end{equation} 
As such, under this conformal transformation, the system of equations \eqref{eq:f-eq-DF-Q}--\eqref{eq:sf-eq-rewritten} takes the following form:
\begin{align}
    \label{eq:Friedmann-Einstein-frame}
    &\hat{H}^2= \frac{1}{3\m^2}\left[\frac{1}{2}\left(\frac{\dif\varphi}{\dif\hat{\tau}}\right)^2+U(\varphi)+\hat{\bar{\rho}}_r\right],\\
    &\frac{\dif\hat{H}}{\dif\hat{\tau}} = -\frac{1}{2\m^2}\left[\left(\frac{\dif\varphi}{\dif\hat{\tau}}\right)^2+\frac{4}{3}\hat{\bar{\rho}}_r\right],\label{eq: Hdot_df} \\
    &\frac{\dif^2\varphi}{\dif\hat{\tau}^2}+\left(3\hat{H}+\frac{\bar{\Upsilon}}{\bar{\mathcal{K}}\bar{F}^{3/2}}\right)\frac{\dif\varphi}{\dif\hat{\tau}}+U_{,\varphi}(\varphi)=0~,\label{eq: varphi_2}\\
    \label{eq:rho_r_2}&\frac{\dif\hat{\bar{\rho}}_r}{\dif\hat{\tau}}=-4\hat{H}\hat{\bar{\rho}}_r+\frac{\bar{\Upsilon}}{\bar{\mathcal{K}}\bar{F}^{3/2}}\left(\frac{\dif\varphi}{\dif\hat{\tau}}\right)^2~.
\end{align}
In the above expression we have also defined the field $\varphi$, which is the background component of the canonically normalised field $\tilde{\Phi}$ in Eq.~\eqref{eq:sf-redefinition}:
\begin{equation}
    \dif \varphi \equiv \sqrt{\bar{\mathcal{K}}}\dif \phi~.
    \label{varphi-def}
\end{equation}

Equations~\eqref{eq:Friedmann-Einstein-frame}--\eqref{eq:rho_r_2} appear to be of a similar form to the analogous warm inflation equations in GR \cite{Berera:1995ie,Berera:2008ar}
, with the one crucial difference being the dissipation term. As can be seen in Eqs.~\eqref{eq: varphi_2} and \eqref{eq:rho_r_2} the dissipation term is divided by a factor whose origin is the modification of the Einstein-Hilbert action. To make later equations more concise we define
\begin{equation}
    \tilde{\bar{\Upsilon}} \equiv \frac{\bar{\Upsilon}}{\bar{\mathcal{K}}\bar{F}^{3/2}}~.\label{eq: tilde_dc}
\end{equation}
It is tempting to interpret this term as the dissipation coefficient in the Einstein frame. That is, if one started from the action in the Einstein frame and computed the dissipation coefficient from first principles, one might expect the resulting $\hat{\bar{\Upsilon}}$ to be related to the defining-frame dissipation term $\bar{\Upsilon}$ through Eq.~\eqref{eq: tilde_dc}, with $\tilde{\bar{\Upsilon}}=\hat{\bar{\Upsilon}}$. However, we do not claim that this must be the case, since in our approach the dissipation coefficient is computed only in the defining frame and subsequently transformed into the Einstein frame. This is the reason why we do not denote the left-hand side of Eq.~\eqref{eq: tilde_dc} by $\hat{\bar{\Upsilon}}$. In fact, the calculation of $\hat{\bar{\Upsilon}}$ might be very difficult. In the current approach, the scalar field $\Phi$ in Eq.~\eqref{eq:the_action_nonminimal} is assumed to have direct couplings to a limited number of matter fields only. This allows us to use the standard techniques of thermal field theory to compute dissipation effects. In contrast, for the action written in the Einstein frame, $\Phi$ (or equivalently $\tilde{\Phi}$) couples to \emph{all} matter fields. Accounting for all these additional interactions in full generality may be particularly challenging, if not impossible. Moreover, such a computation would have to confront the issue of frame equivalence at the quantum level, which is still an unsolved issue \cite{Kamenshchik:2014waa,Pandey:2016unk,Falls:2018olk}. Therefore, we keep (unhatted) $\bar{\Upsilon}$ in Eqs.~\eqref{eq:Friedmann-Einstein-frame}--\eqref{eq:rho_r_2}, and at most define $\tilde{\bar{\Upsilon}}$ as in Eq.~\eqref{eq: tilde_dc}. The original $\bar{\Upsilon}$ is a scalar field on the spacetime manifold, and as such is well defined whichever frame we choose to work with. In spite of this, a crucial point to bear in mind is that $\bar{\Upsilon}$ can be assigned the physical interpretation of a dissipation coefficient only in the defining frame.

\subsection{Slow-Roll Conditions in the Einstein Frame}
\noindent
While thermal effects are more readily computed in the defining frame, the analysis of the inflationary dynamics is more transparent in the Einstein frame. Our approach is similar to the one adopted in Ref.~\cite{Karciauskas:2022jzd}. We start by writing the Hubble-flow parameters \cite{Liddle:1994dx,Leach:2002ar}. In terms of the proper time of the normal observer (see Eq.~\eqref{hattau-def}) they are given by
\begin{equation}
    \hat{\epsilon}_1 \equiv -\frac{\dif\hat{H}/\dif \hat{\tau}}{\hat{H}^2}~, \qquad \hat{\epsilon}_{i+1} \equiv\frac{\dif\hat{\epsilon}_i/\dif\hat{\tau}}{\hat{H}\hat{\epsilon}_i}~, \qquad i=1,2,\ldots~. \label{eq: slow-roll-cond_1}
\end{equation}
In slow-roll inflation, the above parameters are required to be much smaller than unity, at least for the first few values of $i$:
\begin{equation}
    \label{sr-hf}
    \left|\hat{\epsilon}_i\right| \ll 1~.
\end{equation}
If such an evolution is caused by the scalar field dynamics, we can write the slow-roll conditions in terms of the scalar-field potential $U$, as defined in Eq.~\eqref{eq:potentialU}. In that case, Eqs.~\eqref{eq:Friedmann-Einstein-frame}--\eqref{eq:rho_r_2} can also be simplified by neglecting higher-order time derivatives. In particular, upon imposing the conditions
\begin{equation}
    \frac{1}{2}\left(\frac{\dif \varphi}{\dif \hat{\tau}}\right)^2+\hat{\bar{\rho}}_r \ll \bar{U}(\varphi)~, \qquad \left|\frac{\dif^2\varphi/\dif\hat{\tau}^2}{(3\hat{H}+\tilde{\bar{\Upsilon}})\dif \varphi/\dif \hat{\tau}}\right|\ll 1~, \qquad \left|\frac{\dif \hat{\bar{\rho}}_r/\dif \hat{\tau}}{\hat{H}\hat{\bar{\rho}}_r}\right|\ll 1~, \label{eq: slow-roll-cond_2}
\end{equation}
with $\tilde{\bar{\Upsilon}}$ defined as in Eq.~\eqref{eq: tilde_dc}, Eqs.~\eqref{eq:Friedmann-Einstein-frame}--\eqref{eq:rho_r_2} reduce to
\begin{eqnarray}
    \hat{H}^2&\simeq& \frac{\bar{U}}{3\m^2}~,\label{eq: Slow_hubble_1}\\
    \frac{\dif \hat{H}}{\dif \hat{\tau}}&\simeq& -\frac{1}{2\m^2}\left(\frac{\dif\varphi}{\dif\hat{\tau}}\right)^2\left(1+\frac{\tilde{\bar{\Upsilon}}}{3\hat{H}}\right),\label{eq: Slow_dhubble_1} \\
    \left(3\hat{H}+\tilde{\bar{\Upsilon}}\right)\frac{\dif\varphi}{\dif\hat{\tau}} &\simeq& -\bar{U}_{,\varphi}~,\label{eq: Slow_varphi_1}\\
    \hat{\bar{\rho}}_r &\simeq& \frac{\tilde{\bar{\Upsilon}}}{4\hat{H}}\left(\frac{\dif\varphi}{\dif\hat{\tau}}\right)^2~,\label{eq:rad_slowrolllll}
\end{eqnarray}
where Eq.~\eqref{eq:rad_slowrolllll} has been used to eliminate $\hat{\bar{\rho}}_r$ in Eq.~\eqref{eq: Hdot_df}. 
Plugging Eqs.~\eqref{eq: Slow_hubble_1}--\eqref{eq: Slow_varphi_1} into the expression of $\hat{\epsilon}_1$ in Eq.~\eqref{eq: slow-roll-cond_1}, the first Hubble-flow parameter takes on the form
\begin{equation}
    \label{eq:the-below-equation}\hat{\epsilon}_1 \simeq \frac{1}{2\hat{H}^2\m^2}\left(\frac{\dif \varphi}{\dif \hat{\tau}}\right)^2\left(1+\tilde{Q}\right)\simeq \frac{\m^2}{2}\left(\frac{\bar{U}_{,\varphi}}{\bar{U}}\right)^2\frac{1}{1+\tilde{Q}}~,  
\end{equation}
with $\tilde{Q}$ defined as the dimensionless ratio 
\begin{equation}
  \tilde{Q}\equiv \frac{\tilde{\bar{\Upsilon}}}{3\hat{H}}~.  
  \label{tildeQ-def}
\end{equation}
This parameter quantifies the strength of the dissipation. Spatial slices are expanding in an accelerating fashion if $\hat{\epsilon}_1<1$.

For slow-roll inflation, it is convenient to define another slow-roll parameter \cite{Liddle:1994dx}, namely
\begin{equation}
    \label{eq:slow-rollepsu}\epsilon_U \equiv \frac{\m^2}{2}\left(\frac{\bar{U}_{,\varphi}}{\bar{U}}\right)^2~. 
\end{equation}
If the conditions in Eq.~\eqref{sr-hf} are satisfied, then the bound 
\begin{equation}
    \epsilon_U \ll 1+ \tilde{Q}~,
\end{equation}
must also hold. This expression looks the same as the one used in models of slow-roll warm inflation in GR \cite{Berera:2008ar,Bastero-Gil:2011rva,Bartrum:2013fia,Bastero-Gil:2019gao}. Notice that this bound is more stringent than the one obtained from the first condition in Eq.~\eqref{eq: slow-roll-cond_2}. In particular, using the first condition in Eq.~\eqref{eq: slow-roll-cond_2}, which can be written as $(\dif \varphi/\dif \hat{\tau})^2 \ll 2\bar{U}$, we find a less restrictive constraint, $\epsilon_U\ll(1+\tilde{Q})^2$.

Similarly, we can express the second Hubble-flow parameter $\hat{\epsilon}_2$ in terms of the scalar-field potential. Substituting Eq.~\eqref{eq: Slow_dhubble_1} into Eq.~\eqref{eq: slow-roll-cond_1}, one finds
\begin{equation}
    \hat{\epsilon}_2 \equiv \frac{\dif \hat{\epsilon}_1/\dif \hat{\tau}}{\hat{H}\hat{\epsilon}_1} = \frac{\dif^2\hat{H}}{\dif\hat{\tau}^2}\left(\hat{H}\frac{\dif\hat{H}}{\dif\hat{\tau}}\right)^{-1}+2\hat{\epsilon}_1 \simeq 2\frac{\dif^2\varphi}{\dif\hat{\tau}^2}\left(\hat{H}\frac{\dif\varphi}{\dif\hat{\tau}}\right)^{-1}+\frac{1}{1+\tilde{Q}}\left(\frac{1}{\hat{H}}\frac{\dif \tilde{Q}}{\dif \hat{\tau}}+2\epsilon_U\right),
\end{equation}
which, by means of Eq.~\eqref{eq: Slow_varphi_1}, simplifies to
\begin{equation}
     \hat{\epsilon}_2\simeq \frac{4\epsilon_U-2\eta_U}{1+\tilde{Q}}-\frac{\dif\tilde{Q}/\dif\hat{\tau}}{\hat{H}(1+\tilde{Q})}~,\label{eq: the-below-equation-2}
\end{equation}
where the second potential slow-roll parameter is given by
\begin{equation}
    \label{eq:slow-rolletau}\eta_U \equiv \m^2 \frac{\bar{U}_{,\varphi\varphi}}{\bar{U}}~.
\end{equation}
As the bound in Eq.~\eqref{sr-hf} implies, the second Hubble-flow parameter must also satisfy  $\left|\hat{\epsilon}_2\right|\ll1$.\footnote{In contrast to $\hat{\epsilon}_1$, which is positive for universes without phantom energy \cite{Bronnikov:2006pt,Gannouji:2006jm,Ballardini:2023mzm}, $\hat{\epsilon}_2$ can be negative.} This guarantees that inflation lasts long enough to solve the flatness and horizon problems of the Hot Big Bang model. Up to possible cancellations between $\epsilon_U$ and $\eta_U$, which we do not take into account here, the latter bound yields the following two conditions:
\begin{equation}
    |\eta_U| \ll 1+\tilde{Q}~, \qquad \left|\frac{1}{\hat{H}}\frac{\dif \tilde{Q}}{\dif \hat{\tau}}\right| \ll 1+\tilde{Q}~. \label{eq: new_slowroll}
\end{equation}
Combining the third condition in Eq.~\eqref{eq: slow-roll-cond_2} with the slow-roll relation in Eq.~\eqref{eq:rad_slowrolllll}, we can show that
\begin{equation}
    \left|\frac{\dif \hat{\bar{\rho}}_r/\dif \hat{\tau}}{\hat{H}\hat{\bar{\rho}}_r}\right| \simeq \left|\frac{\dif \tilde{\bar{\Upsilon}}/\dif \hat{\tau}}{\hat{H}\tilde{\bar{\Upsilon}}}+2\frac{\dif ^2\varphi}{\dif\hat{\tau}^2}\left(\hat{H}\frac{\dif \varphi}{\dif \hat{\tau}}\right)^{-1}+\hat{\epsilon}_1\right| \ll 1~,\label{eq: expand_drhordt}
\end{equation}
where the conditions in Eq.~\eqref{sr-hf} have been used. The second and third terms in the second equality are already independently constrained by the preceding conditions. This leads to 
\begin{align}
    \left|\frac{\dif \tilde{\bar{\Upsilon}}/\dif \hat{\tau}}{\hat{H}\tilde{\bar{\Upsilon}}}\right|\ll1~.\label{eq: dUpsilon/dtau}
\end{align}
The second condition in Eq.~\eqref{eq: new_slowroll} introduces a new slow-roll parameter specific to warm inflation, which captures the dependence of $\tilde{\bar{\Upsilon}}$ on the inflaton and the temperature $T$ of the thermal bath. In the case where $\tilde{\bar{\Upsilon}} =\tilde{\Upsilon}(\varphi)$, and upon defining the slow-roll parameter \cite{Ramos:2013nsa}
\begin{equation}
    \label{eq:parameter-beta-Upsilon-Einstein}\tilde{\beta}_{\Upsilon} \equiv \m^2\frac{\tilde{\bar{\Upsilon}}_{,\varphi}\bar{U}_{,\varphi}}{\tilde{\bar{\Upsilon}}\bar{U}}~,
\end{equation}
Eq.~\eqref{eq: dUpsilon/dtau} results in
\begin{equation}
|\tilde{\beta}_{\Upsilon}|\ll 1+\tilde{Q}~.
\end{equation}
In fact, the combination
\begin{equation}
    \left|\frac{\dif \tilde{\bar{\Upsilon}}/\dif \hat{\tau}}{\hat{H}\tilde{\bar{\Upsilon}}}+\hat{\epsilon}_1\right| = \left|\frac{\dif \tilde{Q}/\dif \hat{\tau}}{\hat{H}\tilde{Q}}\right|,
\end{equation}
implies a stronger condition than Eq.~\eqref{eq: new_slowroll}:
\begin{equation}
   \left|\frac{\dif \tilde{Q}/\dif \hat{\tau}}{\hat{H}\tilde{Q}}\right|\ll1~.\label{eq: new_slowroll_Qhat}
\end{equation}

\subsection{Slow-Roll Conditions in the Defining Frame}
\noindent
Having derived the slow-roll conditions in the Einstein frame, they can then be translated to the defining frame. These relations allow the inflationary observables to be expressed in terms of homogeneous quantities in the frame in which the action is originally formulated and the effective dissipation is computed. 

The first Hubble-flow parameters in the Einstein and defining frames are related by \cite{Kuusk:2016rso, Karciauskas:2022jzd}
\begin{equation}
    \hat{\epsilon}_1 = \frac{\epsilon_1+\theta_1}{1+\theta_1}-\frac{\theta_1\theta_2}{(1+\theta_1)^2}~,\label{eq: eps_1_hateps_1}
\end{equation}
where the parameters $\theta_1$ and $\theta_2$ characterise the time evolution of the background effective Planck mass $\sqrt{\bar{F}}$:
\begin{equation}
    \theta_1 \equiv \frac{\d{\bar{F}}}{2H\bar{F}\d\tau}~, \qquad \theta_{i+1} \equiv \frac{\d{\theta}_i}{H\theta_i\d\tau}~,\qquad i=1,2,\ldots~.\label{eq: thetas_def}
\end{equation}
The defining-frame Hubble-flow parameters are 
\begin{equation}
    \epsilon_1 \equiv -\frac{\d H}{H^2\d\tau}~, \qquad \epsilon_{i+1} \equiv \frac{\d{\epsilon}_i}{H\epsilon_i\d\tau}~,\qquad i=1,2,\ldots~. \label{eq: hf_2}
\end{equation}
These expressions hold for arbitrary $\n$. However, before proceeding further, an important observation is in order. As discussed in Refs.~\cite{Karciauskas:2022jzd,Diaz:2023tma}, diffeomorphism invariance of the class of gravitational theories considered here implies that the physics is independent of the choice of coordinates. Nevertheless, when studying conformal transformations, it is convenient to fix a particular coordinate system and retain it across frames. This facilitates the comparison of quantities defined in the two frames, since they can then be evaluated at the same point on the manifold. Accordingly, we shall henceforth adopt the slicing of the FLRW spacetime adapted to the comoving observer in the defining frame. Specifically, we fix the lapse function to be
\begin{equation}
\n=1~,
\end{equation}
which guarantees that $\d\tau=\d t$. In order to keep the slicing fixed, according to Eq.~\eqref{eq:transformation-background-lapse-conformal}, the hatted-frame lapse function must therefore satisfy
\begin{equation}
    \hat{\n}=\sqrt{\bar{F}}~.
\end{equation}

Using the definition of parameter $\theta_1$ in Eq.~\eqref{eq: thetas_def} we can write the transformation of the Hubble parameter in Eq.~\eqref{eq:conformal-transformation-Hubbleexprate} as
\begin{equation}
    \hat{H} = \frac{H}{\sqrt{\bar{F}}}\left(1+\theta_1\right).\label{eq: relation_HhatH}
\end{equation}
Similarly, the slow-roll conditions in Eq.~\eqref{eq: slow-roll-cond_2} can be written in the defining frame as
\begin{equation}
    \frac{\bar{F}\bar{\mathcal{K}}\dot \phi^2}{2}+\bar{\rho}_r\ll V(\phi)~, \qquad \left|\frac{\ddot\phi}{3H\dot \phi}+\frac{\theta_{\mathcal{K}}-\theta_1}{3}\right|\ll\left|1+\theta_1+\frac{Q}{\bar{\mathcal{K}}\bar{F}}\right|~, \qquad \left|\frac{\dot{\bar{\rho}}_r}{H\bar{\rho}_r}-4\theta_1\right| \ll \left|1+\theta_1\right|~,\label{eq: slow-roll-cond_3}
\end{equation}
where we have used the field redefinition in Eq.~\eqref{eq:sf-redefinition}, which, for the background fields, reads $\dif \varphi = \sqrt{\mathcal{K}(\phi)}\dif \phi$, together with $\hat{\bar{\rho}}_r = \bar{F}^{-2} \bar{\rho}_r$ from Eq.~\eqref{eq:rescale-rhom}, $\dif \hat{\tau} = \sqrt{\bfnmc}\dif t$ (for $\n=1$), and $\bar{U} = \bar{V}/\bar{F}^2$ from Eq.~\eqref{eq:potentialU}. The dimensionless parameter $\theta_{\mathcal{K}}$ is defined as \cite{Diaz:2023tma} 
\begin{equation}
    \label{eq:def-thetaK}\theta_{\mathcal{K}} \equiv \frac{\dot{\bar{\mathcal{K}}}}{2H\bar{\mathcal{K}}}~,
\end{equation}
and the dissipation ratio in the defining frame is 
\begin{equation}
    Q\equiv \frac{\bar{\Upsilon}}{3H}~,
    \label{Qdef}
\end{equation}
which parametrises the strength of dissipation relative to the defining-frame Hubble rate. The two dissipation ratios are related by (see Eq.~\eqref{eq:the-below-equation}) 
\begin{equation}
    \label{eq:Best-relation-ever}\tilde{Q}=\frac{Q}{\bar{F}\bar{\mathcal{K}}(1+\theta_1)}~.
\end{equation}
The slow-roll equations of motion in the defining frame take on the form (see Eqs.~\eqref{eq: Slow_hubble_1}--\eqref{eq:rad_slowrolllll})
\begin{eqnarray}
    H^2&\simeq& \frac{\bar{V}}{3\m^2\bar{F}(1+\theta_1)^2}\label{eq: hubble-sr_1}~,\\
    3H\left(1+\theta_1+\frac{Q}{\bar{F}\bar{\mathcal{K}}}\right)\dot \phi &\simeq& -\frac{\bar{F}}{\bar{\mathcal{K}}}\left(\frac{\bar{V}}{\bar{F}^2}\right)_{,\phi}~,\\
    \bar{\rho}_r&\simeq& \frac{3Q}{4(1+\theta_1)}\dot \phi^2~.\label{eq: rhor_sr_df}
\end{eqnarray}
Note that, during slow-roll, the inflaton field experiences a weaker effective dissipation than in GR, owing to the factor $\bar{F}\bar{\mathcal{K}}>1$ (see Eq.~\eqref{eq:just-a-reminder}), unless $\theta_1 \rightarrow -1$, a possibility that will be shown below to be excluded. Finally, we introduce the auxiliary parameters 
\begin{equation}
    \gamma_1^2 \equiv \frac{\bar{\mathcal{K}}\dot{\phi}^2}{2H^2\m^2}~, \qquad \gamma_{i+1}\equiv\frac{\dot \gamma_{i}}{H\gamma_i}~, \qquad i=1,2,\ldots~.\label{eq: gamma_slowroll}
\end{equation}
As explained in Refs.~\cite{Karciauskas:2022jzd,Diaz:2023tma}, this is not an independent set of slow-roll parameters. We can trade $\gamma_i$ for $\theta_i$ parameters defined in Eq.~\eqref{eq: thetas_def}. However, many of the equations become more concise when using both sets simultaneously. Equations~\eqref{eq: hubble-sr_1}--\eqref{eq: rhor_sr_df} hold under the slow-roll conditions imposed in the Einstein frame, which translate into the corresponding slow-roll dynamics in the defining frame and thereby constrain its associated slow-roll parameters. 

Plugging Eq.~\eqref{eq: Hdot_df} into Eq.~\eqref{eq: gamma_slowroll}, the first Einstein-frame Hubble-flow parameter can be expressed exactly in terms of defining-frame quantities as
\begin{equation}
    \label{eq: hateps1_df-EXACT}\hat{\epsilon}_1 = \frac{1}{(1+\theta_1)^2}\left(\gamma_1^2+\frac{2\bar{\rho}_r}{3H^2 \m^2 \bar{F}}\right).
\end{equation}
Using Eq.~\eqref{eq: rhor_sr_df}, this reduces to
\begin{equation}
    \hat{\epsilon}_1 \simeq \frac{\gamma^2_1}{(1+\theta_1)^3}\left(1+\theta_1+\frac{Q}{\bar{F}\bar{\mathcal{K}}}\right).\label{eq: hateps1_df}
\end{equation}
Then, we can insert the above result into Eq.~\eqref{eq: eps_1_hateps_1} to relate the defining-frame Hubble-flow parameter to the Einstein frame variables:
\begin{equation}
    \label{eq:approximations-smalleps1defining}\epsilon_1
    \simeq-\theta_1+\frac{\gamma_1^2}{(1+\theta_1)^2}\left(1+\theta_1+\frac{Q}{\bar{F}\bar{\mathcal{K}}}\right)+\frac{\theta_1\theta_2}{1+\theta_1}~.
\end{equation}
Similarly, the second Einstein-frame Hubble-flow parameter follows from Eq.~\eqref{eq: hateps1_df-EXACT} via Eq.~\eqref{eq: slow-roll-cond_1}:
\begin{equation}
    \hat{\epsilon}_2 = \frac{2}{1+\theta_1}\left\{\left(\gamma_1^2+\frac{2\bar{\rho}_r}{3H^2 \m^2 \bar{F}}\right)^{-1}\left[\gamma^2_1\left(\gamma_2+2\frac{Q}{\bar{F}\bar{\mathcal{K}}}\right) -\left(2+\theta_1-\epsilon_1\right)\frac{2\bar{\rho}_r}{3H^2\m^2 \bar{F}}\right]-\frac{\theta_1\theta_2}{1+\theta_1}\right\},
\end{equation}
or, equivalently, from Eq.~\eqref{eq: hateps1_df}:
\begin{equation}
    \label{eq: eps2_approx_1}\hat{\epsilon}_2\simeq \frac{2}{1+\theta_1}\left[\gamma_2-\frac{\theta_1\theta_2}{2(1+\theta_1)}\left(3-\frac{1+\theta_1}{1+\theta_1+\frac{Q}{\bar{F}\bar{\mathcal{K}}}}\right)+\frac{\dot{\left(\frac{Q}{\bar{F}\bar{\mathcal{K}}}\right)}}{2H(1+\theta_1+\frac{Q}{\bar{F}\bar{\mathcal{K}}})}\right].
\end{equation}
Setting $Q=0$ recovers the expressions derived in Ref.~\cite{Karciauskas:2022jzd}, as expected.

The first parameter that we can constrain is $|\theta_1|$. Since $\dot \phi^2<\bar{F}\bar{\mathcal{K}}\dot \phi^2$, as can be seen from Eq.~\eqref{eq:just-a-reminder}, and because the first condition in Eq.~\eqref{eq: slow-roll-cond_3} reads $\bar{F}\bar{\mathcal{K}}\dot \phi^2 \ll 2\bar{V}$, at the leading order in slow-roll approximation Eq.~\eqref{eq:f-eq-DF-Q} reduces to
\begin{equation}
    H^2 \simeq \frac{\bar{V}}{3\m^2\bar{F}}-2H^2\theta_1~.
\end{equation}
Consistency with Eq.~\eqref{eq: hubble-sr_1} then demands
\begin{equation}
    \left|\theta_1\right|\ll 1~,
    \label{theta1_small}
\end{equation}
since $(1+\theta_1)^2\simeq 1+2\theta_1$ in that regime. At the same order in slow-roll, Eqs.~\eqref{eq: hubble-sr_1} and \eqref{eq: rhor_sr_df} consequently simplify to 
\begin{eqnarray}
    H^2 &\simeq& \frac{\bar{V}}{3\m^2 \bar{F}}~,
    \label{eq:Friedmann-eq-modgrav-defining-slowroll}\\
    \bar{\rho}_r &\simeq& \frac{3Q}{4}\dot \phi^2~,
    \label{eq:radiation-eq-modgrav-defining-slowroll}
\end{eqnarray}
and the Hubble rates in the two frames are related at that order by $\hat{H}\simeq H/\sqrt{\bar{F}}$. A further constraint can be drawn from the first condition in Eq.~\eqref{eq: slow-roll-cond_3}, along with Eqs.~\eqref{eq: rhor_sr_df} and \eqref{eq:Friedmann-eq-modgrav-defining-slowroll}:
\begin{equation}
    \label{eq:super-condition}\gamma_1^2 \left(1+\frac{Q}{\bar{F}\bar{\mathcal{K}}}\right) \ll 1~,
\end{equation}
which is consistent with $\hat{\epsilon}_1 \ll 1$ in the Einstein frame in view of Eq.~\eqref{eq: hateps1_df}. The scalar-field slow-roll equation, Eq.~\eqref{eq: Slow_varphi_1}, gives an additional constraint; noting that
\begin{equation}
    \left|\frac{\bar{F}\bar{U}_{,\phi}}{3H\bar{\mathcal{K}}(1+\frac{Q}{\bar{F}\bar{\mathcal{K}}})\dot \phi}\right| \simeq 1 \Rightarrow  \left|\frac{\bar{V}}{3\bar{F}\bar{\mathcal{K}}\dot \phi^2(1+\frac{Q}{\bar{F}\bar{\mathcal{K}}})}\right|\left|\frac{\dot{\bar{V}}}{H\bar{V}}-4\theta_1\right|\simeq \left|\frac{1}{2\gamma_1^2(1+\frac{Q}{\bar{F}\bar{\mathcal{K}}})}\right|\left|\frac{\dot{\bar{V}}}{H\bar{V}}-4\theta_1\right|\simeq 1~,
\end{equation}
the condition in Eq.~\eqref{eq:super-condition} forces 
\begin{equation}
    \theta_V\equiv \left|\frac{\dot{\bar{V}}}{H\bar{V}}\right|\ll1~.
\end{equation}
By analogy with Eq.~\eqref{eq:parameter-beta-Upsilon-Einstein}, we define two further slow-roll parameters:
\begin{equation}
    \label{eq:super-parameters-defining}\left|\beta_F\right| \equiv \frac{\m^2}{2\bar{\mathcal{K}}}\left|\frac{\bar{F}_{,\phi} \bar{U}_{,\phi}}{\bar{F}\bar{U}}\right|\ll 1+\frac{Q}{\bar{F}\bar{\mathcal{K}}}~, \qquad \left|\beta_V\right| \equiv \frac{\m^2}{\bar{\mathcal{K}}}\left|\frac{\bar{V}_{,\phi} \bar{U}_{,\phi}}{\bar{V}\bar{U}}\right|\ll 1+\frac{Q}{\bar{F}\bar{\mathcal{K}}}~,
\end{equation}
which satisfy $\beta_F = \theta_1(1+\frac{Q}{\bar{F}\bar{\mathcal{K}}})$ and $\beta_V = \theta_V(1+\frac{Q}{\bar{F}\bar{\mathcal{K}}})$.

From Eq.~\eqref{eq:Best-relation-ever},
\begin{equation}
    \dot{\left(\frac{Q}{\bar{F}\bar{\mathcal{K}}}\right)}~  = H\theta_1\theta_2\tilde{Q}+(1+\theta_1)\sqrt{\bar{F}}\frac{\dif\tilde{Q}}{\dif \hat{\tau}}~,
\end{equation}
where we have used Eq.~\eqref{eq: thetas_def} and the fact that the normal observer's proper time satisfies $\dif\hat{\tau} =\sqrt{\bar{F}}\dif t$. Hence, dividing the latter by $2H(1+\theta_1+\frac{Q}{\bar{F}\bar{\mathcal{K}}}) =2\hat{H}\sqrt{\bar{F}}(1+\tilde{Q})$, we obtain
\begin{equation}
    \frac{\dot{\left(\frac{Q}{\bar{F}\bar{\mathcal{K}}}\right)}}{2H(1+\theta_1+\frac{Q}{\bar{F}\bar{\mathcal{K}}})} = \frac{\theta_1\theta_2\frac{Q}{\bar{F}\bar{\mathcal{K}}}}{2(1+\theta_1+\frac{Q}{\bar{F}\bar{\mathcal{K}}})(1+\theta_1)}+(1+\theta_1)\frac{\dif \tilde{Q}/\dif \hat{\tau}}{2\hat{H}(1+\tilde{Q})}~,
\end{equation}
corresponding to the third term of Eq.~\eqref{eq: eps2_approx_1}. Applying Eq.~\eqref{eq: new_slowroll} to $\dif \tilde{Q}/\dif \hat{\tau}$, together with the condition $|\theta_1|\ll 1$ just derived, the second Hubble-flow parameter in Eq.~\eqref{eq: eps2_approx_1} becomes
\begin{equation}
    \hat{\epsilon}_2 \simeq \frac{2}{1+\theta_1}\left[\gamma_2-\frac{\theta_1\theta_2}{1+\theta_1}+(1+\theta_1)\frac{\dif \tilde{Q}/\dif \hat{\tau}}{2\hat{H}(1+\tilde{Q})}\right],\label{eq: epshat2-2}
\end{equation}
which, as can be seen, yields the condition
\begin{equation}
    |\gamma_2-\theta_1\theta_2| \ll 1~.
\end{equation}
The same relation is also satisfied by the cold-inflation scenario \cite{Karciauskas:2022jzd}. If we exclude the possibility of fine-tuned cancellations, the above inequality must be satisfied by each term separately:
\begin{eqnarray}
|\theta_1\theta_2|&\ll& 1~,
\label{theta1theta2_small}\\
|\gamma_2|&\ll& 1~.
\label{gamma2_small}
\end{eqnarray}
Plugging Eqs.~\eqref{theta1_small}, \eqref{eq:super-condition} and \eqref{theta1theta2_small} into Eq.~\eqref{eq:approximations-smalleps1defining} leads to
\begin{equation}
    \epsilon_1 \ll 1~.
    \label{e1_small}
\end{equation}

We can find more slow-roll conditions by expanding the $\gamma_2$ parameter as
\begin{equation}
    \gamma_2 = \frac{{(\bar{\mathcal{K}}\dot \phi^2)}^{\boldsymbol{\cdot}}}{2H\bar{\mathcal{K}}\dot \phi^2}+\epsilon_1 = \frac{\ddot \phi}{H\dot \phi}+\theta_{\mathcal{K}}+\epsilon_1~,
    \label{eq: gamma_2_expanded}
\end{equation}
where $\theta_{\mathcal{K}}$ is defined in Eq.~\eqref{eq:def-thetaK}. As shown in Eq.~\eqref{gamma2_small}, the left-hand side of the above expression is small, and the same holds for the last term on the right-hand side (see Eq.~\eqref{e1_small}). It then follows that the sum of the remaining terms must also be small:
\begin{equation}
    \left|\frac{\ddot \phi}{H\dot \phi}+\theta_{\mathcal{K}}\right|\ll 1\label{eq:phiprimeprime_condition_1}~.
\end{equation}
Using the slow-roll parameters $\theta_{\mathcal{K}}$ and $\theta_1$ defined above, Eq.~\eqref{eq: new_slowroll_Qhat} can be written as
\begin{equation}
    \left|\frac{\dif \tilde{Q}/\dif \hat{\tau}}{\hat{H}\tilde{Q}}\right| = \frac{1}{1+\theta_1}\left|\frac{\dot{Q}}{HQ}-2\theta_\mathcal{K}-2\theta_1\right|.
\end{equation}
In the defining frame, the counterpart of Eq.~\eqref{eq: new_slowroll_Qhat} is defined as
\begin{equation}
    |\theta_Q|\equiv\left|\frac{\dot Q}{2HQ}\right| \ll 1~.\label{eq: def_thetaQ}
\end{equation}

Finally, the potential slow-roll parameters in Eqs.~\eqref{eq:slow-rollepsu} and \eqref{eq:slow-rolletau} (which will be useful for the computation of inflationary observables) can be recast in terms of defining-frame quantities as 
\begin{align}
    \label{eq:epsU-intermsof-phi}&\epsilon_U(\phi) = \frac{\m^2}{2\bar{\mathcal{K}}}\left(\frac{\bar{V}_{,\phi}}{\bar{V}}-2\frac{\bar{F}_{,\phi}}{\bar{F}}\right)^2~,\\
    \label{eq:etaU-intermsof-phi}&\eta_U(\phi) = \frac{\m^2}{\bar{\mathcal{K}}}\left[\frac{\bar{V}_{,\phi\phi}}{\bar{V}}-2\frac{\bar{F}_{,\phi\phi}}{\bar{F}}-4\frac{\bar{F}_{,\phi}\bar{V}_{,\phi}}{\bar{F}\bar{V}}+6\left(\frac{\bar{F}_{,\phi}}{\bar{F}}\right)^2-\frac{\bar{\mathcal{K}}_{,\phi}}{2\bar{\mathcal{K}}}\left(\frac{\bar{V}_{,\phi}}{\bar{V}}-2\frac{\bar{F}_{,\phi}}{\bar{F}}\right)\right].
\end{align}
To conclude this section, the relevant slow-roll conditions in both the Einstein and defining frames are summarised in Table~\ref{tab:slow-roll-conditions}.

\begin{table}[h!]
    \centering
    \renewcommand{\arraystretch}{1.6}
    \begin{tabular}{@{}c@{\hspace{1.5cm}}c@{}}
        \toprule
        \textbf{Einstein Frame} & \textbf{Defining Frame} \\
        \midrule
        $\hat\epsilon_{1} \ll 1$, \ \  $\epsilon_{U} \ll 1+\tilde Q$ 
            & $|\theta_{1}| \ll 1$, \ \ $\gamma_{1}^{2}\!\left(1+\dfrac{Q}{\bar F\bar{\mathcal K}}\right) \ll 1$ \\[4pt]
        $|\hat\epsilon_{2}| \ll 1$, \ \ $|\eta_{U}| \ll 1+\tilde Q$ 
            & $|\theta_{1}\theta_{2}| \ll 1$, \ \ $|\gamma_{2}| \ll 1$ \\[4pt]
        $|\tilde\beta_{\Upsilon}| \ll 1+\tilde Q$ 
            & $|\beta_{F}| \ll 1+\dfrac{Q}{\bar F\bar{\mathcal K}}$, \ \ $|\beta_{V}| \ll 1+\dfrac{Q}{\bar F\bar{\mathcal K}}$ \\
        \bottomrule
    \end{tabular}
    \caption{Slow-roll conditions for non-minimally coupled warm inflation in the Einstein and defining frames. $\tilde{Q}$ and $Q$ are related through Eq.~\eqref{eq:Best-relation-ever}. The Hubble-flow parameters $\hat{\epsilon}_1$ and $\hat{\epsilon}_2$ are defined in Eq.~\eqref{eq: slow-roll-cond_1}, while $\theta_1$ and $\theta_2$ are introduced in Eq.~\eqref{eq: thetas_def}. The potential parameters $\epsilon_U$ and $\eta_U$ follow from Eqs.~\eqref{eq:epsU-intermsof-phi} and \eqref{eq:etaU-intermsof-phi}, respectively. The auxiliary parameters $\gamma_1^2$ and $\gamma_2$ are given in Eq.~\eqref{eq: gamma_slowroll}. The $\beta$ parameters are presented in Eqs.~\eqref{eq:parameter-beta-Upsilon-Einstein} and \eqref{eq:super-parameters-defining}.}
    \label{tab:slow-roll-conditions}
\end{table}

\section{\label{sec:cosmoperturb-einsteinframe}Cosmological Perturbations}
\noindent
Continuing the discussion of frames from the previous sections, we formulate the perturbation equations in the Einstein frame for the scalar-tensor theory with a non-minimal coupling function,\footnote{See Refs.~\cite{Hwang:1990re,Hwang:1990jh,Hwang:1991aj,Fakir:1992cg,Hwang:1996xh,Hwang:2001qk,Diaz:2023tma} for earlier studies of cosmological perturbations in the Jordan and Einstein frames in the absence of dissipation.} and subsequently calculate the power spectrum of the curvature perturbation. The primary inflationary observables, namely the scalar spectral index and the tensor-to-scalar ratio, are first determined in the Einstein frame and then expressed in terms of defining-frame variables, following Refs.~\cite{Karciauskas:2022jzd,Diaz:2023tma}, thereby allowing direct comparison with the frame in which the effective thermal quantities are computed. The starting point is the perturbed line element in the hatted frame:
\begin{equation}
    \label{eq:perturbed-line-element}\textrm{d}\hat{s}^2 = -\left(1+2\hat{A}\right)\hat{\mathcal{N}}^2 \textrm{d}t^2-2\hat{a}\hat{\mathcal{N}}\partial_i \hat{B} \textrm{d}x^{i} \textrm{d}t +\hat{a}^2\left[\left(1-2\hat{\psi}\right) \delta_{ij}+2\partial_i \partial_j \hat{E}+\hat{h}_{ij}\right]\textrm{d}x^{i}\textrm{d}x^{j}~,
\end{equation}
where, as everywhere in this work, only scalar and tensor perturbations are retained on a spatially flat background. Latin indices are lowered and raised with the Euclidean metric $\delta_{ij}$ and its inverse $\delta^{ij}$, respectively. We refer the reader to Appendix~\ref{sec:app-linearperturbations} for the linear perturbations of the relevant geometric quantities and of the EMT on a background with arbitrary lapse function $\mathcal{N}$, and to Appendix~\ref{sec:app-lineargaugeconf} for the gauge and conformal transformations that act on these perturbations. 

\subsection{\label{sec:perturbation-equations-Longitudinal-gauge}Perturbation Equations and the Longitudinal Gauge}
\noindent 
Linearising the metric field equations \eqref{eq:METRIC-eq-rescaledandredefined} yields the energy and momentum constraint equations:\footnote{\label{footnote:an-interesting-footnote-butwementionthisintheappendix}Equation~\eqref{eq:momentum-equation-Einsteiframe} is obtained by dropping an overall spatial derivative $\partial_i$, which yields a scalar relation. A spatially constant (but generally time-dependent) integration term is missing. The reason for dropping the spatially homogeneous mode is explained in footnote~\ref{footnote:explanation-no-one-thought-of}; it is consistent with the splitting between homogeneous background and small inhomogeneities.}
\begin{eqnarray}
    2\m^2 \left(\hat{H} \hat{\kappa}+\hat{a}^{-2}\partial_i \partial^{i} \hat{\psi}\right) &=& \delta \hat{\rho}_{r} +\delta \rho_{\tilde{\Phi}}~,
    \label{eq:energy-equation-Einsteiframe}\\
    -2\m^2\left(\frac{\dot{\hat{\psi}}}{\hat{\mathcal{N}}}+\hat{H}\hat{A}\right) &=& \hat{\Psi}_{r}+\Psi_{\tilde{\Phi}}~,
    \label{eq:momentum-equation-Einsteiframe}
\end{eqnarray}
respectively, with $\kappa$ defined in any frame by Eq.~\eqref{eq:kappa-expansion}, and where we recall that the matter sector is taken to be radiation, and $\hat{\cal N} = \sqrt{F(\varphi)}$. The canonical Einstein-frame field $\tilde{\Phi}$ is decomposed as in Eq.~\eqref{phi-dec}. The corresponding scalar-field energy density and momentum perturbations can be written as
\begin{eqnarray}
    \delta \rho_{\tilde{\Phi}} &=& \frac{\dot \varphi \dot{\delta \varphi}}{\hat{\cal N}^2}-\frac{\dot \varphi^2}{\hat{\cal N}^2}\hat{A}+\bar{U}_{,\varphi}\delta \varphi~,\\
    \Psi_{\tilde{\Phi}} &=& -\frac{\dot \varphi}{\hat{\cal N}}\delta \varphi~.
    \label{eq:scalar-field-momentum-canonical-field-Einstein}
\end{eqnarray}
Equations~\eqref{eq:energy-equation-Einsteiframe} and \eqref{eq:momentum-equation-Einsteiframe} are the $\left(0,0\right)$ and $\left(0,i\right)$ components of Eq.~\eqref{eq:METRIC-eq-rescaledandredefined}, with the first (second) index contravariant (covariant). Alternatively, the momentum equation can be derived from the $\left(i,0\right)$ components upon using Eq.~\eqref{eq: Hdot_df}. Further details are given in Appendix~\ref{sec:app-linearperturbations}. 

From the trace-free part of the spatial $(i,j)$ components of the metric field equations one obtains the shear-propagation equation (again dropping the term that is constant on spacetime slices; see footnote~\ref{footnote:an-interesting-footnote-butwementionthisintheappendix}), 
\begin{equation}
    \dot{\hat{\chi}}+\frac{\dot{\hat{\mathcal{N}}}}{\hat{\cal N}}\hat{\chi}+\hat{\cal N}\hat{H}\hat{\chi} = \hat{A}-\hat{\psi}~,
\end{equation}
and the equation for the tensor modes,
\begin{equation}
    \label{eq:tensor-modes-TT-equationEinsteinframe}\ddot{\hat{h}}_{ij}+\left(3\hat{\cal N}\hat{H}-\frac{\dot{\hat{\mathcal{N}}}}{\hat{\cal N}}\right)\dot{\hat{h}}_{ij} -\frac{\hat{\cal N}^{2}}{\hat{a}^2}\partial_k \partial^{k} \hat{h}_{ij}=0~. 
\end{equation}
Contributions from the matter anisotropic stress have been dropped in both equations.\footnote{In the warm-inflation scenario, in which the matter sector is strongly interacting and the interacting species have short mean free paths, the contribution from free-streaming species is not significant. The effect of the radiation-fluid shear has nevertheless been treated in GR case \cite{Bastero-Gil:2011rva}, as has the damping of the squared amplitude of the tensor modes due to this anisotropic inertia \cite{Weinberg:2003ur,Dicus:2005rh,Watanabe:2006qe}.} The shear potential is defined in Eq.~\eqref{eq:shear-potential}. The Raychaudhuri equation follows from the difference between the trace of the spatial $(i,j)$ components and the temporal $(0,0)$ component of the metric field equations: 
\begin{equation}
    \label{eq:Raychaudhuri-eq-Einstein-frame}\frac{\dot{\hat{\kappa}}}{\hat{\cal N}}+2\hat{H}\hat{\kappa}-\left(\hat{a}^{-2}\partial_i \partial^{i}+3\frac{\dot{\hat{H}}}{\hat{\cal N}}\right)\hat{A}=-\m^{-2}\left[\delta \hat{\rho}_r+\frac{1}{2}\left(\delta \rho_{\tilde{\Phi}}+3\delta P_{\tilde{\Phi}}\right)\right].
\end{equation}
The pressure perturbation $\delta P_{\tilde{\Phi}}$ is given by
\begin{equation}
    \delta P_{\tilde{\Phi}} = \frac{\dot \varphi \dot{\delta \varphi}}{\hat{\cal N}^2}-\frac{\dot \varphi^2}{\hat{\cal N}^2}\hat{A}-\bar{U}_{,\varphi}\delta \varphi~.
\end{equation}
Combining Eq.~\eqref{eq:Raychaudhuri-eq-Einstein-frame} with the energy constraint \eqref{eq:energy-equation-Einsteiframe} eliminates the radiation perturbation:
\begin{equation}
    \label{eq:perturbed-trace-Einstein-frame}\frac{\dot{\hat{\kappa}}}{\hat{\cal N}}+4\hat{H}\hat{\kappa}-\left(\hat{a}^{-2}\partial_i \partial^{i}+3\frac{\dot{\hat{H}}}{\hat{\cal N}}\right)\hat{A}+2\hat{a}^{-2}\partial_i \partial^{i} \hat{\psi} = \frac{1}{2\m^2}\left(\delta \rho_{\tilde{\Phi}}-3\delta P_{\tilde{\Phi}}\right).
\end{equation}
The left-hand side can be obtained from the linear perturbation $\delta \hat{R}/2$ of the Ricci scalar (cf.~Eq.~\eqref{eq:Ricci-scalar-pert}). Equation~\eqref{eq:perturbed-trace-Einstein-frame} is the perturbed trace of the metric field equations \eqref{eq:METRIC-eq-rescaledandredefined}.

Lastly, we linearise the scalar-field equation \eqref{eq:sf-eq-rescaledandredefined}, together with the continuity and Euler equations, Eqs.~\eqref{eq:continuity-eq-rescaledandredefined} and \eqref{eq:Euler-eq-rescaledandredefined}, respectively, including the source $\mathcal{J}$ of Eq.~\eqref{eq: source1}. By virtue of the fluctuation-dissipation theorem \cite{Bartrum:2014fla}, the dynamics of the scalar-field inhomogeneous modes are governed by the perturbed scalar-field equation alongside the combined effect of the stochastic \emph{quantum} and \emph{thermal} noises, denoted by $\tilde{\xi}_q$ and $\tilde{\xi}_T$, respectively. These introduce a degree of randomness in the evolution equation of perturbations. The source $\mathcal{J}$ is decomposed as $\bar{\mathcal{J}}(t)+\delta \mathcal{J}(t,\mathbf{x})$, with the homogeneous part already computed in Sec.~\ref{sec:nonminimally-coupled-warm-inflation}:
\begin{eqnarray}
    \frac{\bar{\mathcal{J}}}{\bar{F}^2 \sqrt{\bar{\mathcal{K}}}}&=& \tilde{\bar{\Upsilon}}\frac{\dot \varphi}{\hat{\mathcal{N}}}~, \\
    \delta\left(\frac{\mathcal{J}}{F^2 \sqrt{\mathcal{K}}}\right)&=&\delta \tilde{\Upsilon}\frac{\dot \varphi}{\hat{\mathcal{N}}}+\frac{\tilde{\bar{\Upsilon}}}{\hat{\mathcal{N}}}\left(\dot{\delta \varphi}-\dot \varphi\hat{A}\right)-\tilde{\xi}_T-\tilde{\xi}_q~,
\end{eqnarray}
where $\tilde{\bar{\Upsilon}}$ was defined in Eq.~\eqref{eq: tilde_dc} and $\hat{T}$ and $T$ are related through Eq.~\eqref{eq:hatted temperature}. $\delta \tilde{\Upsilon}(x)$ denotes the perturbation of the effective dissipation coefficient.

At this point, it is important to emphasise that we assume throughout this work that statistical distributions describing local equilibrium at temperature $T$ in the defining frame correspond to local equilibrium distributions at temperature $\hat{T}$ in the Einstein frame, with $\hat{T}$ given by Eq.~\eqref{eq:hatted temperature}. Then, under the adiabatic conditions \eqref{eq:thermal-domination-cond} and \eqref{eq:some-conditions-keepinmind} above, $F$ and $H$ evolve on (proper) timescales much longer than those of thermal processes, and thus all computations can be performed assuming approximate \emph{global} equilibrium for the radiation fluid at each instant of time $t$. A rigorous proof of the equivalence of local equilibrium between frames is beyond the scope of the present investigation, and we leave this question for future work. Nonetheless, we present an argument supporting the validity of this assumption.\footnote{It is worth mentioning that other authors, \emph{e.g.}~Ref.~\cite{Faraoni:2023gqg}, have likewise assumed that local thermal equilibrium is preserved under frame transformations, while acknowledging the lack, to date, of a rigorous proof of this statement.} In the defining frame, a bosonic gas in local thermodynamic equilibrium follows the Bose-Einstein distribution:
\begin{equation} \label{eq:defining bose-einstein}
    f(x,p)=\left[e^{p_\mu\beta^\mu(x)}-1\right]^{-1}~,
\end{equation}
in order to solve Boltzmann's equation with vanishing collision term (notice that, for simplicity, we neglect the contribution from the chemical potential). Here, $p_\mu$ is the four-momentum one-form associated to each of the microscopic particles, while $\beta^\mu$ is the (time-like) temperature four-vector:
\begin{equation} \label{eq:defining u and T}
    \beta^\mu=\dfrac{u^\mu}{T}~, \qquad T=(-g_{\mu\nu}\beta^\mu\beta^\nu)^{-1/2}~,
\end{equation}
where $u^\mu$ is the four-velocity of the radiation fluid and $T$ its temperature, both as seen from the perspective of the unhatted frame. An entirely analogous construction can be carried out in the Einstein frame: the bosonic distribution function satisfying the hatted-frame Boltzmann equation with vanishing collision term (and hence describing local thermodynamic equilibrium) is of the form
\begin{equation} \label{eq:einstein bose-einstein}
    \hat{f}(x,p)=\left[e^{p_\mu\hat{\beta}^\mu(x)}-1\right]^{-1}~,
\end{equation}
with
\begin{equation} \label{eq:einstein u and T}
    \hat{\beta}^\mu=\dfrac{\hat{u}^\mu}{\hat{T}}~, \qquad \hat{T}=(-\hat{g}_{\mu\nu}\hat{\beta}^\mu\hat{\beta}^\nu)^{-1/2}~.
\end{equation}
$\hat{u}^{\mu}$ and $\hat{T}$ represent the four-velocity and temperature of the fluid in the hatted frame, and we have taken into account that the four-momentum one-form $p_\mu$, being entirely independent of the metric tensor, is frame-invariant (\emph{i.e.}~$\hat{p}_\mu=p_\mu$). The hatted and unhatted distributions in Eqs.~\eqref{eq:defining bose-einstein} and \eqref{eq:einstein bose-einstein} are manifestly of the same form. Moreover, using the transformation rule \eqref{eq:rescale-velocity} for the four-velocity and
\begin{equation}
    \hat{T} = F^{-1/2}T\label{eq:hatted temperature}
\end{equation}
for the temperature, one finds that $\beta^\mu$ is conformally-invariant: $\hat{\beta}^\mu=\beta^\mu$. Therefore, both $f$ and $\hat{f}$ are conformally-invariant, and local equilibrium distributions in one frame should correspond to local equilibrium distributions in the other, albeit with different temperature and resulting fluid four-velocity in each frame. In order to achieve not only local, but instead global equilibrium, $\beta^\mu$ must be a (time-like) Killing vector in each frame, which in turn requires the conformal factor to be compatible with the time-translation symmetry generated by $\beta^\mu$:
\begin{equation}
    \beta^\mu\partial_\mu F=0~.
\end{equation}
FLRW is non-stationary, and therein the previous condition reduces to
\begin{equation}
    \dot{\bar{F}}=0~,
\end{equation}
since $\beta^\mu\propto(\partial_t)^\mu$. Now, notice that if the adiabaticity conditions in Eqs.~\eqref{eq:thermal-domination-cond} and \eqref{eq:some-conditions-keepinmind} hold, spacetime becomes effectively Minkowski, and $\dot{\bar{F}}\simeq 0$.\footnote{Observe that the condition $\dot{\bar{F}}\simeq 0$ $\Rightarrow$ $\theta_1\simeq 0$ is guaranteed to hold in slow roll (see Eq.~\eqref{theta1_small}). Moreover, $\dot{\bar{F}}$ can become very small through $\bar{F}_{,\phi}\simeq 0$ rather than $\dot{\phi}\simeq 0$, as implied by Eq.~\eqref{eq:some-conditions-keepinmind}. However, the conditions $\bar{F}_{,\phi}\simeq 0$ and $\dot{\phi}\simeq 0$ have different implications, the main difference being that derivatives of $\bar{F}$ with respect to the scalar field contribute to the computation of $\bar{\Upsilon}$.} Hence, on such short thermal timescales, $\beta^\mu$ is an approximate Killing vector in both frames, so that approximate instantaneous global equilibrium can be assumed. The equilibrium state then evolves smoothly in time, with the corresponding Einstein-frame temperature given by $\hat{T}$.

With this clarification in place, let us return to the linearised equations of motion. Combining all the above, the perturbed equations following from Eqs.~\eqref{eq:sf-eq-rescaledandredefined}--\eqref{eq:Euler-eq-rescaledandredefined} read (recall that $\hat{P}_{\textrm{m}} = \hat{\rho}_{\textrm{m}}/3$)
\begin{align}
    \nonumber&\frac{\ddot{\delta \varphi}}{\hat{\mathcal{N}}^2}+\left[3\hat{H}\left(1+\tilde{Q}\right)-\frac{\dot{\hat{\mathcal{N}}}}{\hat{\mathcal{N}}^2}\right]\frac{\dot{\delta \varphi}}{\hat{\mathcal{N}}}+\left(\bar{U}_{,\varphi\varphi}-\hat{a}^{-2}\partial_i \partial^{i}\right)\delta \varphi =\frac{\dot \varphi}{\hat{\mathcal{N}}}\frac{\dot{\hat{A}}}{\hat{\mathcal{N}}}+\left[2\left(\frac{\ddot \varphi}{\hat{\mathcal{N}}^2}-\frac{\dot{\hat{\mathcal{N}}}}{\hat{\mathcal{N}}^2}\frac{\dot \varphi}{\hat{\mathcal{N}}}\right)+3\hat{H}\left(2+\tilde{Q}\right)\frac{\dot \varphi}{\hat{\mathcal{N}}}\right]\hat{A}\\
    &\phantom{---------------------------------}-\hat{a}^{-2}\dot \varphi\partial_i \partial^{i} \hat{\chi}-\delta \tilde{\Upsilon}\frac{\dot \varphi}{\hat{\mathcal{N}}}+\tilde{\xi}_T+\tilde{\xi}_q~,\\
    \label{eq:energy-conservation-radiation-Einstein}&\frac{\dot{\delta\hat{\rho}}_r}{\hat{\mathcal{N}}}+4\hat{H}\delta \hat{\rho}_r = -\hat{a}^{-2}\partial_i \partial^{i} \hat{\Psi}_r +4\hat{\bar{\rho}}_r\left(\frac{\dot{\hat{\psi}}}{\hat{\mathcal{N}}}-\frac{1}{3}\hat{a}^{-2}\hat{\mathcal{N}}\partial_i \partial^{i}\hat{\chi}\right)+\frac{\dot \varphi^2}{\hat{\mathcal{N}}^2}\left(\delta \tilde{\Upsilon}-\tilde{\bar{\Upsilon}}\hat{A}\right)+2\tilde{\bar{\Upsilon}}\frac{\dot \varphi}{\hat{\mathcal{N}}}\frac{\dot{\delta \varphi}}{\hat{\mathcal{N}}}-\frac{\dot \varphi}{\hat{\mathcal{N}}}\left(\tilde{\xi}_T+\tilde{\xi}_q\right),\\
    \label{eq:momentum-conservation-radiation-Einstein}&\frac{\dot{\hat{\Psi}}_r}{\hat{\mathcal{N}}}+3\hat{H}\hat{\Psi}_r = -\frac{1}{3}\left(\delta \hat{\rho}_r +4\hat{\rho}_r \hat{A}\right)-\tilde{\bar{\Upsilon}}\frac{\dot \varphi}{\hat{\mathcal{N}}}\delta \varphi~.
\end{align}
As can be seen, the radiation momentum equation is unaffected by the noise terms, with $\hat{\Psi}_r$ given by (cf. Eq.~\eqref{eq:momentum-matter})
\begin{equation}
    \hat{\Psi}_r \equiv \frac{4}{3}\hat{a}\hat{\bar{\rho}}_r \left(\hat{v}-\hat{B}\right),
\end{equation}
where $\hat{v}$ is the matter velocity potential in the hatted frame. The absence of noise terms in the radiation Euler equation reflects the fact that, at linear order in perturbations, momentum transfer between the inflaton and the thermal bath arises solely from gradients of the inflaton field, rather than from the local stochastic source (see footnote~\ref{footnote:an-interesting-footnote-butwementionthisintheappendix} regarding the spatially constant term that is removed upon dropping the overall spatial derivative).

The resulting equations have the same structure as in GR \cite{Bastero-Gil:2011rva,Kamali:2023lzq}, with the rescaled proper time $\hat{\tau}$ replacing the coordinate time $t$, and with the corresponding effective quantities carrying tildes. The diffeomorphism invariance of the scalar-tensor gravity theory under consideration permits a gauge choice that simplifies the system of equations. As discussed in Appendix~\ref{sec:app-lineargaugeconf}, the longitudinal gauge, in which the shear potential vanishes, is an example of a frame-insensitive gauge fixing \cite{Brown:2011eh} (see Eq.~\eqref{eq:frame-transformation-shear-potential}). The shear-propagation equation then enforces $\hat{A}^{\hat{\chi}} = \hat{\psi}^{\hat{\chi}}$,\footnote{When a perturbation variable carries another perturbation quantity as a superscript, the superscripted variable denotes the corresponding perturbation evaluated in a gauge in which the superscript quantity vanishes (see Ref.~\cite{Hwang:1991aj}).\label{footnote:superscript-perturbation}} and the scalar-field equation can be written as ($\d \hat{\tau}$ is given in Eq.~\eqref{hattau-def})\footnote{No superscript is attached to the noise terms below because the noise sources only the perturbations and has vanishing one-point function. It therefore contributes nothing to the homogeneous evolution. By the Stewart-Walker lemma \cite{Stewart:1974uz,Lyth:2009zz}, a perturbation whose background counterpart vanishes is gauge-invariant at linear order on an FLRW background.}
\begin{align}
    \nonumber&\frac{\dif^2 \delta \varphi^{\hat{\chi}}}{\dif \hat{\tau}^2}+3\hat{H}\left(1+\tilde{Q}\right)\frac{\dif \delta \varphi^{\hat{\chi}}}{\dif \hat{\tau}}+\left(\bar{U}_{,\varphi\varphi}-\hat{a}^{-2}\partial_i \partial^{i}\right)\delta \varphi^{\hat{\chi}} = -\delta \tilde{\Upsilon}^{\hat{\chi}} \frac{\dif \varphi}{\dif \hat{\tau}}+\left[2\frac{\dif^2 \varphi}{\dif\hat{\tau}^2}+3\hat{H}\left(2+\tilde{Q}\right) \frac{\dif \varphi}{\dif \hat{\tau}}\right]\hat{\psi}^{\hat{\chi}}+\frac{\dif \varphi}{\dif \hat{\tau}}\frac{\dif \hat{\psi}^{\hat{\chi}}}{\dif \hat{\tau}}+\\
    \label{eq:sf-perturbation-eq-hatted-proper-time}&\phantom{------------------------------------------}+\tilde{\xi}_T+\tilde{\xi}_q~.
\end{align}
The momentum constraint \eqref{eq:momentum-equation-Einsteiframe}, the radiation continuity \eqref{eq:energy-conservation-radiation-Einstein} and Euler \eqref{eq:momentum-conservation-radiation-Einstein} equations complete the closed system:
\begin{align}
    \label{eq:curvature-perturbation-Einstein-Longitudinal}&\frac{\dif \hat{\psi}^{\hat{\chi}}}{\dif \hat{\tau}}=-\hat{H}\hat{\psi}^{\hat{\chi}}-\frac{\hat{\Psi}_r^{\hat{\chi}}}{2\m^2}+\frac{1}{2\m^2}\frac{\dif \varphi}{\dif \hat{\tau}}\delta \varphi^{\hat{\chi}}~,\\
    \label{eq:radiationn-perturbation-eq-hatted-proper-time}&\frac{\dif \delta \hat{\rho}_r^{\hat{\chi}}}{\dif \hat{\tau}}+4\hat{H} \delta \hat{\rho}_r^{\hat{\chi}} = -\hat{a}^{-2}\partial_i \partial^{i} \hat{\Psi}^{\hat{\chi}}_r+\left(\frac{\dif \varphi}{\dif \hat{\tau}}\right)^2 \delta \tilde{\Upsilon}^{\hat{\chi}}+4\hat{\bar{\rho}}_r \frac{\dif \hat{\psi}^{\hat{\chi}}}{\dif \hat{\tau}}+\tilde{\bar{\Upsilon}}\frac{\dif \varphi}{\dif \hat{\tau}}\left(\frac{\dif \delta \varphi^{\hat{\chi}}}{\dif \hat{\tau}}-\frac{\dif \varphi}{\dif \hat{\tau}}\hat{\psi}^{\hat{\chi}}\right)-\frac{\dif\varphi}{\dif\hat{\tau}}\left(\tilde{\xi}_T+\tilde{\xi}_q\right),\\
    &\frac{\dif \hat{\Psi}_r^{\hat{\chi}}}{\dif \hat{\tau}}+3\hat{H}\hat{\Psi}^{\hat{\chi}}_r = -\frac{1}{3}\left(\delta \hat{\rho}_r^{\hat{\chi}}+4\hat{\bar{\rho}}_r \hat{\psi}^{\hat{\chi}}\right) -\tilde{\bar{\Upsilon}}\frac{\dif \varphi}{\dif \hat{\tau}}\delta \varphi^{\hat{\chi}}~.\label{eq: dif_rad_pertur}
\end{align}
Equation~\eqref{eq:curvature-perturbation-Einstein-Longitudinal} follows from \eqref{eq:momentum-equation-Einsteiframe} upon using $\hat{A}^{\hat{\chi}} = \hat{\psi}^{\hat{\chi}}$ and Eq.~\eqref{eq:scalar-field-momentum-canonical-field-Einstein}. Combining the energy \eqref{eq:energy-equation-Einsteiframe} and momentum \eqref{eq:momentum-equation-Einsteiframe} constraints fixes the radiation momentum to \cite{Ballesteros:2023dno} 
\begin{equation}
    \hat{\Psi}^{\hat{\chi}}_r = -\left[\left(\frac{\dif \varphi}{\dif\hat{\tau}}\right)^2+2\m^{2}\hat{a}^{-2}\partial_i \partial^{i}\right]\frac{\hat{\psi}^{\hat{\chi}}}{3\hat{H}}+\frac{\dif \varphi}{\dif \hat{\tau}}\delta \varphi^{\hat{\chi}}+\frac{1}{3\hat{H}}\left(\delta \hat{\rho}_r^{\hat{\chi}}+\frac{\dif \varphi}{\dif \hat{\tau}}\frac{\dif \delta \varphi^{\hat{\chi}}}{\dif \hat{\tau}}+\bar{U}_{,\varphi}\delta \varphi^{\hat{\chi}}\right).
\end{equation}

\subsection{Power Spectra and Inflationary Observables}
\noindent
In this subsection, we use the approximate expressions obtained in Sec.~\ref{sec:nonminimally-coupled-warm-inflation} to compute the scalar and tensor power spectra during slow-roll inflation. The former in particular can be determined analytically by assuming that the dissipation coefficient depends solely on the scalar field value, thereby decoupling the scalar-field perturbation equation \eqref{eq:sf-perturbation-eq-hatted-proper-time} from the radiation-energy-density perturbation equation \eqref{eq:radiationn-perturbation-eq-hatted-proper-time}. This is a well-known procedure in the literature \cite{2009JCAP...07..013G,Ramos:2013nsa,Montefalcone:2023pvh}, where the dissipation coefficient is generally parametrised as \cite{Bastero-Gil:2011rva,Kamali:2023lzq} 
\begin{equation}
    \Upsilon(\Phi,T) = \frac{C_{\Upsilon}}{M^{c-d-1}}\frac{T^{c}}{\Phi^{d}}~,\label{eq: upsilon_param}
\end{equation}
with $C_{\Upsilon}$ being a dimensionless constant encoding microphysical aspects, and $M$ is the mass scale of the particle-physics model under consideration. 

The frame-independent inflationary observables are derived next and then expressed in terms of defining-frame quantities, including the effective dissipation coefficient computed in the defining frame following Appendix~\ref{app:dissipative-Minkowski-dyn} and \textit{e.g.} Refs.~\cite{Berera:2008ar,Kamali:2023lzq}. Before proceeding further, we need to address the issue of quantisation. As with the background quantities, we work out the perturbations in the Einstein frame, where the equations take their familiar form. In this frame, the field $\varphi$ becomes the primary variable. Nevertheless, our theory is fundamentally formulated in the defining frame, where the quantum and thermal effects are properly defined. It is therefore reasonable to quantise the fields therein, thereby avoiding possible ambiguities associated with the conformal transformation to the Einstein frame. However, even in this case, we deal with two, in principle, distinct (yet related) quantum fields, namely, the perturbations of $\phi$ and $\varphi$, which satisfy
\begin{align} \label{eq:delta varphi delta phi}
    \delta\varphi = \sqrt{\bar{\mathcal{K}}}\delta \phi~,
\end{align}
where $\sqrt{\bar{\mathcal{K}}}$ depends only on the background field $\phi$. As shown in Appendix \ref{app:quantisation}, the quantisation schemes based on $\delta\varphi$ and $\delta\phi$ are equivalent, thus ensuring the consistency of our results.

In general, the coupled system of Eqs.~\eqref{eq:sf-perturbation-eq-hatted-proper-time}--\eqref{eq: dif_rad_pertur} does not admit an analytical solution and must be solved numerically \cite{Bastero-Gil:2014jsa,Ramos:2013nsa,Montefalcone:2022owy,Rodrigues:2025neh}. This requires specifying a particular model. For simplicity, we focus on the case in which an analytical result can be obtained at zeroth order in the slow-roll parameters, namely, a temperature-independent dissipation coefficient, $\tilde{\bar{\Upsilon}} \equiv \tilde{\Upsilon}(\varphi)$, and negligible metric perturbation $\hat{\psi}^{\hat{\chi}}$.\footnote{\label{just-one-thing-that-happens-to-be-relevant-later}It can be shown that $\hat{\psi}^{\hat{\chi}}\propto \hat{\epsilon}_1$ on super-Hubble scales. Hence, $\hat{\psi}^{\hat{\chi}}$ can be consistently neglected at zeroth order in the slow-roll parameters \cite{Riotto:2002yw,Ramos:2013nsa,Bastero-Gil:2019rsp,Kamali:2023lzq}.} Thus, we only need to consider Eq.~\eqref{eq:sf-perturbation-eq-hatted-proper-time}, written as
\begin{align}
 \frac{\dif^2 \delta \varphi^{\hat{\chi}}}{\dif \hat{\tau}^2}+3\hat{H}\left(1+\tilde{Q}\right)\frac{\dif \delta \varphi^{\hat{\chi}}}{\dif \hat{\tau}}+\left(-\frac{\delta^{ij}\partial_i\partial_j}{\hat{a}^2}+\bar{U}_{,\varphi\varphi}+\tilde{\bar{\Upsilon}}_{,\varphi}\frac{\dif\varphi}{\dif\hat{\tau}}\right)\delta \varphi^{\hat{\chi}} = \tilde{\xi}_T+\tilde{\xi}_q~,\label{eq: simplified_deltaphi_pert_1}
\end{align}
with the dissipation coefficient perturbation
\begin{align}
\delta \tilde{\Upsilon} = \frac{\tilde{\bar{\Upsilon}}_{,\phi}}{\sqrt{\bar{\mathcal{K}}}}\delta \varphi^{\hat{\chi}}~,
\end{align}
and the noise term $\tilde{\xi}_T = \xi_T/(\bar{F}^2\sqrt{\mathcal{\bar{K}}})$ (see Eq.~\eqref{eq:rescaling-noise-term-whywasthisnotlabelled}); the correlator of $\xi_T$ satisfies Eq.~\eqref{eq:correlator-noise-term-stilldontunderstandwhythiswasnotlabelled}, while the two-point correlation function of the quantum noise term $\tilde{\xi}_q$ is given by \cite{Ramos:2013nsa,Kamali:2023lzq,Montefalcone:2023pvh} 
\begin{equation}
\langle \tilde{\xi}_q(t,\mathbf{x})\tilde{\xi}_q(t',\mathbf{x}')\rangle =\frac{\hat{H}^2}{\pi}\frac{(1+2\tilde{n})\sqrt{9+12\pi \tilde{Q}}}{\hat{a}^3\hat{\mathcal{N}}}\delta(t-t')\delta^{(3)}(\mathbf{x}-\mathbf{x}')~,\label{eq: quantumnoise2point}
\end{equation}
and captures the backreaction of short-wavelength quantum modes on super-Hubble ones, which can be treated classically.\footnote{Strictly speaking, whether a mode is super- or sub-Hubble is frame-dependent. The physically relevant quantity is instead the curvature scale at which quantum perturbations become effectively classical. However, $\hat{a}\hat{H}=aH(1+\theta_1)$ implies that the horizon-crossing conditions in the two frames coincide up to first order in the slow-roll hierarchy \cite{Karciauskas:2022jzd}.} Here $\tilde{n}$ is the thermal distribution function of the sub-Hubble inflaton modes, conventionally taken to be of Bose-Einstein form \cite{Ramos:2013nsa}:
\begin{equation} \label{eq:bose-einstein hatted}
    \tilde{n}(k) \equiv \left[\exp\left(\frac{k}{\hat{a}\hat{T}}\right)-1\right]^{-1}~,
\end{equation}
with $k=\hat{k}$ denoting the magnitude of the comoving momentum labelling each mode, which is a conformally invariant quantity (see Appendix \ref{app:comoving momenta}). Notice that, since our theory is quantised in the defining frame, the thermal distribution function appearing in Eq.~\eqref{eq: quantumnoise2point} is in fact
\begin{equation}
    \label{eq:Bose-and-Einstein-dont-care-about-frames}n(k) = \left[\exp\left(\frac{k}{aT}\right)-1\right]^{-1},
\end{equation}
as defined via the annihilation and creation operators in the defining frame or, more precisely, through the expectation value of the number operator in a thermal state at temperature $T$ in the unhatted frame, which is given by
\begin{equation}
    \langle \mathtt{a}^\dagger_{\mathbf{k}}\mathtt{a}_{-\mathbf{k}'}\rangle=n(k)\,\delta^{(3)}(\mathbf{k}+\mathbf{k}')~.
\end{equation}
However, since the comoving wavenumber satisfies $k = \hat{k}$ and the combination $aT = \hat{a}\hat{T}$ is invariant under conformal rescalings, it follows that $\tilde{n}(k) = n(k)$. 

Writing Eq.~\eqref{eq:sf-perturbation-eq-hatted-proper-time} in terms of the variable $\hat{z}\equiv k/(\hat{a}\hat{H})$, we have
\begin{align}
 \frac{\dif^2 \delta \varphi^{\hat{\chi}}(\hat{z},\mathbf{k})}{\dif \hat{z}^2}-\frac{1}{\hat{z}}\left(2+3\tilde{Q}\right)\frac{\dif \delta \varphi^{\hat{\chi}}(\hat{z},\mathbf{k})}{\dif \hat{z}}+\left(1+3\frac{\eta_U-\tilde{\beta}_\Upsilon \tilde{Q}/(1+\tilde{Q})}{\hat{z}^2}\right)\delta \varphi^{\hat{\chi}}(\hat{z},\mathbf{k}) = \frac{1}{\hat{H}^2\hat{z}^2}\left[\tilde{\xi}_{T}(\hat{z},\mathbf{k})+\tilde{\xi}_{q}(\hat{z},\mathbf{k})\right]~,\label{eq: simplified_deltaphi_pert}
\end{align}
with the slow-roll parameters $\eta_U$ and $\tilde{\beta}_{\Upsilon}$ defined in Eqs.~\eqref{eq:etaU-intermsof-phi} and \eqref{eq:parameter-beta-Upsilon-Einstein}, respectively. $\tilde{Q}$ is given by Eq.~\eqref{eq:Best-relation-ever}. The Fourier modes of the noise term become \cite{Ramos:2013nsa} (see Sec.~\ref{app:flat-spacetime-subsection})
\begin{equation}
    \langle\tilde{\xi}_{T}(\hat{z},\mathbf{k})\tilde{\xi}_{T}(\hat{z}',\mathbf{k}')\rangle =2\tilde{\bar{\Upsilon}}\hat{T}\frac{\hat{H}^4\hat{z}^4}{k^3}\delta(\hat{z}-\hat{z}')\delta^{(3)}(\mathbf{k}+\mathbf{k}')~,\label{eq: thermal_noise_in_z}
\end{equation} 
where we used
\begin{equation}
    \delta(t-t') \simeq \hat{\mathcal{N}}\hat{H}\hat{z}\, \delta(\hat{z}-\hat{z}')~.
\end{equation}
The solution to Eq.~\eqref{eq: simplified_deltaphi_pert} can be written in terms of a Green function constructed from Bessel functions $J_{\alpha}$ and $Y_{\alpha}$ \cite{Ramos:2013nsa}, namely
\begin{align}
\delta\varphi^{\hat{\chi}}(\hat{z},\mathbf{k}) = \int_{\hat{z}}^\infty\dif \hat{z}'\, G(\hat{z},\hat{z}')\frac{(\hat{z}')^{-1-2\nu}}{\hat{H}^2}\Big[\tilde{\xi}_{T}(\hat{z}',\mathbf{k})+\tilde{\xi}_{q}(\hat{z}',\mathbf{k})\Big]~,
\end{align}
where
\begin{align}
G(\hat{z},\hat{z}') = \frac{\pi}{2}\hat{z}^\nu \hat{z}'^\nu\left[J_\alpha(\hat{z})Y_\alpha(\hat{z}')-J_\alpha(\hat{z}')Y_\alpha(\hat{z})\right],
\end{align}
with $\hat{z}'>\hat{z}$, and coefficients 
\begin{align}
&\nu \equiv \frac{3}{2}(1+\tilde{Q})~,\\
&\alpha \equiv\sqrt{\nu^2+\frac{3\tilde{\beta}_\Upsilon \tilde{Q}}{1+\tilde{Q}}-3\eta_U}~.
\end{align}

Since the two stochastic noise terms have different origins, we assume that they are statistically uncorrelated, \textit{i.e.} $\langle \tilde{\xi}_q \tilde{\xi}_T\rangle = 0$. The two-point correlation function of the $\varphi$-perturbations can therefore be written as the sum of the quantum and thermal contributions:
\begin{align}
\langle \delta \varphi^{\hat{\chi}}(\hat{z},\mathbf{k})\delta \varphi^{\hat{\chi}}(\hat{z}',\mathbf{k}')\rangle = \langle \delta \varphi^{\hat{\chi}}(\hat{z},\mathbf{k})\delta \varphi^{\hat{\chi}}(\hat{z}',\mathbf{k}')\rangle_q+\langle\delta \varphi^{\hat{\chi}}(\hat{z},\mathbf{k})\delta \varphi^{\hat{\chi}}(\hat{z}',\mathbf{k}')\rangle_T~.\label{eq: total_deltaphi_ave}
\end{align} 
These expectation values are evaluated in the defining frame, where the theory is quantised. As discussed in Ref.~\cite{Ballesteros:2023dno}, the quantum and thermal averaging procedures commute, namely $\left\langle\langle ... \rangle_q \right\rangle_T =\left\langle\langle ... \rangle_T \right\rangle_q $, so that one may first evaluate the quantum expectation value in the vacuum state and subsequently average over the realisations of the thermal noise, or vice versa. 

The dimensionless power spectrum of the inflaton perturbations is defined as
\begin{align}
    \mathcal{P}_{\delta\varphi} \equiv \frac{k^3}{2\pi^2}\int \dif^3\mathbf{k}'\,\langle \delta \varphi^{\hat{\chi}}(\hat{z},\mathbf{k})\delta \varphi^{\hat{\chi}}(\hat{z},\mathbf{k}')\rangle = \mathcal{P}_{\delta\varphi}^{\textrm{(qu)}}+\mathcal{P}_{\delta\varphi}^{\textrm{(th)}}~,
\end{align}
which is the sum of quantum and thermal contributions. The quantum contribution is \cite{Ramos:2013nsa,Montefalcone:2022owy}
\begin{align}
    \mathcal{P}_{\delta\varphi}^{\textrm{(qu)}} &\equiv \frac{k^3}{2\pi^2\hat{H}^4}\int \dif^3\mathbf{k}' \int_{\hat{z}}^\infty\dif \hat{z}_1\int_{\hat{z}}^\infty\dif \hat{z}_2\, G(\hat{z},\hat{z}_1)G(\hat{z},\hat{z}_2)(\hat{z}_1)^{-1-2\nu}(\hat{z}_2)^{-1-2\nu}\langle\tilde{\xi}_q(\hat{z}_1,\mathbf{k})\tilde{\xi}_q(\hat{z}_2,\mathbf{k}')\rangle\nonumber\\
    &\hspace{22.14em}=\frac{\hat{H}^2\sqrt{9+12\pi\tilde{Q}}}{32\pi^3}(1+2\tilde{n})\int_{\hat{z}}^{\infty}\dif \hat{z}'\,\hat{z}'^{2-4\nu}\left[G(\hat{z},\hat{z}')\right]^2~,
\end{align}
while the thermal contribution reads
\begin{align}
    \mathcal{P}_{\delta\varphi}^{\textrm{(th)}} &\equiv \frac{k^3}{2\pi^2\hat{H}^4}\int \dif^3\mathbf{k}' \int_{\hat{z}}^\infty\dif \hat{z}_1\int_{\hat{z}}^\infty\dif \hat{z}_2\, G(\hat{z},\hat{z}_1)G(\hat{z},\hat{z}_2)(\hat{z}_1)^{-1-2\nu}(\hat{z}_2)^{-1-2\nu}\langle\tilde{\xi}_T(\hat{z}_1,\mathbf{k})\tilde{\xi}_T(\hat{z}_2,\mathbf{k}')\rangle\nonumber\\
    &\hspace{30.21em}=\frac{\tilde{\bar{\Upsilon}}\hat{T}}{16\pi^2}\int_{\hat{z}}^{\infty}\dif \hat{z}'\hat{z}'^{2-4\nu}\left[G(\hat{z},\hat{z}')\right]^2~,
\end{align}
where we have used the expressions for the two-point correlation functions of the quantum and thermal noises given in Eqs.~\eqref{eq: quantumnoise2point} and \eqref{eq: thermal_noise_in_z}, respectively, and assumed that $\hat{H}$, $\hat{T}$, and $\tilde{\bar{\Upsilon}}$ are approximately constant with respect to $\hat{z}$.

\begin{figure}[h]
         \centering
\includegraphics[width=0.74\textwidth]{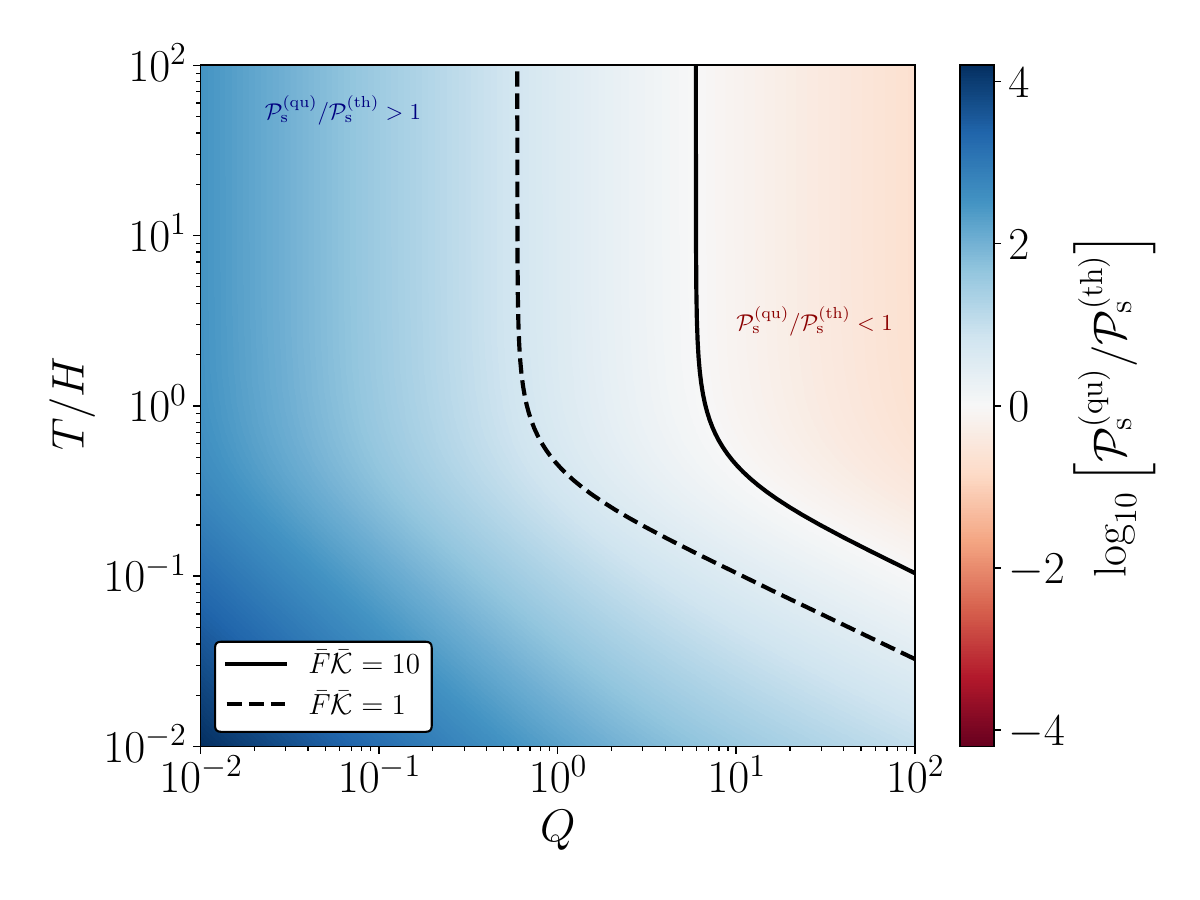}
     \caption{Ratio of the quantum to thermal contributions to the scalar power spectrum (see Eqs.~\eqref{how-can-this-not-be-labelled-is-beyond-me} and \eqref{another-equation-which-is-not-labelled-and-should-be}), as a function of the dissipative ratio $Q$, and the temperature-to-Hubble ratio $T/H$. Blue and red regions correspond to quantum- and thermal-dominated regimes, respectively. Dashed and solid curves represent the respective GR and modified-gravity cases. As can be seen, the gravity modifications considered here enlarge the region of parameter space in which quantum perturbations dominate the scalar spectrum.}
     \label{fig:sps_DEMO}
\end{figure}

We can already identify a new feature arising from the combined consideration of modified gravity and warm inflation. The ratio of the quantum and thermal contributions can be written as (see Eq.~\eqref{eq:bose-einstein hatted})
\begin{align}
    \frac{\mathcal{P}_{\delta\varphi}^{\textrm{(qu)}}}{\mathcal{P}_{\delta\varphi}^{\textrm{(th)}}} = \frac{\hat{H}}{\hat{T}}\frac{\sqrt{9+12\pi\tilde{Q}}}{6\pi\tilde{Q}}\left\{1+2\left[\exp\left(\frac{\hat{H}}{\hat{T}}\right)-1\right]^{-1}\right\}~,\label{how-can-this-not-be-labelled-is-beyond-me}
\end{align}
when evaluated at horizon crossing, $k = \hat{a}\hat{H}$. Depending on the values of $\hat{T}/\hat{H}$ and $\tilde{Q}$, the scalar power spectrum of the inflaton perturbations may be dominated either by thermal fluctuations or by quantum fluctuations. In Ref.~\cite{Ramos:2013nsa}, it was shown that, for $\hat{T}/\hat{H}>1$, quantum perturbations dominate for $\tilde{Q}\lesssim 0.5$, whereas thermal fluctuations dominate otherwise. These conclusions remain valid in the Einstein frame, and the dominant contribution is therefore the same in both frames. However, the corresponding parameter regimes need not coincide between the two frames. As illustrated in Fig.~\ref{fig:sps_DEMO}, the system can occupy different dissipative regimes in the defining and Einstein frames, owing to the relation $Q \simeq \tilde{Q}\bar{F}\bar{\mathcal{K}}>\tilde{Q}$ for positive $\bar{F}$ (see Eq.~\eqref{eq:Best-relation-ever}). Thus, a system that lies in the weak-dissipative regime in the Einstein frame can simultaneously lie in the strong-dissipative regime in the defining frame. Some care is required with terminology at this point, however. Warm inflation is defined by $T/H>1$, a condition that holds in either dissipative regime \cite{Berera:2008ar,Ramos:2013nsa}, and in either conformal frame during slow roll, given that $\hat{H}\simeq H/\sqrt{\bar{F}}$ (see Eqs.~\eqref{eq: relation_HhatH} and \eqref{theta1_small}). What distinguishes the two regimes is the dissipation ratio. We therefore reserve the terms `weak' and `strong' dissipation for the value of the dissipation ratio alone: $Q$ in the defining frame and $\tilde{Q}$ in the Einstein frame. Furthermore, the defining frame can be in the strong-dissipative regime while quantum perturbations remain the dominant contribution to the scalar power spectrum, even when $T/H\simeq \hat{T}/\hat{H}>1$. This behaviour has no analogue in the standard warm-inflation dynamics of GR and therefore represents a distinctive feature of the combined effects of modified gravity and dissipation. 

To compute the full scalar power spectrum, we use the asymptotic form of the Bessel functions for $\hat{z}\ll1$ (see Sec.~10.7 of Ref.~\cite{NIST:DLMF}):
\begin{align}
    Y_\alpha(\hat{z}) \simeq -\left(\frac{2}{\hat{z}}\right)^{\alpha}\frac{\Gamma(\alpha)}{\pi}~,
\end{align}
since superhorizon scales are the observationally relevant ones. Working to zeroth order in the slow-roll parameters gives $\alpha\simeq \nu$. The integral can then be approximated as \cite{Montefalcone:2023pvh}
\begin{align}
    \int_{\hat{z}}^{\infty}\dif\hat{z}'\,\hat{z}'^{2-4\nu}[G(\hat{z},\hat{z}')]^2\simeq \frac{8\pi}{\sqrt{9+12\pi\tilde{Q}}}~,\quad \textrm{for}\quad \hat{z}\ll1~,
\end{align}
yielding the full scalar power spectrum
\begin{align}
    \mathcal{P}_{\delta\varphi}\Big|_{k=\hat{a}\hat{H}} =\frac{\hat{H}^2}{4\pi^2}\left(1+2n+\frac{\hat{T}}{\hat{H}}\frac{2\pi\sqrt{3}\tilde{Q}}{\sqrt{3+4\pi\tilde{Q}}}\right),
\end{align}
evaluated at horizon crossing.

The power spectrum for the comoving curvature perturbation is computed using the following definition in the Einstein frame \cite{Weinberg:2003sw,Bastero-Gil:2011rva,Ballesteros:2023dno}: 
\begin{equation}
    \hat{\mathcal{R}}\equiv \hat{\psi}-\frac{\hat{H}}{\hat{\bar{\rho}}+\hat{\bar{P}}}\left(\hat{\Psi}_r -\frac{\dif \varphi}{\dif \hat{\tau}}\delta \varphi\right)=\hat{\psi}^{\hat{\chi}}-\frac{\hat{H}}{\hat{\bar{\rho}}+\hat{\bar{P}}}\left(\hat{\Psi}_r^{\hat{\chi}} -\frac{\dif \varphi}{\dif \hat{\tau}}\delta \varphi^{\hat{\chi}}\right),\label{eq: comoving_R_perturbation}
\end{equation}
where
\begin{equation}
    \hat{\bar{\rho}}+\hat{\bar{P}} = \left(\frac{\dif \varphi}{\dif \hat{\tau}}\right)^2+\frac{4}{3}\hat{\bar{\rho}}_r~.
\end{equation}
Two considerations are in order. First, the metric perturbation is first order in the slow-roll parameters, $\hat{\psi}^{\hat{\chi}}\propto\hat{\epsilon}_1$, and can therefore be consistently neglected when working to zeroth order in slow roll (as noted in footnote~\ref{just-one-thing-that-happens-to-be-relevant-later}). Second, at zeroth order in the slow-roll parameters, the radiation momentum perturbation is related to the inflaton momentum perturbation (see Eq.~\eqref{eq:scalar-field-momentum-canonical-field-Einstein}) as given in Ref.~\cite{Bastero-Gil:2019rsp}:
\begin{align}
\hat{\Psi}_r^{\hat{\chi}}\simeq\tilde{Q}\hat{\Psi}_\varphi^{\hat{\chi}}~.
\end{align}
Then, we have
\begin{align}
    \mathcal{P}_\s \equiv \frac{k^3}{2\pi^2}\int \dif^3\mathbf{k}'\,\langle \hat{\mathcal{R}}(\hat{z},\mathbf{k})\hat{\mathcal{R}}(\hat{z},\mathbf{k}')\rangle \simeq \frac{\hat{H}^2}{\left[\left(\frac{\dif \varphi}{\dif \hat{\tau}}\right)^2+\frac{4}{3}\hat{\bar{\rho}}_r\right]^2}\left(1+\tilde{Q}\right)^2\left(\frac{\dif \varphi}{\dif \hat{\tau}}\right)^2\mathcal{P}_{\delta\varphi} \simeq \frac{\hat{H}^2}{\left(\frac{\dif \varphi}{\dif \hat{\tau}}\right)^2}\mathcal{P}_{\delta\varphi}~,\label{another-equation-which-is-not-labelled-and-should-be}
\end{align}
where we employed $\frac{4}{3}\hat{\bar{\rho}}_r\simeq \tilde{Q}(\frac{\dif\varphi}{d\hat{\tau}})^2$. Using the slow-roll equation for the field $\varphi$, Eq.~\eqref{eq: Slow_varphi_1}, the scalar power spectrum can finally be written as
\begin{align}
    \mathcal{P}_\s = \frac{\hat{H}^2(1+\tilde{Q})^2}{8\pi^2\epsilon_U \m^2}\left(1+2n+\frac{\hat{T}}{\hat{H}}\frac{2\pi\sqrt{3}\tilde{Q}}{\sqrt{3+4\pi\tilde{Q}}}\right)~,\label{eq: fullpowerspectrum}
\end{align}
with $\epsilon_U$ given by Eq.~\eqref{eq:epsU-intermsof-phi}.

The above expression is valid only for a temperature-independent dissipation coefficient ($c=0$), and when the metric perturbations can be neglected. In addition, we have assumed that $\hat{T}$, $\tilde{\bar{\Upsilon}}$, and $\hat{H}$ remain approximately constant with respect to $\hat{z}$, as pointed out above Eq.~\eqref{how-can-this-not-be-labelled-is-beyond-me}. It was shown in Ref.~\cite{Montefalcone:2023pvh} that relaxing the latter two assumptions does not significantly affect the scalar power spectrum, whereas allowing for a temperature-dependent dissipation coefficient ($c\neq 0$) can lead to substantial changes in its behaviour. To parametrise these corrections, it is customary to write 
\begin{align}
    \mathcal{P}_\s\Big|_{\rm numerical} = \mathcal{P}_\s\Big|_{\rm analytic}\times f(\tilde{Q})~,
\end{align}
where $\mathcal{P}_\s\Big|_{\rm analytic}$ is given by Eq.~\eqref{eq: fullpowerspectrum}, and the function $f$ is obtained by numerically solving the coupled perturbation equations. The function $f$ has been computed numerically for several models of warm inflation employing the parametrisation of the dissipation coefficient given in Eq.~\eqref{eq: upsilon_param}, and for a range of inflaton potentials \cite{Bastero-Gil:2011rva, Montefalcone:2023pvh, Rodrigues:2025neh}. Analytical results can also be obtained in the strong-dissipation regime, $\tilde{Q}\gg 1$ \cite{Graham_2009}. Ref.~\cite{Montefalcone:2023pvh} found that $f(\tilde{Q})$ exhibits a universal behaviour that is independent of the inflaton potential when the noise term in the radiation-perturbation equation is neglected and the inflaton is thermalised with the radiation bath. This result was later confirmed in Ref.~\cite{Rodrigues:2025neh} even when the inflaton is \emph{not} thermalised with the bath. However, including the noise term in the radiation-perturbation equation breaks this universality. Previous analyses have also found that $f(\tilde{Q})$ depends only weakly on the inflaton field. Nonetheless, these analyses assumed the simple parametrisation of Eq.~\eqref{eq: upsilon_param}, which need not apply in modified gravity due to the additional $\bar{F}\bar{\mathcal{K}}$ contribution to the effective dissipation coefficient. The presence of this extra factor can therefore introduce a non-trivial field dependence that is not captured by the standard power-law parametrisation. A related example in the context of a non-canonical scalar field in GR can be found in Ref.~\cite{Ballesteros:2022hjk}.\footnote{For instance, one may have a certain function $\bar{F}$ such that $\bar{F}^{3/2}\bar{\mathcal{K}} = \phi^2$. In this case, a defining-frame dissipation coefficient $\bar{\Upsilon} \propto T^3$ gives rise to an Einstein-frame dissipation coefficient $\tilde{\bar{\Upsilon}}\propto T^3/\phi^2$. These correspond to two well-studied forms of dissipation for which $f$ has been computed for several inflaton potentials \cite{Bastero-Gil:2011rva,Montefalcone:2023pvh,Rodrigues:2025neh}. However, $\phi$ is not the canonical field in the Einstein frame. Consequently, the relevant parametrisation of the dissipation coefficient in terms of the canonical field $\varphi$ takes the form $\tilde{\Upsilon}\propto \hat{T}^3/h(\varphi)$ instead, for some non-trivial function $h(\varphi)$.} For these reasons, a more detailed numerical investigation of such effects is left for future work. The remainder of this work therefore focuses on the case of a temperature-independent dissipation coefficient. We then take $c=0$ in Eq.~\eqref{eq: upsilon_param} such that
\begin{equation}
    \label{eq:our-upsilon}\Upsilon(\Phi) =MC_{\Upsilon} \left(\frac{M}{\Phi}\right)^d~. 
\end{equation}     

In order to determine the tensor-to-scalar ratio $r$, we first need to compute the tensor power spectrum. For this purpose, Eq.~\eqref{eq:tensor-modes-TT-equationEinsteinframe} is rewritten in terms of derivatives with respect to the rescaled proper time, following the same procedure as for the scalar perturbation equations above: 
\begin{equation}
    \frac{\dif^2 \hat{h}_{ij}}{\dif \hat{\tau}^2}+3\hat{H}\frac{\dif \hat{h}_{ij}}{\dif \hat{\tau}}-\hat{a}^{-2}\partial_k \partial^{k} \hat{h}_{ij} = 0~.
    \label{eq:TTtensor-perturbation-eq-hatted-proper-time}
\end{equation}
As with the scalar perturbation equations discussed at the end of Sec.~\ref{sec:perturbation-equations-Longitudinal-gauge}, Eq.~\eqref{eq:TTtensor-perturbation-eq-hatted-proper-time} has the same form as its GR counterpart, with derivatives now taken with respect to the Einstein-frame proper time. We next choose the spacetime slicing for which $\n=a$. The corresponding time variable is the conformal time, conventionally denoted by $\eta$, and the associated slicing is referred to as the `conformal slicing'. As noted earlier, we keep the coordinates fixed under the conformal transformation. That is, the same conformal-time coordinate is also used in the hatted frame, implying $\hat{\n}=\hat{a}$. Indeed, the relation between an arbitrary slicing and the conformal one can be written as
\begin{equation}
    \dif \eta = \frac{\mathcal{N}}{a}\dif t = \frac{\hat{\mathcal{N}}}{\hat{a}}\dif t~,
\end{equation}
where $\n$ and $\hat{\n}$ denote the lapse functions associated with the slicings defined with respect to $t$ in the unhatted and hatted frames, respectively. In terms of the conformal time $\eta$, Eq.~\eqref{eq:TTtensor-perturbation-eq-hatted-proper-time} becomes
\begin{equation}
    \hat{h}_{ij}^{''}+2\hat{\mathcal{H}} \hat{h}_{ij}^{'}-\partial_k \partial^{k} \hat{h}_{ij} = 0~,
\end{equation}
where $\hat{\mathcal{H}}\equiv \hat{a}\hat{H}$ is the conformal Hubble parameter in the Einstein frame. Notice that $\mathcal{H}$ is frame-dependent because of the non-trivial transformation of the Hubble parameter $H$ given in Eq.~\eqref{eq:conformal-transformation-Hubbleexprate}.

The dimensionless tensor power spectrum is \cite{Baumann:2009ds,Diaz:2023tma}
\begin{equation}
    \mathcal{P}_{\textrm{t}} \equiv \frac{k^3}{2\pi^2}\sum_{p=+,\cross}\int \dif^3\mathbf{k}'\,\langle \hat{h}^p(\eta,\mathbf{k})\hat{h}^p(\eta,\mathbf{k}')\rangle \simeq \frac{2}{\pi^2}\left(\frac{\hat{H}}{\m}\right)^2~,\label{eq: tensor_power_spectrum}
\end{equation}
which is equivalent to the result for cold inflation. We decompose the tensor perturbation as
\begin{equation}
    \hat{h}_{ij}(\eta,\mathbf{k}) = \sum_{p=+,\cross} \hat{h}^p(\eta,\mathbf{k}) \varepsilon_{ij}^{p}(\mathbf{k})~,
\end{equation}
with $\varepsilon_{ij}$ the polarisation tensors satisfying 
\begin{equation}
    k^{i}\varepsilon_{ij}^{p}=0~, \qquad \delta^{ij}\varepsilon_{ij}^{p}=0~, \qquad \varepsilon_{ij}^{p}(\mathbf{k})\varepsilon^{p'*}_{ij}(\mathbf{k}) = 2\delta_{pp'}~.
\end{equation}
For transverse-traceless tensor perturbations, we follow the standard warm-inflation literature and assume that gravitons (spin-2 modes) do not thermalise with the radiation bath, so that their initial state remains the Bunch-Davies vacuum \cite{Ramos:2013nsa,Bartrum:2014fla,Bastero-Gil:2016qru,Levy:2020zfo}. In the presence of a thermal environment, if the gravitons were in thermal equilibrium, stimulated emission could contribute to the tensor power spectrum \cite{Bhattacharya:2006dm,Qiu:2021ytc}. This would introduce a factor $1+n$ in Eq.~\eqref{eq: tensor_power_spectrum}, analogous to the corresponding factor in the scalar power spectrum. However, if the couplings to the graviton were strong enough to generate such a thermal-like spectrum, the temperature required for this to occur, $T\sim \m$, would exceed the usual values during warm inflation. For this reason, we do not include this factor in the tensor power spectrum. Non-standard scenarios in which the thermal bath contributes to the tensor power spectrum have also been considered in Ref.~\cite{Kamali:2023lzq} and references therein.

We can therefore see that only the curvature power spectrum is significantly affected by dissipation. Consequently, the scalar spectral index $n_\s$ and tensor-to-scalar ratio $r$ differ from their values in standard cold-inflation scenarios, which we compute next:
\begin{align}
    n_\s-1 &= \frac{\dif \log \mathcal{P}_\s}{\dif \log k} \simeq \frac{\dif \log \mathcal{P}_\s}{\dif \hat{N}}~,\\
    r &= \frac{\mathcal{P}_{\rm t}}{\mathcal{P}_\s}  = \frac{16\epsilon_U}{(1+\tilde{Q})^2}\left(1+\frac{2}{e^{H/T}-1}+\frac{T}{H}\frac{2\pi\sqrt{3}\tilde{Q}}{\sqrt{3+4\pi \tilde{Q}}}\right)^{-1}\equiv \frac{16\epsilon_U}{\Delta_T}~,\label{eq: str_r}
\end{align}
where the number of e-folds is defined by $\dif \hat{N} = \hat{H}\dif\hat{\tau}$, and
\begin{equation}
    \dif \log k = \dif \hat{N}(1-\hat{\epsilon}_1)\simeq \dif\hat{N}~,
\end{equation}
at horizon crossing and to zeroth order in slow roll. The function $\Delta_T$ incorporates all thermal contributions to the power spectrum:
\begin{align}
    \Delta_T\equiv\left(1+\tilde{Q}\right)^2\left(1+2n+\frac{\hat{T}}{\hat{H}}\frac{2\pi\sqrt{3}\tilde{Q}}{\sqrt{3+4\pi\tilde{Q}}}\right).
\end{align} 
Taking into account Eq.~\eqref{eq: fullpowerspectrum}, we have
\begin{align}
    \frac{\dif \log \mathcal{P}_\s}{\dif \hat{N}}
    \simeq 2\frac{\dif \hat{H}}{\hat{H}\dif \hat{N}}-\frac{\dif \epsilon_U}{\epsilon_U\dif \hat{N}}+\frac{\dif \Delta_T}{\Delta_T\dif \hat{N}}=\frac{2\eta_U-6\epsilon_U}{1+\tilde{Q}}+\frac{\dif \Delta_T}{\Delta_T\dif \hat{N}}~,\label{eq: dlogPrdN}
\end{align}
where Eq.~\eqref{eq:the-below-equation} was employed, together with
\begin{align}
    \frac{\dif \epsilon_U}{\dif\hat{N}} = \frac{2\epsilon_U}{1+\tilde{Q}}(2\epsilon_U-\eta_U)~.
\end{align}
The last term in Eq.~\eqref{eq: dlogPrdN} can then be written as
\begin{align}
    \frac{\dif \Delta_T}{\Delta_T\dif \hat{N}}&=\frac{(1+\tilde{Q})^2}{\Delta_T}\left\{\frac{\dif (\hat{T}/\hat{H})}{ \dif \hat{N}}\left[\frac{2e^{\hat{H}/\hat{T}}}{(\hat{T}/\hat{H})^2}n^2+\frac{2\pi\sqrt{3}\tilde{Q}}{\sqrt{3+4\pi\tilde{Q}}}\right]+\frac{\dif\tilde{Q}}{ \dif\hat{N}}\frac{\hat{T}}{\hat{H}}\frac{2\pi\sqrt{3}}{\sqrt{3+4\pi\tilde{Q}}}\left(\frac{3+2\pi\tilde{Q}}{3+4\pi\tilde{Q}}\right)\right\}+\frac{2\dif\tilde{Q}}{(1+\tilde{Q})\dif\hat{N}}~.
\end{align}
Applying the slow-roll equations~\eqref{eq: Slow_varphi_1} and \eqref{eq:rad_slowrolllll} for the radiation and $\varphi$ fields, we obtain
\begin{align}
    \left(\frac{\hat{T}}{\hat{H}}\right)^4 &= \frac{90}{2\pi^2g_{\textrm{eff}}}\frac{\epsilon_U}{1+\tilde{Q}}\frac{\tilde{Q}}{1+\tilde{Q}}\left(\frac{\m}{\hat{H}}\right)^2~,\label{eq: THforth}\\
    \frac{\dif \hat{T}/\hat{H}}{\dif\hat{N}} &=\frac{\hat{T}}{4\hat{H}}\left[\frac{2(2\epsilon_U-\eta_U)}{1+\tilde{Q}}+\left(\frac{1-\tilde{Q}}{1+\tilde{Q}}\right)\frac{\dif\tilde{Q}}{\tilde{Q}\dif\hat{N}}+\frac{2\epsilon_U}{1+\tilde{Q}}\right],\\
    \frac{1}{\tilde{Q}}\frac{\dif \tilde{Q}}{\dif \hat{N}} &= \frac{1}{\tilde{Q}}\frac{\dif}{\dif \hat{N}}\left(\frac{\tilde{\Upsilon}}{3\hat{H}}\right) = -\frac{\tilde{\beta}_\Upsilon}{1+\tilde{Q}}+\frac{\epsilon_U}{1+\tilde{Q}}~.
\end{align}
$g_{\textrm{eff}}$ is the number of relativistic degrees of freedom such that
\begin{equation}
    \label{eq:radiation-energy-density-in-geff}\rho_r =\frac{\pi^2}{30}g_{\textrm{eff}}T^4~.
\end{equation}
The full expression for the scalar spectral index is then
\begin{align}
    \nonumber& n_{\s} \simeq 1+ \frac{1}{1+\tilde{Q}}\left[-6\epsilon_U +2\eta_U +\frac{2\tilde{Q}}{1+\tilde{Q}}\left(\epsilon_U -\tilde{\beta}_{\Upsilon}\right)\right]+\frac{T}{4H(1+\tilde{Q})}\left[1+\frac{2}{e^{H/T}-1}\right.\\
    \nonumber&\left.+\frac{T}{H}\frac{2\pi\sqrt{3}\tilde{Q}}{\sqrt{3+4\pi\tilde{Q}}}\right]^{-1}\left\{\left[6\epsilon_U-2\eta_U+\frac{\left(1-\tilde{Q}\right)\left(\epsilon_U -\tilde{\beta}_{\Upsilon}\right)}{1+\tilde{Q}}\right]\left[\frac{2\pi\sqrt{3}\tilde{Q}}{\sqrt{3+4\pi \tilde{Q}}}+\frac{2e^{H/T}}{\left(e^{H/T}-1\right)^2}\frac{H^2}{T^2}\right]\right.\\
    &\left.\hspace{27.6em}+\frac{8\pi\sqrt{3}\tilde{Q}}{\sqrt{3+4\pi \tilde{Q}}}\left(\epsilon_U -\tilde{\beta}_{\Upsilon}\right)\left(1-\frac{2\pi \tilde{Q}}{3+4\pi \tilde{Q}}\right)\right\}~.\label{eq: full_ns}
\end{align}

We can now analyse the different regimes in which warm inflation may take place, along with their corresponding predictions for the scalar spectral index and tensor-to-scalar ratio. Table~\ref{tab: inflationary_observables} summarises the inflationary observables and their corresponding predictions. Since the regime $\hat{T}/\hat{H}\gg1$ is the relevant one for warm inflation, only this one is displayed in the table.

\begin{table}[ht!]
    \centering
    \renewcommand{\arraystretch}{2.5}
    \begin{tabular}{@{}c c c c@{}}
        \toprule
        \textbf{Observable} & \hspace{0.6cm} \textbf{Regime} \hspace{0.6cm} & \textbf{Einstein Frame} & \textbf{Defining Frame} \\
        \midrule

        $\mathcal{P}_{\s}$
        & --
        & $\displaystyle \frac{\bar{U}(1+\tilde{Q})}{24\pi^2\hat{\epsilon}_1\m^4}\left(1+2n+\frac{\hat{T}}{\hat{H}}\frac{2\pi\sqrt{3}\tilde{Q}}{\sqrt{3+4\pi\tilde{Q}}}\right)$
        & $\displaystyle \frac{\bar{V}}{24\pi^2\bar{F}^2\gamma_1^2\m^4}\left(1+2n+\frac{T}{H}\frac{2\pi\sqrt{3}\frac{Q}{\bar{F}\bar{\mathcal{K}}}}{\sqrt{3+4\pi\frac{Q}{\bar{F}\bar{\mathcal{K}}}}}\right)$ \\

        \midrule

        $n_{\s}$
        & \hspace{0.3cm} $\tilde{Q} \gg 1$ \hspace{0.3cm}
        & $\displaystyle 1+\frac{1}{\tilde{Q}}\left(-\frac{9}{4}\epsilon_U-\frac{9}{4}\tilde{\beta}_\Upsilon+\frac{3}{2}\eta_U\right)$
        & $\displaystyle 1-\frac{3}{2}\left(\gamma_1^2\frac{Q}{\bar{F}\bar{\mathcal{K}}}+\gamma_2+\theta_1-\theta_1\theta_2+\theta_\mathcal{K}-\theta_Q\right)$ \\

        \phantom{blank}
        & \hspace{0.3cm} $\tilde{Q} \ll 1$ \hspace{0.3cm}
        & $\displaystyle 1+\frac{1}{4}\left(6\eta_U-17\epsilon_U-\tilde{\beta}_\Upsilon\right)$
        & $\displaystyle 1-\frac{3}{2}\left(\gamma_1^2+\gamma_2-\theta_1\theta_2\right)+\frac{1}{2}\left(\theta_Q-\theta_\mathcal{K}-\theta_1\right)$ \\

        \midrule

        $r$
        & \hspace{0.3cm} $\tilde{Q} \gg 1$ \hspace{0.3cm}
        & $\displaystyle \frac{16\epsilon_U}{\sqrt{3\pi}\tilde{Q}^{5/2}}\frac{\hat{H}}{\hat{T}}$
        & $\displaystyle \frac{16\gamma_1^2}{\sqrt{3\pi}\left(\frac{Q}{\bar{F}\bar{\mathcal{K}}}\right)^{1/2}}\frac{H}{T}$ \\

        \phantom{blank}
        & \hspace{0.3cm} $\tilde{Q} \ll 1$ \hspace{0.3cm}
        & $\displaystyle 8\epsilon_U\frac{\hat{H}}{\hat{T}}$
        & $\displaystyle 8\gamma_1^2\frac{H}{T}$ \\

        \bottomrule
    \end{tabular}
    \caption{Scalar power spectrum $\mathcal{P}_\s$, scalar spectral index $n_\s$, and tensor-to-scalar ratio $r$ in the Einstein and defining frames. $\tilde{Q}$ and $Q$ are related by Eq.~\eqref{eq:Best-relation-ever}, and $n$ is the Bose-Einstein distribution function presented in Eq.~\eqref{eq:Bose-and-Einstein-dont-care-about-frames}. The parameters $\theta_1$ and $\theta_2$ are introduced in Eq.~\eqref{eq: thetas_def}, while the potential parameters $\epsilon_U$ and $\eta_U$ follow from Eqs.~\eqref{eq:epsU-intermsof-phi} and \eqref{eq:etaU-intermsof-phi}, respectively. The auxiliary parameters $\gamma_1^2$ and $\gamma_2$ are given in Eq.~\eqref{eq: gamma_slowroll}. The $\tilde{\beta}_{\Upsilon}$ parameter is presented in Eq.~\eqref{eq:parameter-beta-Upsilon-Einstein}, and $\theta_{\mathcal{K}}$ and $\theta_Q$ are defined in Eqs.~\eqref{eq:def-thetaK} and \eqref{eq: def_thetaQ}, respectively. Only the regime $\hat{T}\gg\hat{H}$ is considered for $n_\s$ and $r$.}
    \label{tab: inflationary_observables}
\end{table}

\subsubsection{Cold Limit}
\noindent
In the regime $\tilde{Q}\ll1$ and $\hat{T}/\hat{H}\ll1$, we recover, as expected, the cold-inflation results with modified-gravity contributions \cite{Karciauskas:2022jzd}:
\begin{align}
    \mathcal{P}_\s &\simeq \frac{\bar{U}}{24\pi^2 \epsilon_U \m^4}~,\\
    n_\s-1&\simeq 2\eta_U-6\epsilon_U~,\\
    r &\simeq 16\epsilon_U~.
\end{align}
In this regime, the Hubble-flow parameters reduce to
\begin{align}
    \hat{\epsilon}_1\simeq \epsilon_U\simeq \gamma_1^2~,\qquad \hat{\epsilon}_2\simeq 2(\gamma_2-\theta_1\theta_2)\simeq 4\epsilon_U-2\eta_U~,
\end{align}
allowing us to express the scalar spectral index and tensor-to-scalar ratio in terms of the defining-frame quantities:
\begin{align}
    n_\s-1&\simeq-2(\gamma_1^2+\gamma_2-\theta_1\theta_2)~,\\
    r&\simeq 16\gamma_1^2~.\label{eq: r_cold_1}
\end{align}

We can still compute the leading-order corrections by noting that\footnote{From Eq.~\eqref{eq: THforth}, we have $\hat{T}/\hat{H}\sim(\epsilon_U\tilde{Q})^{1/4}$. Then $\Delta_T/(1+\tilde{Q})^2 \sim1+2\pi\epsilon_U^{1/4}\tilde{Q}^{5/4}$, and the approximation adopted here remains valid at least up to $\mathcal{O}(\tilde{Q})$.} $\Delta_T\simeq (1+\tilde{Q})^2$, and
\begin{align}
    n_\s-1&\simeq 2\eta_U-6\epsilon_U+(8\epsilon_U-2\eta_U-2\tilde{\beta}_\Upsilon)\tilde{Q}~,\\
    r&\simeq \frac{16\epsilon_U}{(1+\tilde{Q})^2}~.
\end{align}
Using Eqs.~\eqref{eq:the-below-equation}, \eqref{eq: the-below-equation-2}, \eqref{eq: hateps1_df}, and \eqref{eq: epshat2-2}, we obtain
\begin{align}
    \epsilon_U &\simeq (1+\tilde{Q})^2\gamma_1^2~,\\
    \hat{\epsilon}_2 &\simeq \frac{4\epsilon_U-2\eta_U}{1+\tilde{Q}}-\frac{1}{1+\tilde{Q}}\frac{\dif \tilde{Q}}{\dif \hat{N}}\simeq 2(\gamma_2-\theta_1\theta_2)+\frac{1}{1+\tilde{Q}}\frac{\dif \tilde{Q}}{\dif \hat{N}}~,
\end{align}
together with
\begin{align}
    &\frac{\dif \tilde{Q}}{\dif \hat{N}}= \tilde{Q}\left(2\theta_Q-2\theta_\mathcal{K}-2\theta_1\right),\\
    &(1+\tilde{Q})^2\simeq 1+2\tilde{Q}~,
\end{align}
the slow-roll parameters in the Einstein frame can be expressed in terms of the defining-frame slow-roll parameters as
\begin{align}
    \epsilon_U&\simeq \gamma_1^2(1+2\tilde{Q})~,\label{eq: relation1_smallQ}\\
    \eta_U &\simeq2\gamma_1^2(1+2\tilde{Q)}-(\gamma_2-\theta_1\theta_2)(1+\tilde{Q})-2\tilde{Q}\left(\theta_Q-\theta_\mathcal{K}-\theta_1\right)~,\label{eq: relation2_smallQ}\\
    \tilde{\beta}_\Upsilon &\simeq \gamma_1^2(1+2\tilde{Q})-2(1+\tilde{Q})(\theta_Q-\theta_{\mathcal{K}}-\theta_1)~.\label{eq: relation3_smallQ}
\end{align}
Thus, the scalar spectral index and the tensor-to-scalar ratio in the defining frame become
\begin{align}
    n_\s-1&\simeq -2(\gamma_1^2+\gamma_2-\theta_1\theta_2)-2\frac{Q}{\bar{F}\bar{\mathcal{K}}}\gamma_1^2~,\\
    r&\simeq 16\gamma_1^2~,
\end{align}
where we retain terms up to $\mathcal{O}(\tilde{Q})$.

\subsubsection{Warm Inflation: Strong Dissipation}
\noindent
In the regime $\tilde{Q}\gg 1$ and $\hat{T}\gg \hat{H}$, the function $\Delta_T$ is given by $\Delta_T \simeq \sqrt{3\pi}\tilde{Q}^{5/2}\hat{T}/\hat{H}$, such that the scalar power spectrum can be approximated as
\begin{align}
    \mathcal{P}_\s\simeq \frac{\hat{H}^2\tilde{Q}^{5/2}}{8\pi^2\epsilon_U\m^2}\sqrt{3\pi}\frac{\hat{T}}{\hat{H}}~.
\end{align}
Consequently, the inflationary observables are
\begin{align}
    n_\s-1&\simeq \frac{1}{\tilde{Q}}\left(-\frac{9}{4}\epsilon_U-\frac{9}{4}\tilde{\beta}_\Upsilon+\frac{3}{2}\eta_U\right),\label{eq: nslargeQlargeT}\\
    r&\simeq \frac{16\epsilon_U}{\sqrt{3\pi}\tilde{Q}^{5/2}}\frac{\hat{H}}{\hat{T}}~.
\end{align}
The Hubble-flow parameters can also be written in this regime. Using Eqs.~\eqref{eq:the-below-equation}, \eqref{eq: the-below-equation-2}, \eqref{eq: hateps1_df}, and \eqref{eq: epshat2-2}, we find
\begin{align}
    &\hat{\epsilon}_1 \simeq \frac{\epsilon_U}{\tilde{Q}} \simeq \tilde{Q}\gamma_1^2~,\\
    &\hat{\epsilon}_2 \simeq \frac{4\epsilon_U-2\eta_U}{\tilde{Q}}-\frac{1}{\tilde{Q}}\frac{\dif \tilde{Q}}{\dif \hat{N}}\simeq 2(\gamma_2-\theta_1\theta_2)+\frac{1}{\tilde{Q}}\frac{\dif \tilde{Q}}{\dif \hat{N}}~,\label{eq: eps2_IVB}
\end{align}
where we have used the approximation $1+\tilde{Q}\simeq \tilde{Q}$. Combining the results above, we can express all quantities in terms of the defining-frame slow-roll parameters:
\begin{align}
    \eta_U &=2\tilde{Q}^2\gamma_1^2-\tilde{Q}\left(\gamma_2-\theta_1\theta_2+2\theta_Q-2\theta_\mathcal{K}-2\theta_1\right),\\
    \tilde{\beta}_\Upsilon &= \tilde{Q}^2\gamma_1^2-2\tilde{Q}(\theta_Q-\theta_\mathcal{K}-\theta_1)~.
\end{align}
Finally, substituting the last relations into Eqs.~\eqref{eq: nslargeQlargeT} and \eqref{eq: str_r}, we obtain
\begin{align}
    n_\s-1&\simeq -\frac{3}{2}\left(\gamma_1^2\frac{Q}{\bar{F}\bar{\mathcal{K}}}+\gamma_2+\theta_1-\theta_1\theta_2+\theta_\mathcal{K}-\theta_Q\right)~,\label{eq: ns_strongdiss}\\
    r&\simeq  \frac{16\gamma_1^2}{\sqrt{3\pi}\left(\frac{Q}{\mathcal{K}F}\right)^{1/2}}\frac{H}{T}~.
\end{align}

\subsubsection{Warm Inflation: Weak Dissipation}
\noindent
For $\tilde{Q}\ll1$ and $\hat{T}/\hat{H}\gg1$, we have $\Delta_T\simeq 2\hat{T}/\hat{H}$, resulting in the scalar power spectrum
\begin{align}
    \mathcal{P}_\s\simeq \frac{\hat{H}\hat{T}}{4\pi^2\epsilon_U\m^2}~.
\end{align}
The inflationary observables are then given by
\begin{align}
    n_\s-1&\simeq \frac{1}{4}\left(6\eta_U-17\epsilon_U-\tilde{\beta}_\Upsilon\right),\\
    r&\simeq 8\epsilon_U\frac{\hat{H}}{\hat{T}}~.
\end{align}
Using the relations in Eqs.~\eqref{eq: relation1_smallQ}--\eqref{eq: relation3_smallQ} and retaining only terms of $\mathcal{O}(\tilde{Q}^0)$, we obtain, in terms of the defining-frame quantities,
\begin{align}
    n_\s-1&\simeq -\frac{3}{2}\left(\gamma_1^2+\gamma_2-\theta_1\theta_2\right)+\frac{1}{2}\left(\theta_Q-\theta_\mathcal{K}-\theta_1\right),\\
    r&\simeq 8\gamma_1^2\frac{H}{T}~.
\end{align}
It is worth mentioning that this last regime is the most common one for single-field warm-inflation models. Compared with Eq.~\eqref{eq: r_cold_1}, the presence of a temperature larger than the Hubble scale during inflation, even in the weak-dissipative regime, significantly suppresses the tensor-to-scalar ratio. This suppression can, however, be compensated if the gravitons thermalise with the thermal bath, as discussed above. Moreover, it is possible in the last regime to have $Q\gg 1$ and $T/H\gg1$. This means that a strong-dissipative regime in the defining frame can give rise to inflationary observables characteristic of a weak-dissipative regime in GR.

\subsection{The Case of Non-Minimally Coupled Quartic Warm Inflation}
\noindent
Monomial cold inflation in GR is by now strongly constrained. The $\lambda\Phi^{4}$ potential is excluded by Planck~\cite{Planck:2018jri} and BICEP/Keck~\cite{BICEP:2021xfz} at high confidence, whilst even the milder $\lambda\Phi^{2}$ case is in increasing tension with the latest CMB data from ACT DR6~\cite{AtacamaCosmologyTelescope:2025blo,AtacamaCosmologyTelescope:2025nti}. Two physically distinct mechanisms have been shown to bring monomial potentials back into agreement with the data: the introduction of a non-minimal coupling, which flattens the effective Einstein-frame potential at large field values~\cite{Spokoiny:1984bd,Bezrukov:2007ep}, and the warm-inflation mechanism itself, in which the dissipative friction suppresses both the field excursion and the tensor-to-scalar ratio~\cite{Bartrum:2013fia,Kamali:2023lzq}. The combination of the two is therefore well motivated phenomenologically, while remaining simple enough to admit a fully analytical slow-roll treatment.

Thus, as an application of the formalism developed in the preceding sections, we consider a monomial inflationary potential, namely the quartic potential
\begin{equation}
    \label{eq:the-potentials-we-consider}V(\Phi) = \frac{\lambda}{4}\Phi^{4}~,
\end{equation}
with a dimensionless self-coupling constant $\lambda$, supplemented by a non-minimal coupling function of the form
\begin{equation}
    F(\Phi) = 1+\xi \left(\frac{\Phi}{\m}\right)^2~,
\end{equation}
where $\xi$ is the dimensionless non-minimal coupling constant, which we take to be positive (a well-known negative value is the conformal coupling $\xi=-1/6$ \cite{Birrell:1982ix,Faraoni:1996rf}). The quadratic non-minimal coupling, combined with a $\lambda \Phi^4$ potential, reproduces the Higgs-inflation setup of Bezrukov and Shaposhnikov \cite{Bezrukov:2007ep,Rubio:2018ogq,Cheong:2021vdb}, and provides the prototype for $\xi$-attractor models, reviewed in Ref.~\cite{Kallosh:2025ijd} (see also Ref.~\cite{Kallosh:2013tua}). In what follows, we estimate the inflationary observables during slow roll for two cases: a constant and a quadratic temperature-independent dissipation coefficient ($d=0$ and $d=-2$ in Eq.~\eqref{eq:our-upsilon}). To confront these predictions with observational data, however, one must specify the number of e-folds before the end of inflation at which the inflationary observables are evaluated. It is customary to consider $50$--$60$ e-folds, in line with the Planck team~\cite{Planck:2018jri}. We associate the observational benchmark with the Einstein-frame e-folding, consistent with the fact that the Einstein frame is the one in which the slow-roll conditions are interpreted and the end of inflation is subsequently determined.

The Einstein-frame number of e-folds between horizon crossing of the pivot mode and the end of inflation is defined as
\begin{equation}
    \label{eq:efolds-Einstein-def}
    \hat{N}_{\star} \equiv \ln \left(\frac{\hat{a}_{\textrm{end}}}{\hat{a}_{\star}}\right)=\int_{\hat{\tau}_{\star}}^{\hat{\tau}_{\mathrm{end}}}\mathrm{d}\hat{\tau}\,\hat{H} = -\int_{\varphi_{\mathrm{end}}}^{\varphi_{\star}}\mathrm{d}\varphi\,\frac{\hat{H}}{\mathrm{d}\varphi/\mathrm{d}\hat{\tau}}~,
\end{equation}
where $\varphi_{\star}$ and $\varphi_{\mathrm{end}}$ denote the values of the canonical Einstein-frame inflaton at pivot-mode crossing and at the end of the slow-roll phase, respectively. The latter is determined by the condition $\epsilon_U = 1+\tilde{Q}$ (see Eq.~\eqref{eq:slow-rollepsu}). Substituting the Einstein-frame slow-roll equations \eqref{eq: Slow_varphi_1} and \eqref{eq: Slow_hubble_1}, we obtain
\begin{equation}
    \hat{N}_{\star} \simeq \m^{-2}\int^{\varphi_\star}_{\varphi_{\textrm{end}}}\mathrm{d}\varphi\,\frac{\bar{U}}{\bar{U}_{,\varphi}}\left(1+\tilde{Q}\right).
\end{equation}
In terms of the defining-frame variables (see Eqs.~\eqref{eq:non-canonical-function}, \eqref{eq:potentialU}, and \eqref{eq:Best-relation-ever}), this becomes
\begin{equation}
    \hat{N}_{\star} = \hat{N}(\phi_{\star}) \simeq \m^{-2} \int^{\phi_{\star}}_{\phi_{\textrm{end}}} \dif \phi\,\bar{\mathcal{K}}\left(1+\frac{Q}{\bar{F}\bar{\mathcal{K}}}\right)\left(\frac{\bar{V}_{,\phi}}{\bar{V}}-2\frac{\bar{F}_{,\phi}}{\bar{F}}\right)^{-1}~.\label{eq: Ne_sec4}
\end{equation}
Taking 
\begin{equation}
    V(\phi) = \frac{\lambda}{4}\phi^4~, \qquad \Upsilon(\phi) = \frac{M^{1+d}C_{\Upsilon}}{\phi^{d}}~, \qquad F(\phi) = 1+\xi \left(\frac{\phi}{\m}\right)^2~,\label{eq: VUPF_sec4}
\end{equation}
with $\phi$ being the homogeneous scalar field, one obtains
\begin{equation}
    \bar{\mathcal{K}} = \frac{\m^2}{\left(\m^2+\xi \phi^2\right)^2}\left[\m^2+\left(1+6\xi\right)\xi \phi^2\right] \Rightarrow \bar{F}\bar{\mathcal{K}} = \frac{\m^2+(1+6\xi)\xi \phi^2}{\m^2+\xi \phi^2}~.
\end{equation}
In this way, we can write the dimensionless parameter
\begin{equation}
   \frac{Q}{\bar{F}\bar{\mathcal{K}}}=\frac{M^{1+d}C_{\Upsilon}}{\phi^{2+d}}\frac{2}{\sqrt{3\lambda}}\frac{\left(\m^2+\xi \phi^2\right)^{3/2}}{\m^2+\left(1+6\xi\right)\xi \phi^2}~,\label{eq: KQ_sec4}
\end{equation}
where we recall that $C_\Upsilon$ is dimensionless (see Eq.~\eqref{eq:our-upsilon}). The ratio between the temperature and the Hubble parameter is
\begin{align}
    \frac{T}{H}\simeq \frac{\m}{\phi}\left[\frac{18}{\lambda C_r}\frac{Q}{\bar{F}\bar{\mathcal{K}}}\left(1+\frac{Q}{\bar{F}\bar{\mathcal{K}}}\right)^{-2}\epsilon_U(\phi) \left(1+\xi \frac{\phi^2}{\m^2}\right)^2\right]^{1/4}~.\label{eq: TH_sec4}
\end{align}
The constant 
\begin{equation}
    C_r \equiv \frac{\pi^2 g_{\textrm{eff}}}{30}
\end{equation}
accounts for the effective number of relativistic degrees of freedom in the thermal bath, such that $\rho_r = C_r T^4$ (see Eq.~\eqref{eq:radiation-energy-density-in-geff}), for which we may take $g_{\textrm{eff}} = 106.75$ \cite{Husdal:2016haj}.\footnote{The exact value for $g_{\textrm{eff}}$ depends on the specific warm-inflation model and the particle content of the thermal bath during inflation \cite{Berera:2008ar,Bastero-Gil:2009sdq,Kamali:2023lzq}. However, this only introduces a dependence on $g_{\textrm{eff}}^{1/4}$ in the dynamics (see Eq.~\eqref{eq: TH_sec4}).} This value is used throughout the viability maps of Secs.~\ref{subsub:constant-dissipation} and \ref{subsub:inversesquared-dissipation}, which therefore display the dependence on the parameters relevant to the inflaton dynamics for a fixed and representative particle content. For the benchmark points collected in the tables therein, by contrast, we allow $g_{\textrm{eff}}$ to vary within the range $[10,200]$, spanning particle contents from a minimal bath of a few light species up to supersymmetric extensions of the Standard Model \cite{Bastero-Gil:2009sdq}, treating $C_r$ as an additional free parameter. Substituting Eqs.~\eqref{eq: VUPF_sec4} and \eqref{eq: KQ_sec4} into Eq.~\eqref{eq: Ne_sec4}, we obtain
\begin{equation}
    \hat{N}(\phi_{\star}) \simeq (2\m)^{-2} \int^{\phi_{\star}}_{\phi_{\textrm{end}}} \dif \phi\,\left\{\frac{\m^2[\m^2+(1+6\xi)\xi\phi^2]}{(\m^2+\xi \phi^2)^2}+\frac{M^{1+d}C_{\Upsilon}}{\phi^{2+d}}\frac{2}{\sqrt{3\lambda}}\frac{\m^{2}}{(\m^2+\xi \phi^2)^{1/2}}\right\}\left(\frac{1}{\phi}-\frac{\xi \phi}{\m^2+\xi \phi^2}\right)^{-1}.
\end{equation}
A general solution can be written as
\begin{equation}
    \hat{N}(\phi_\star) \simeq \left[\mathcal{V}(\phi)+\mathcal{D}(\phi)\right]^{\phi_\star}_{\phi_{\textrm{end}}}~,
\end{equation}
where $\mathcal{V}$ is independent of $d$:
\begin{equation}
        \mathcal{V}(\phi) =-\frac{3}{4}\ln \left(1+\xi \frac{\phi^2}{\m^2}\right)+\frac{(1+6\xi)\phi^2}{8\m^2}~,
    \end{equation}
while $\mathcal{D}$ depends on the choice of $d$ and takes the following forms:
\begin{itemize}
\item $d =0$:
    \begin{equation}
        \label{eq:case-d-zero}\mathcal{D}(\phi) = \frac{M C_{\Upsilon}}{2\m \sqrt{3\lambda}}\left[\ln\left(\frac{\phi}{\m+\sqrt{\m^2+\xi \phi^2}}\right)+\sqrt{1+\xi \frac{\phi^2}{\m^2}}\right].
    \end{equation}
    \item $d \neq 0, 2, 4, \ldots\,$:
    \begin{equation}
        \label{eq:case-d-nonzero}\mathcal{D}(\phi) =-\frac{M^{1+d}C_{\Upsilon}}{2d\sqrt{3\lambda}\m^{1+d}}\left(\frac{\m}{\phi}\right)^d\,{}_2F_1\left(-\frac{1}{2},-\frac{d}{2};1-\frac{d}{2};-\frac{\xi \phi^2}{\m^2}\right)~.
    \end{equation}
\end{itemize}
The function ${}_2F_1$ denotes the hypergeometric function (see Sec.~15.1 of Ref.~\cite{NIST:DLMF}). Note that Eq.~\eqref{eq:case-d-nonzero} diverges as $d\rightarrow0$, owing to the $1/d$ prefactor. The divergent piece is nevertheless independent of $\phi$, so that it cancels in $\left[\mathcal{D}\right]^{\phi_\star}_{\phi_{\textrm{end}}}$ and Eq.~\eqref{eq:case-d-zero} is correctly recovered in that limit. The values $d = 2,4,\ldots$ are excluded because $1-d/2$ is then a non-positive integer, making the hypergeometric function singular: the divergence is again $\phi$-independent, but the resulting limits contain logarithmic contributions that must be evaluated case by case. We refrain from quoting the corresponding expressions here, since we restrict ourselves to $d=0$ and $d=-2$ in what follows.

The inflationary slow-roll parameters relevant to this analysis are (see Eqs.~\eqref{eq:epsU-intermsof-phi}, \eqref{eq:etaU-intermsof-phi}, and \eqref{eq:parameter-beta-Upsilon-Einstein})
\begin{align}
    \epsilon_U(\phi) &= \frac{8\m^4}{\phi^2\left[\m^2+\left(1+6\xi\right)\xi \phi^2\right]}~,\label{eq: slowrolleps_sec5}\\
    \eta_U(\phi) &= \frac{4\m^2\left[3\m^4+\m^2\left(1+12\xi\right)\xi \phi^2 -2\left(1+6\xi\right)\xi^2 \phi^4\right]}{\phi^2\left[\m^2+\left(1+6\xi\right)\xi \phi^2\right]^2}~,\label{eq: slowrolleta_sec5}\\
    \tilde{\beta}_{\Upsilon}(\phi) &= \frac{-4\m^2\left\{\left[\m^2\left(1+12\xi\right)+\left(1+6\xi\right)\xi \phi^2\right]\xi \phi^2+d\left(\m^2+\xi \phi^2\right)\left[\m^2+\left(1+6\xi\right)\xi \phi^2\right]\right\}}{\phi^2\left[\m^2+\left(1+6\xi\right)\xi \phi^2\right]^2}~,\label{eq: slowrollbeta_sec5}
\end{align}
where we note that $\tilde{\beta}_\Upsilon$ depends on $d$. Assuming $\xi \phi^2\gg \m^2$ and $d\neq -1$, these expressions can be approximated as
\begin{align}
    \epsilon_U(\phi) &\simeq \frac{8\m^4}{(1+6\xi)\xi \phi^4}~,\\
    \eta_U(\phi) &\simeq -\frac{8\m^2}{\left(1+6\xi\right)\phi^2}~,\\
    \tilde{\beta}_{\Upsilon}(\phi) &\simeq -\frac{4\m^2\left(1+d\right)}{\left(1+6\xi\right)\phi^2}\simeq\frac{1+d}{2}\eta_U(\phi)~.
\end{align}
Finally, the scalar power spectrum is given by (see Eq.~\eqref{eq: fullpowerspectrum})
\begin{align}
    \mathcal{P}_{\textrm{s}} \simeq \frac{\lambda \phi^4}{96\pi^2 \m^4 \epsilon_U}\left(1+\xi \frac{\phi^2}{\m^2}\right)^{-2}\left(1+\frac{Q}{\bar{F}\bar{\mathcal{K}}}\right)^2\left(1+\frac{2}{e^{H/T}-1}+\frac{T}{H}\frac{2\pi\sqrt{3}}{\sqrt{3+4\pi \frac{Q}{\bar{F}\bar{\mathcal{K}}}}}\frac{Q}{\bar{F}\bar{\mathcal{K}}}\right)~,
\end{align}
and the scalar spectral index and tensor-to-scalar ratio are obtained from Eqs.~\eqref{eq: full_ns} and \eqref{eq: str_r}, respectively.

\subsubsection{\label{subsub:constant-dissipation}Constant Dissipation Coefficient}
\noindent
We first consider the case of a constant dissipation coefficient, $d=0$, for which the number of e-folds is
\begin{align}
    \hat{N}(\phi_\star) \simeq \left\{-\frac{3}{4}\ln \left(1+\xi \frac{\phi^2}{\m^2}\right)+\frac{(1+6\xi)\phi^2}{8\m^2}+\frac{M C_{\Upsilon}}{2\m \sqrt{3\lambda}}\left[\ln\left(\frac{\phi}{\m+\sqrt{\m^2+\xi \phi^2}}\right)+\sqrt{1+\xi \frac{\phi^2}{\m^2}}\right]\right\}^{\phi_\star}_{\phi_{\textrm{end}}}~,
\end{align}
and the slow-roll parameter $\tilde{\beta}_{\Upsilon}$ reads ($\epsilon_U$ and $\eta_U$ are given by Eqs.~\eqref{eq: slowrolleps_sec5} and \eqref{eq: slowrolleta_sec5}, respectively, as they are independent of $d$)
\begin{align}
    \tilde{\beta}_{\Upsilon}(\phi) = \frac{-4\xi\m^2\left[\m^2\left(1+12\xi\right)+\left(1+6\xi\right)\xi\phi^2\right]}{\left[\m^2+\left(1+6\xi\right)\xi \phi^2\right]^2}~.
\end{align}
Notice that in the minimally coupled case ($\xi=0$), $\tilde{\beta}_{\Upsilon}=0$ for a constant dissipation coefficient. Moreover,
\begin{align}
    \frac{Q}{\bar{F}\bar{\mathcal{K}}}=\frac{2}{\sqrt{3\lambda}}\frac{M C_{\Upsilon}}{\phi^{2}}\frac{\left(\m^2+\xi \phi^2\right)^{3/2}}{\m^2+\left(1+6\xi\right)\xi \phi^2}~,
\end{align}
which, in contrast to the minimally coupled case, does \emph{not} scale simply as $Q(\phi)\propto \phi^{-2}$. As can be seen, this model has four independent parameters: $\{\xi,\lambda,MC_{\Upsilon}/\m, C_r\}$ (the mass scale $M$ and the coupling $C_{\Upsilon}$ enter only through their product). The inflaton background dynamics depends solely on $\xi$ and on the combination
\begin{equation}
    \sigma_0 \equiv \frac{M C_{\Upsilon}}{\m \sqrt{3\lambda}}~,
\end{equation} 
while the temperature of the thermal bath additionally requires the product $\lambda C_r$. The observables $n_\s$ and $r$ are therefore functions of $\{\xi,\sigma_0,\lambda C_r\}$ only, with $\lambda$ subsequently fixed by the normalisation of the scalar power spectrum $\mathcal{P}_\s$. 

\begin{figure}[h]
         \centering
         \includegraphics[width=0.95\textwidth]{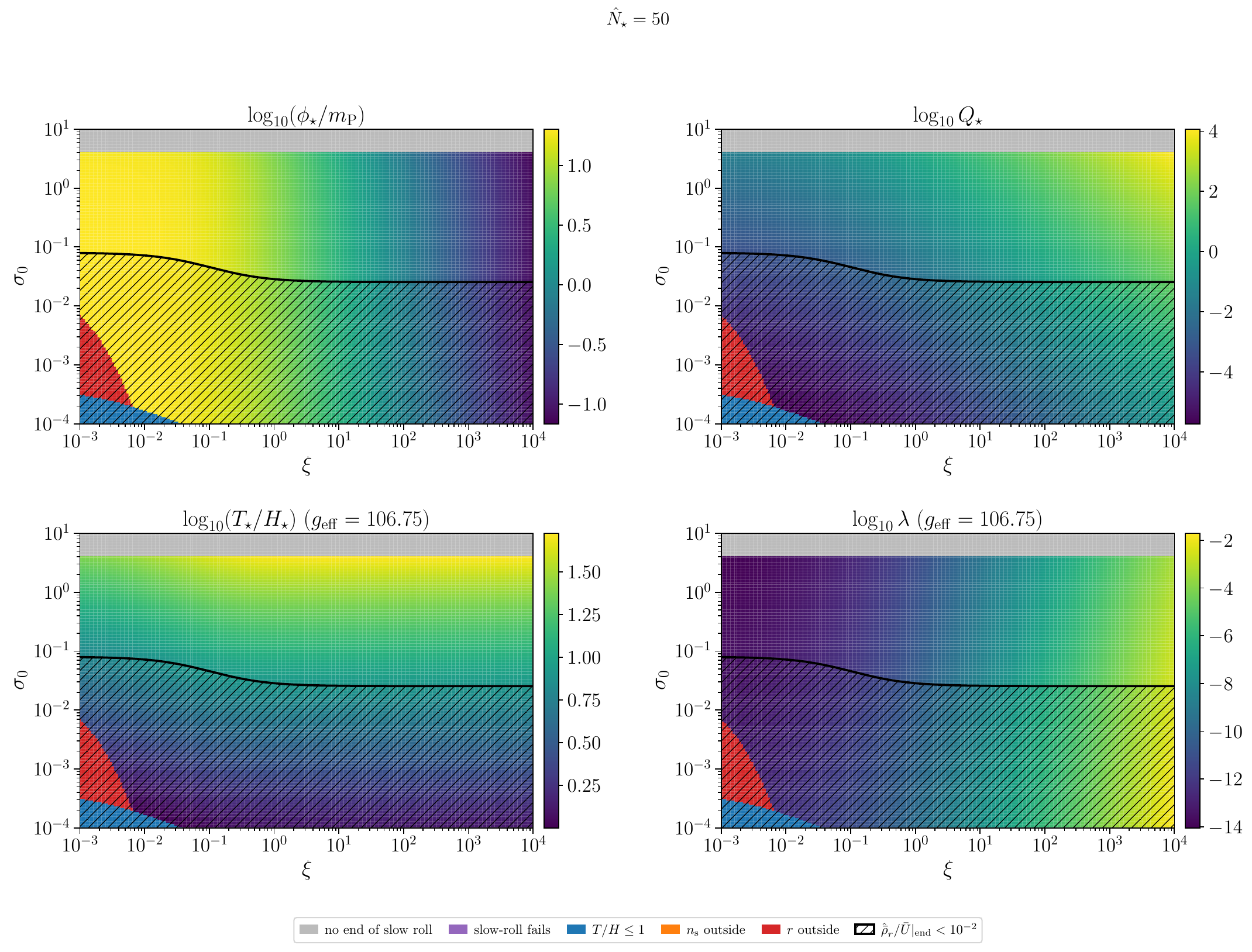}
     \caption{Viability maps for constant $\bar{\Upsilon}$ and $\hat{N}_\star=50$. The reference ranges are $n_{\s}\in [0.957,0.985]$ and $r<0.04$. The scalar power spectrum on CMB scales, $\mathcal{P}_{\textrm{s}}\sim 2.1\times 10^{-9}$, is also taken as a reference. In the majority of cases, we find that warm inflation occurs in the weak-dissipative regime at $\phi_\star$ in the defining frame ($Q_\star\ll1$).}
     \label{fig:viability_constUps_N50}
\end{figure}

\begin{figure}[h!]
         \centering
         \includegraphics[width=0.95\textwidth]{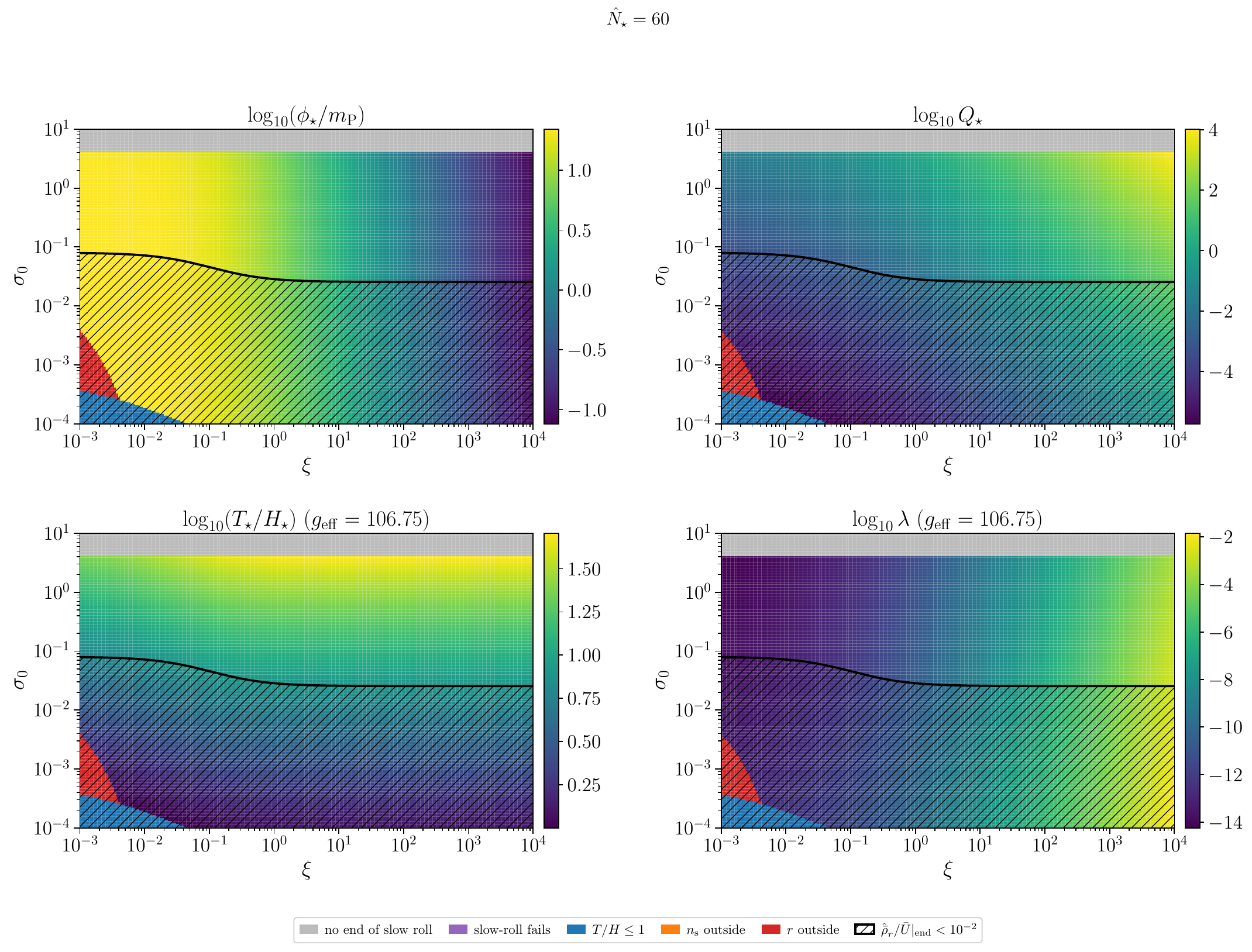}
     \caption{Viability maps for constant $\bar{\Upsilon}$ and $\hat{N}_\star=60$. The same exclusion criteria as for $\hat{N}_\star =50$ apply here. The exclusion boundaries are largely unchanged, although the region excluded by $r$ recedes towards smaller $\xi$. This is partly compensated by a slight widening of the $T/H\leq1$ wedge.}
     \label{fig:viability_constUps_N60}
\end{figure}

Figures~\ref{fig:viability_constUps_N50} and \ref{fig:viability_constUps_N60} display $\phi_\star$, $Q_\star$, $(T/H)_\star$, and $\lambda$ over $\xi \in [10^{-3},10^4]$ and $\sigma_0 \in [10^{-4},10]$, for $\hat{N}_\star = 50$ and $60$, respectively, with $g_{\rm eff}=106.75$. Inflation ends when $\epsilon_U=1+\tilde{Q}$, while the horizon-crossing value is determined from $\hat{N}_\star=\hat{N}(\phi_\star)$. We then impose the slow-roll condition, $\textrm{max}\left(\epsilon_U, |\eta_U|, |\tilde{\beta}_{\Upsilon}|\right)<1+\tilde{Q}$, at $\phi_\star$. Since $\mathcal{P}_\s$ depends on $\lambda$ both explicitly and implicitly through $T/H\propto \lambda^{-1/4}$, the quartic coupling cannot be obtained in closed form. We therefore determine $\lambda$ numerically by imposing $\mathcal{P}_\s\simeq 2.1\times 10^{-9}$ at fixed $C_r$. Finally, we require $(T/H)_\star >1$ to ensure a thermal regime during inflation, and demand that the predicted $(n_\s,r)$ lie within the observationally allowed window. Points failing any of these conditions are classified according to the first condition they violate, so that the four maps share a common
exclusion region that can be read off from any single panel.

Three excluded regions appear. A horizontal band at $\sigma_0 > 4$ admits no end of slow roll, namely for constant $\bar{\Upsilon}$ one finds that $\epsilon_U=1+\tilde{Q}$ reduces to
\begin{equation}
    8\m^4-2\sigma_0\m(\m^2+\xi\phi^2)^{3/2}=\phi^2[\m^2+(1+6\xi)\xi\phi^2]~.
\end{equation}
A solution exists and is unique when $\sigma_0<4$. This bound is specific to $d=0$. The hatched region indicates the parameter space for which the energy density in radiation at the end of inflation satisfies $\hat{\bar{\rho}}_{r}/\bar{U}|_{\mathrm{end}}<10^{-2}$, relative to the Einstein-frame background potential $\bar{U}$. A wedge in the lower-left corner is excluded by $(T/H)_\star \leq 1$, as discussed above. A narrow strip at $\xi \lesssim 5\times10^{-3}$ is excluded by the tensor-to-scalar ratio $r$ exceeding its observational bound, since the non-minimal flattening of the potential is insufficient to suppress the tensor amplitude in this region. The slow-roll conditions are satisfied everywhere a trajectory exists, leaving this category empty. We keep it in the legend for completeness. No points are excluded by the $n_\s$ window (which we take to be $n_\s\in[0.957,0.985]$, encompassing the Planck, BICEP/Keck, and ACT DR6 constraints), for either value of $\hat{N}_\star$.  

\begin{table}[ht]
\centering
\renewcommand{\arraystretch}{1.25}
\setlength{\tabcolsep}{3pt}
\resizebox{\textwidth}{!}{%
\begin{tabular}{c cccccccccccc}
\hline\hline
Cases &
$\hat{N}_{\star}$ & $\xi$ & $\lambda$ & $MC_{\Upsilon}/\m$ &
$C_r$ & $\phi_{\textrm{end}}/\m$ &
$\phi_\star/\m$ & $Q_\star$ & $T_\star/H_\star$ &
$\mathcal{P}_{\s}$ & $n_{\s}$ & $r$ \\
\hline
A &
$50$ & $0.00144$ & $1.30\times10^{-14}$ & $7.70\times10^{-7}$ &
$65.80$ & $0.4455$ &
$18.4333$ & $0.028$ & $20.26$ &
$2.1011\times10^{-9}$ & $0.96115$ & $5.44\times10^{-3}$ \\
\hline
B &
$50$ & $0.00733$ & $2.60\times10^{-14}$ & $1.09\times10^{-6}$ &
$65.80$ & $0.4228$ &
$18.0789$ & $0.044$ & $23.63$ &
$2.0994\times10^{-9}$ & $0.96683$ & $1.94\times10^{-3}$ \\
\hline
C &
$50$ & $0.0217$ & $5.51\times10^{-14}$ & $1.59\times10^{-6}$ &
$65.80$ & $0.3766$ &
$17.2196$ & $0.0719$ & $27.60$ &
$2.0947\times10^{-9}$ & $0.96962$ & $7.07\times10^{-4}$ \\
\hline
D &
$60$ & $267$ & $1.92\times10^{-6}$ & $8.10\times10^{-5}$ &
$65.80$ & $0.0653$ &
$0.5628$ & $1.97$ & $7.00$ &
$2.1046\times10^{-9}$ & $0.97791$ & $2.11\times10^{-4}$ \\
\hline
E &
$50$ & $2.35\times10^{3}$ & $2.15\times10^{-4}$ & $7.35\times10^{-4}$ &
$65.80$ & $0.0220$ &
$0.1740$ & $16.2$ & $6.86$ &
$2.0993\times10^{-9}$ & $0.97363$ & $3.05\times10^{-4}$ \\
\hline
F &
$60$ & $1.00\times10^{4}$ & $2.57\times10^{-3}$ & $2.97\times10^{-3}$ &
$58.00$ & $0.0107$ &
$0.0920$ & $74.0$ & $7.31$ &
$2.0990\times10^{-9}$ & $0.97792$ & $2.02\times10^{-4}$ \\
\hline\hline
\end{tabular}%
}
\caption{Representative viable points for a constant dissipation coefficient $\bar{\Upsilon}$, all yielding a quantum-dominated spectrum. Cases A--C satisfy $0.960\leq n_{\mathrm{s}}\leq0.970$ and lie in the weak-dissipative regime at $\phi_\star$, whereas cases D--F satisfy $Q_\star>1$ and therefore realise strong dissipation in the defining frame. The two conditions cannot be simultaneously satisfied for the parameter space considered here. All cases satisfy $g_{\mathrm{eff}}\in[10,200]$ and reproduce the observed scalar amplitude $\mathcal{P}_{\mathrm{s}}\simeq2.1\times10^{-9}$.}
\label{tab:data-constUps-nsQ}
\end{table}

To illustrate the different regimes identified in the viability maps, we select several representative points in parameter space and collect their corresponding inflationary observables in Table~\ref{tab:data-constUps-nsQ}. In all cases, the spectrum is quantum-dominated $\mathcal{P}^{(\textrm{qu})}_\s>\mathcal{P}^{(\textrm{th})}_\s$. Cases A--C correspond to models with a scalar spectral index in the range $0.960\leq n_{\s}\leq0.970$ and weak dissipation in the defining frame at $\phi_\star$, \emph{i.e.} $Q_\star<1$. In contrast, cases D--F realise the strong-dissipative regime in the defining frame, with $Q_\star>1$. An important feature is that the two requirements cannot be simultaneously satisfied within the parameter space explored: models with $Q_\star>1$ do not yield a spectral index in the range selected above, while models with $0.960\leq n_{\s}\leq0.970$ remain weakly dissipative at $\phi_\star$. Nevertheless, all six representative points reproduce the observed scalar amplitude, $\mathcal{P}_{\s}\simeq2.1\times10^{-9}$, while remaining within the assumed range of relativistic degrees of freedom, $g_{\mathrm{eff}}\in[10,200]$ \cite{Husdal:2016haj}.

\begin{figure}[h!]
         \centering
         \includegraphics[width=0.8\textwidth]{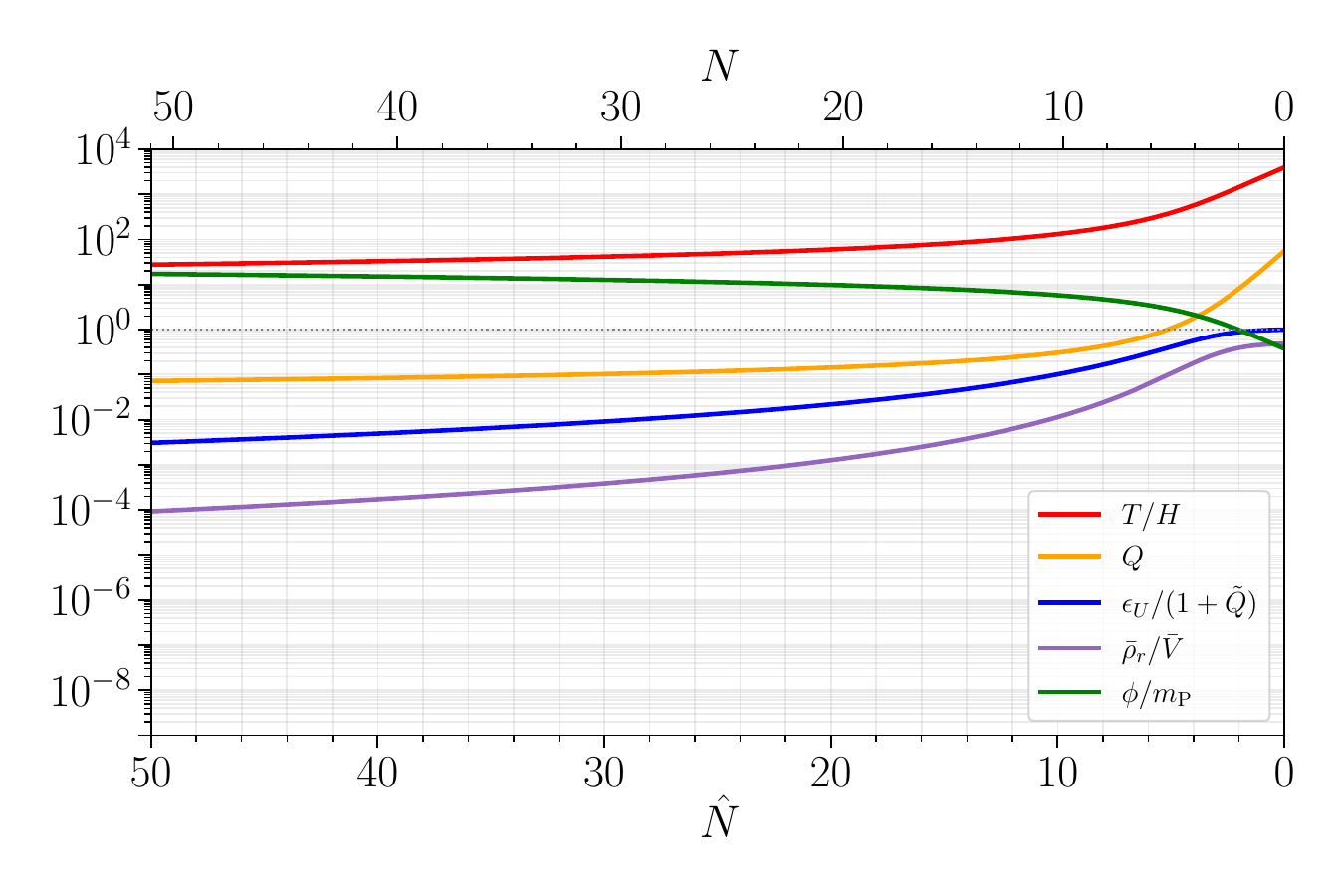}
     \caption{Evolution of the relevant background quantities for case C of Table~\ref{tab:data-constUps-nsQ}, with $\hat{N}_\star=50$. The horizontal axes denote the number of e-folds remaining until the end of inflation in the defining frame (upper) and Einstein frame (lower), respectively, so that $\hat{N}=50$ corresponds approximately to pivot-mode crossing and $\hat{N}=0$ to the end of inflation. The plotted quantities are the temperature-to-Hubble ratio $T/H$, the dissipation ratio $Q$, the slow-roll parameter combination $\epsilon_U/(1+\tilde{Q})$, the radiation-to-potential energy-density ratio $\bar{\rho}_r/\bar{V}$, and the inflaton field value in Planck units $\phi/\m$. Although the model is weakly dissipative at $\phi_\star$, with $Q_\star\ll1$, the dissipation ratio grows rapidly towards the end of inflation and eventually enters the strong-dissipative regime. The thermal condition $T/H>1$ is maintained throughout the evolution.
}
     \label{fig:caseC_evolution_constant_upsilon}
\end{figure}

It is instructive to follow the evolution of the background quantities for one of the representative viable points. Figure~\ref{fig:caseC_evolution_constant_upsilon} shows the evolution of case C in Table~\ref{tab:data-constUps-nsQ} from $\hat{N}_\star=50$ until the end of slow-roll inflation in the Einstein frame. At $\hat{N}_\star=50$, the model is in the weak-dissipative regime in the defining frame, with $Q_\star\simeq0.072$, while remaining in the thermal regime, with $(T/H)_\star\simeq27.6$. As inflation proceeds, the dissipation ratio increases significantly and eventually exceeds unity, so that the system enters the strong-dissipative regime before the end of inflation. At the same time, $T/H$ remains larger than unity and grows. Thus, the dissipative regime characterising the field value $\phi_\star$ need not persist throughout the subsequent evolution, namely a model that is weakly dissipative when the observable modes cross the horizon can evolve into a strongly dissipative regime towards the end of inflation. This behaviour is also found during standard warm inflation (see \emph{e.g.} Ref.~\cite{Bastero-Gil:2016qru}).

\subsubsection{\label{subsub:inversesquared-dissipation}Quadratic Field-Dependent Dissipation Coefficient}
\noindent
For a quadratic field-dependent dissipation coefficient ($d=-2$), the number of e-folds is 
\begin{align}
    \hat{N}(\phi_\star) \simeq \left\{-\frac{3}{4}\ln \left(1+\xi \frac{\phi^2}{\m^2}\right)+\frac{(1+6\xi)\phi^2}{8\m^2}+\frac{\m C_{\Upsilon}}{6M \sqrt{3\lambda}\xi}\left(1+\xi\frac{\phi^2}{\m^2}\right)^{3/2}\right\}^{\phi_\star}_{\phi_{\textrm{end}}}~,
\end{align}
and the slow-roll parameter $\tilde{\beta}_{\Upsilon}$ becomes
\begin{align}
   \tilde{\beta}_{\Upsilon}(\phi) = \frac{4\m^2\left[\m^2\left(2\m^2+3\xi \phi^2\right)+\left(1+6\xi\right)\xi^2\phi^4\right]}{\phi^2\left[\m^2+\left(1+6\xi\right)\xi \phi^2\right]^2}~.
\end{align}
Also, 
\begin{align}
    \frac{Q}{\bar{F}\bar{\mathcal{K}}}=\frac{2}{\sqrt{3\lambda}}\frac{C_{\Upsilon}}{M}\frac{\left(\m^2+\xi \phi^2\right)^{3/2}}{\m^2+\left(1+6\xi\right)\xi \phi^2}~,
\end{align}
meaning that $Q$ is field-dependent, whereas it remains constant in the minimally coupled case ($\xi=0$). For $d=-2$, the reduction variable reads
\begin{equation}
    \sigma_{-2} \equiv \frac{ \m C_{\Upsilon}}{M \sqrt{3\lambda}}~,
\end{equation}
and the analysis proceeds exactly as in the case $d=0$. The inflaton background depends only on $\xi$ and $\sigma_{-2}$, while the bath temperature additionally relies on $\lambda C_r$. The value of $\lambda$ is then determined by numerically solving $\mathcal{P}_{\s} \simeq 2.1\times10^{-9}$ at fixed $C_r$. 

\begin{figure}[h]
         \centering
         \includegraphics[width=0.95\textwidth]{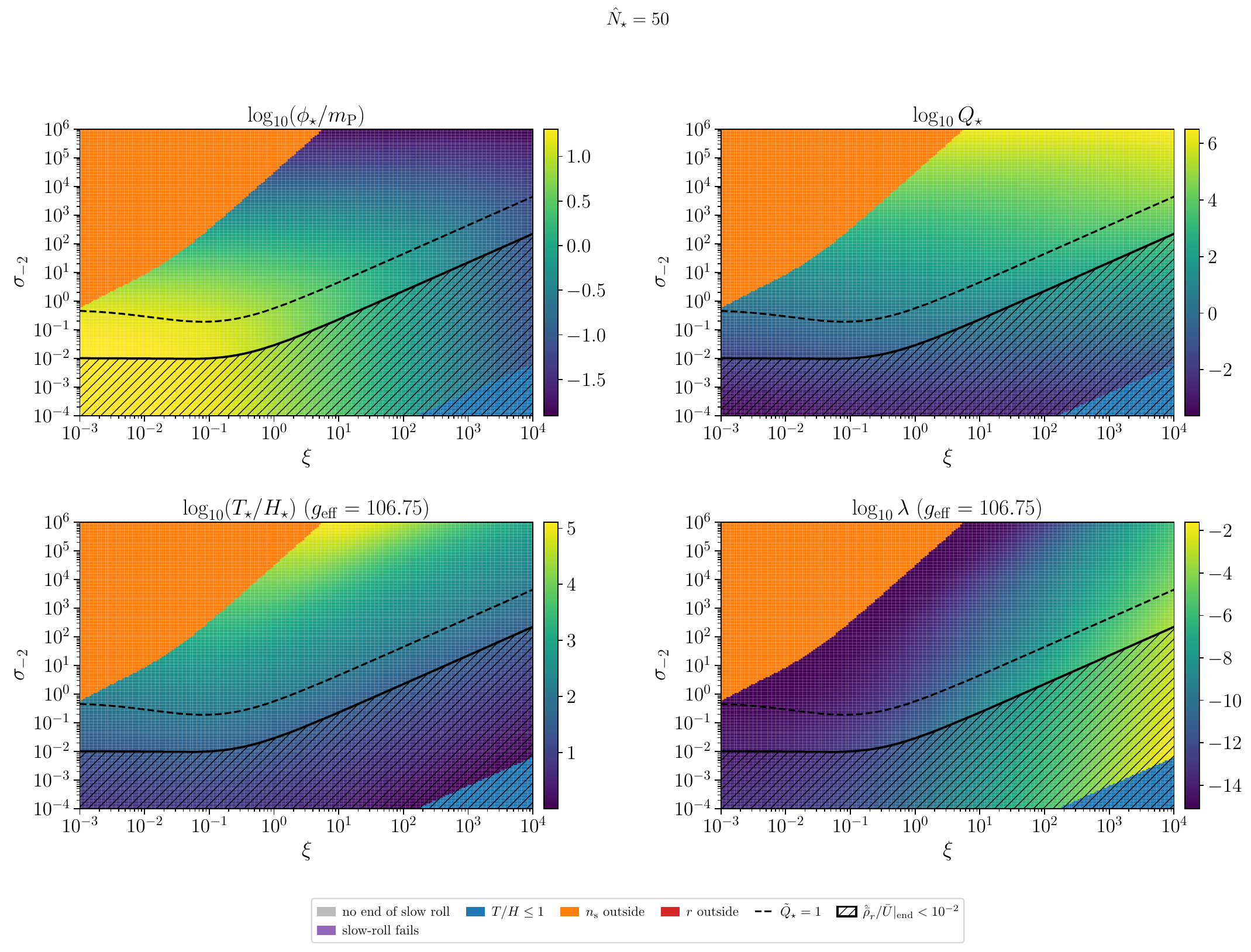}
     \caption{Viability maps for quadratic field-dependent $\bar{\Upsilon}$ and $\hat{N}_\star=50$. The reference ranges are again $n_{\s}\in [0.957,0.985]$ and $r<0.04$, and the scalar power spectrum takes on the CMB-normalised value $\mathcal{P}_{\textrm{s}}\sim 2.1\times 10^{-9}$. The dashed black curve marks $\tilde{Q}=1$, separating the weakly ($\tilde{Q}<1$) from the strongly ($\tilde{Q}>1$) dissipative regime. There is no graceful-exit exclusion for $d=-2$, and no region is excluded by $r$. The orange wedge at small $\xi$ and large $\sigma_{-2}$ is excluded by $n_{\s}$ exceeding the upper bound of the observational window.}
     \label{fig:viability_quadUps_N50}
\end{figure}

\begin{figure}[h]
         \centering
         \includegraphics[width=0.95\textwidth]{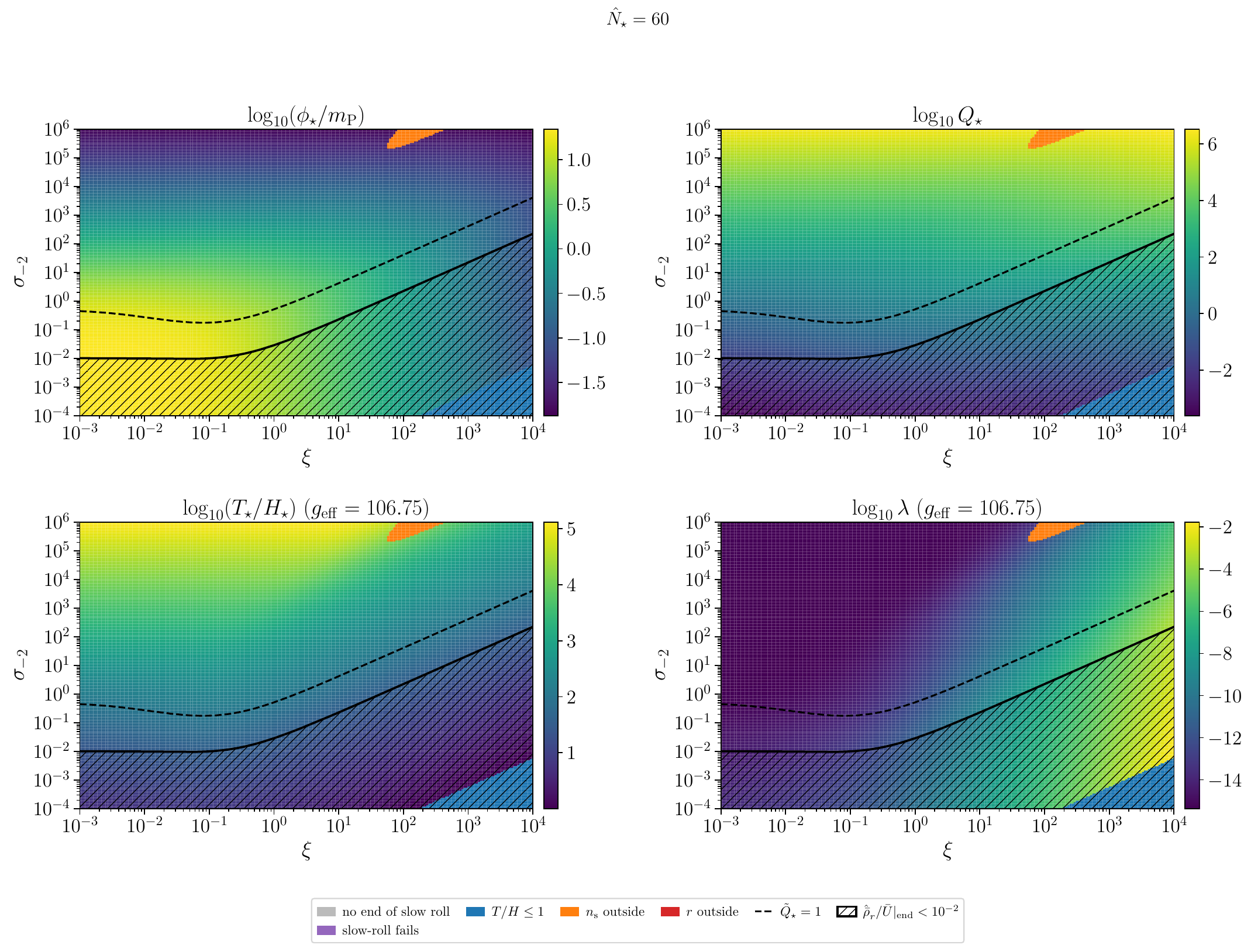}
     \caption{Viability maps for a quadratic field-dependent $\bar{\Upsilon}$ with $\hat{N}_\star=60$. The reference ranges are the same as those used in Fig.~\ref{fig:viability_quadUps_N50}. The region excluded by $n_{\s}$ is substantially reduced, leaving only a small viable domain around $\xi\sim10^{2}$ and $\sigma_{-2}\sim3\times10^{5}$. The $(T/H)_\star\leq1$ wedge remains essentially unchanged, while the fraction of viable points satisfying $\tilde{Q}>1$ increases.}
     \label{fig:viability_quadUps_N60}
\end{figure}

Figures~\ref{fig:viability_quadUps_N50} and~\ref{fig:viability_quadUps_N60} display the same four quantities over $\xi\in[10^{-3},10^{4}]$ and $\sigma_{-2}\in[10^{-4},10^{6}]$, with $g_{\textrm{eff}} = 106.75$. As can be seen, the no-end-of-slow-roll bound disappears entirely. $\epsilon_U = 1+\tilde{Q}$ yields
\begin{equation}
    8\m^4=\phi^2\left[\m^2+(1+6\xi)\xi \phi^2+\frac{2\sigma_{-2}}{\m}(\m^2+\xi \phi^2)^{3/2}\right].
\end{equation}
The right-hand side vanishes as $\phi\to0$ and increases monotonically thereafter, each term in the bracket being positive and non-decreasing, so that $\epsilon_U/(1+\tilde{Q}) = 1$ always admits a unique solution. Slow-roll inflation ends for arbitrarily large dissipation, and $\sigma_{-2}$ is unbounded from above. This is why the vertical range extends to $\sigma_{-2} = 10^{6}$. On the other hand, for $d=-2$ no point of the plane is excluded by $r$, whereas a substantial wedge is excluded by $n_{\s}$ (the opposite of the $d=0$ case). This follows from the sign of $\tilde{\beta}_{\Upsilon}$, which is positive here and negative for $d=0$, and enters $n_{\s}$ through the combination $\epsilon_U-\tilde{\beta}_{\Upsilon}$. The spectral index consequently reaches $n_{\s}=0.9834$ at $\hat{N}_\star=50$ and $0.9850$ at $\hat{N}_\star=60$, running off the upper edge of the observational window in the upper-left corner of the plane, at small $\xi$ and large $\sigma_{-2}$. Notice that the hatched region expands as $\xi$ increases.
 
The most significant difference is that strong
dissipation in the Einstein frame is genuinely realised, such that $\tilde{Q}$ spans roughly from $10^{-6}$ to $10^{6}$. The dashed curve in each panel marks $\tilde{Q}=1$. For constant $\bar{\Upsilon}$, the same curve is absent because $\tilde{Q}<1$ throughout that plane, and the combination of the $\sigma_0<4$ bound and the additional $\phi^{-2}$ dependence of $\tilde{Q}$ keeps the model weakly dissipative everywhere. Over the viable region, $\phi_\star$ ranges from $0.012\,\m$ to $\sim20\,\m$, $(T/H)_\star$ from unity to about $10^{5}$, and $\lambda$ from $10^{-16}$ to $2\times10^{-2}$. The tensor-to-scalar ratio reaches down to $r\sim10^{-16}$ in the strongly dissipative corner, far below the sensitivity of any foreseeable experiment.

\begin{table}[ht]
\centering
\renewcommand{\arraystretch}{1.25}
\setlength{\tabcolsep}{3pt}
\resizebox{\textwidth}{!}{%
\begin{tabular}{c cccccccccccc}
\hline\hline
Cases &
$\hat{N}_{\star}$ & $\xi$ & $\lambda$ & $\m C_{\Upsilon}/M$ &
$C_r$ & $\phi_{\textrm{end}}/\m$ &
$\phi_\star/\m$ & $Q_\star$ & $T_\star/H_\star$ &
$\mathcal{P}_{\textrm{s}}$ & $n_{\textrm{s}}$ & $r$ \\
\hline
A &
$50$ & $0.0174$ & $2.90\times10^{-14}$ & $4.19\times10^{-9}$ &
$65.80$ & $2.6199$ &
$19.0006$ & $0.0767$ & $28.48$ &
$2.1001\times10^{-9}$ & $0.96546$ & $5.73\times10^{-4}$ \\
\hline
B &
$50$ & $22$ & $3.43\times10^{-9}$ & $6.37\times10^{-5}$ &
$65.80$ & $0.2268$ &
$1.7479$ & $10.4$ & $29.83$ &
$2.0935\times10^{-9}$ & $0.96726$ & $5.55\times10^{-5}$ \\
\hline
C &
$60$ & $74.7$ & $3.11\times10^{-9}$ & $3.59\times10^{-3}$ &
$65.80$ & $0.1127$ &
$0.7955$ & $516$ & $93.27$ &
$2.1022\times10^{-9}$ & $0.97101$ & $4.29\times10^{-6}$ \\
\hline
D &
$60$ & $1.30\times10^{3}$ & $7.95\times10^{-6}$ & $0.181$ &
$65.80$ & $0.0296$ &
$0.2483$ & $668$ & $30.87$ &
$2.0985\times10^{-9}$ & $0.97249$ & $3.69\times10^{-5}$ \\
\hline
E &
$60$ & $5.42\times10^{3}$ & $1.52\times10^{-5}$ & $19.8$ &
$65.80$ & $0.0131$ &
$0.0920$ & $4.01\times10^{4}$ & $96.38$ &
$2.1000\times10^{-9}$ & $0.97102$ & $3.99\times10^{-6}$ \\
\hline
F &
$50$ & $1.00\times10^{4}$ & $7.01\times10^{-4}$ & $13.1$ &
$65.80$ & $0.0107$ &
$0.0823$ & $4.73\times10^{3}$ & $29.93$ &
$2.0994\times10^{-9}$ & $0.96725$ & $5.48\times10^{-5}$ \\
\hline\hline
\end{tabular}%
}
\caption{Benchmark points for $\bar{\Upsilon}\propto\phi^2$, with $\lambda$ fixed by imposing $\mathcal{P}_{\textrm{s}}\simeq 2.1\times10^{-9}$ and $g_{\textrm{eff}}\in[10,200]$ in all cases. $\tilde{Q}_\star$ exceeds unity only for cases C and E ($\tilde{Q}_\star=1.17$ and $1.26$, with $(T/H)_\star\simeq90$), yielding a thermal-dominated spectrum. These have no counterpart for constant $\bar{\Upsilon}$, where $\tilde{Q}_\star<1$ throughout the viable region. Cases B, D, and F instead have $Q_\star \gg 1$ with $\tilde{Q}_\star=0.08$--$0.09$, and are therefore strongly dissipative in the defining frame with a quantum-dominated spectrum.}
\label{tab:data-quadUps}
\end{table}

We again present a selection of viable benchmark points in Table~\ref{tab:data-quadUps}, this time for $d=-2$. In all cases, the coupling $\lambda$ is fixed by requiring $\mathcal{P}_{\s}\simeq2.1\times10^{-9}$, and the effective number of relativistic degrees of freedom is restricted to $g_{\mathrm{eff}}\in[10,200]$. The selected points reproduce a scalar spectral index in the range $0.965\lesssim n_{\mathrm{s}}\lesssim0.973$ and satisfy the current bound on the tensor-to-scalar ratio. The field dependence of the dissipation coefficient $\bar{\Upsilon}$ substantially enlarges the range of dissipative regimes compatible with the observed scalar spectrum. 

A notable difference with respect to the constant-dissipation case is the appearance of viable points with $\tilde{Q}_{\star}>1$, as pointed out already. In particular, cases C and E have $\tilde{Q}_\star\simeq1.17$ and $1.26$, respectively, while remaining in the high-temperature regime, with $(T/H)_\star\simeq90$, and a thermal-dominated spectrum. These benchmark points therefore realise a strongly dissipative and thermal regime while satisfying the observational constraints. No analogous points were found for a constant dissipation coefficient, for which $\tilde{Q}_\star<1$ throughout the viable parameter space under consideration. Cases B, D, and F illustrate the complementary situation, for which $Q_\star \gg 1$ but $\tilde{Q}_\star \simeq 0.08$--$0.09$ in all three. Dissipation is therefore strong in the defining frame while the spectrum remains quantum-dominated.
 
\begin{figure}[h!]
         \centering
         \includegraphics[width=0.8\textwidth]{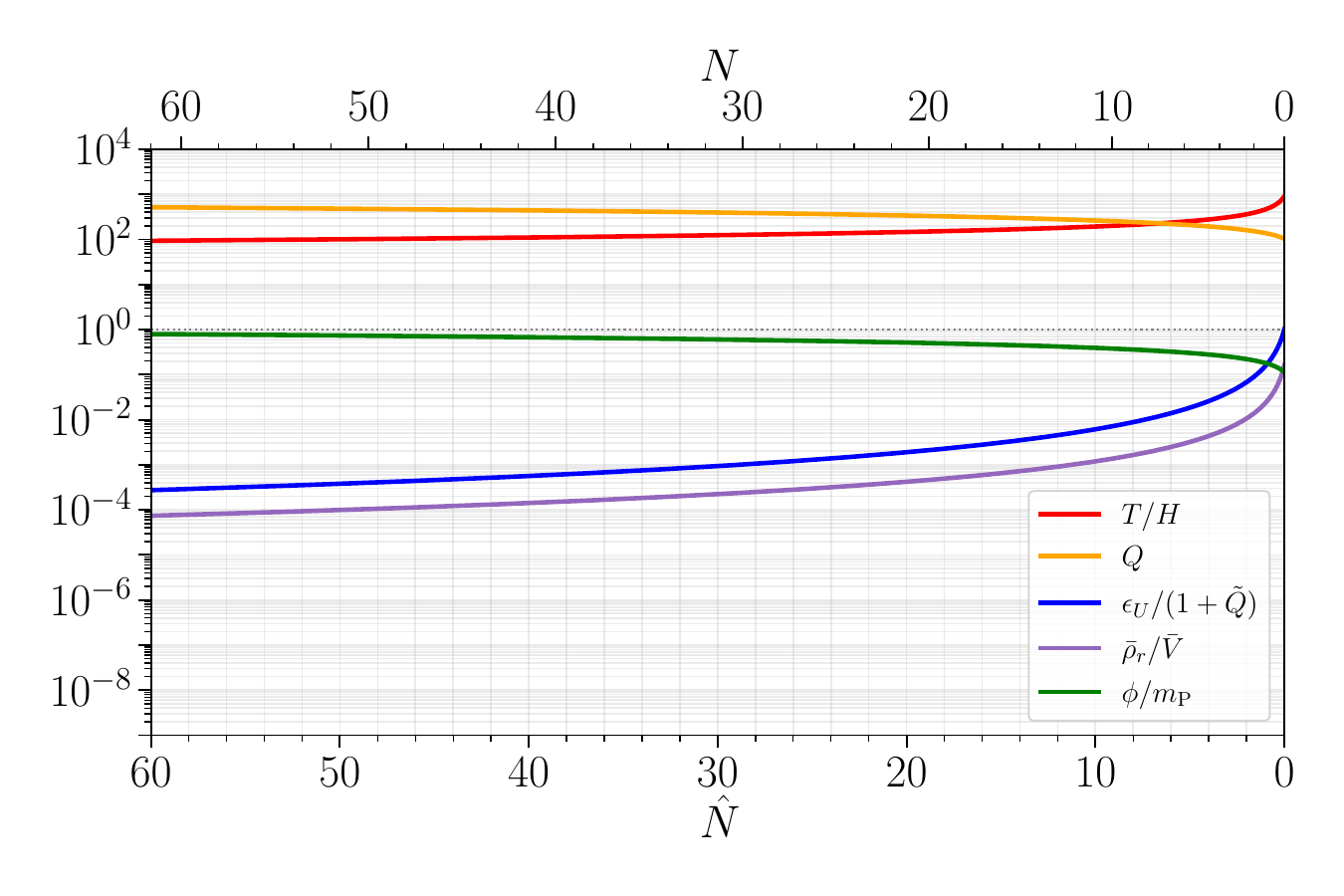}
     \caption{Evolution of $T/H$, $Q$, $\epsilon_U/(1+\tilde{Q})$, $\bar{\rho}_r/\bar{V}$, and $\phi/\m$ for benchmark case C of Table~\ref{tab:data-quadUps} ($\bar{\Upsilon}\propto\phi^2$, $\hat{N}_\star=60$, $\xi=74.7$). The lower axis gives the Einstein-frame number of e-folds $\hat{N}$ before the end of inflation, while the upper axis corresponds to the defining-frame one $N$. The dotted line marks unity, namely inflation ends where $\epsilon_U=1+\tilde{Q}$, which defines $\phi_{\textrm{end}}$ and fixes the origin of both axes. The dissipation ratio decreases during inflation, from $Q_\star=516$ to $104$, since $\bar{\Upsilon}\propto\phi^2$ and the field is rolling down. The same behaviour takes $\tilde{Q}$ from $1.17$ at $\phi_\star$ to $0.43$ at the end, so that the observable scales exit in the strongly dissipative regime while the final e-folds are weakly dissipative. $T/H$ rises from $93$ to $903$, and $\bar{\rho}_r/\bar{V}$ reaches $0.16$ at the end of inflation, well above the $10^{-2}$ level demarcated in Figs.~\ref{fig:viability_quadUps_N50} and \ref{fig:viability_quadUps_N60}.}
     \label{fig:caseC_evolution_quadratic_upsilon}
\end{figure}

The background evolution of a representative benchmark is shown in
Fig.~\ref{fig:caseC_evolution_quadratic_upsilon}, for case C of Table~\ref{tab:data-quadUps}. Inflation begins at $\phi_\star=0.7955\,\m$ and ends at $\phi_{\textrm{end}}=0.1127\,\m$, the whole excursion being sub-Planckian. A characteristic feature of the $d=-2$ case is the evolution of the dissipation ratio: whereas for $d=0$ the dissipation ratio grows as inflation proceeds, here $\bar{\Upsilon}\propto\phi^2$ couples $Q$ directly to the rolling field, so that $Q$ falls monotonically from $Q_\star=516$ to $Q_{\rm end}=104$. Consequently, $\tilde{Q}$ decreases from $1.17$ to $0.43$, crossing unity during the evolution. Thus, the CMB modes leave the horizon in the Einstein-frame strong-dissipative regime, while the final stages of inflation are weakly dissipative. This behaviour has no counterpart in the case of constant $\bar{\Upsilon}$.

The thermal quantities evolve accordingly. The ratio $\epsilon_U/(1+\tilde{Q})$ increases monotonically and reaches unity at $\hat{N}=0$, as required by the end-of-inflation condition. Meanwhile, $T/H$ grows from $93$ to $903$, indicating that thermal perturbations, rather than vacuum ones, dominate the sourcing of perturbations throughout the observable range. The radiation energy density grows from $\bar{\rho}_r/\bar{V}\simeq7\times10^{-5}$ to $0.16$ by the end of inflation. This benchmark therefore lies outside the hatched region of Figs.~\ref{fig:viability_quadUps_N50} and \ref{fig:viability_quadUps_N60}. The dissipative dynamics bring the radiation component close to, but still below, the inflationary energy density by the end of inflation, suggesting that no substantial separate reheating stage is required.

\section{Summary and Conclusions\label{sec: summary}}
\noindent
Non-minimal couplings are generic and are therefore expected to arise in UV-complete realisations of inflation. In this work, we have developed an effective description of the interactions between an extra gravitational scalar degree of freedom and the matter sector, and explored the case in which such interactions give rise to dissipative effects. The result is a unified framework encompassing both scalar-tensor theories of gravity and warm inflation.

Although the classical equivalence between frames is well established, its quantum counterpart remains debated. We adopt the approach in which all quantum effects are computed in the defining frame, where the model is specified, including interactions and the renormalisation conditions. The resulting effective classical equations are then transformed consistently to the Einstein frame, where the analysis of inflation is more intuitive, allowing us to establish the relation between defining- and Einstein-frame quantities and to express the inflationary bounds in terms of defining-frame variables.

In warm inflation, the main quantities of interest are the dissipation coefficient and the stochastic noise term. We began by deriving the effective equations of warm inflation in the defining frame, with a dissipation coefficient $\bar{\Upsilon}$ and a noise term $\xi_T$. In computing the dissipation coefficient, an additional condition had to be imposed due to the time dependence of the effective Planck mass, ensuring that Planck-suppressed contributions remain negligible throughout the dynamics. In order to analyse the slow-roll dynamics of the defining frame, we conformally transformed to the Einstein frame, where the inflaton field undergoes effective dissipation with coefficient $\tilde{\bar{\Upsilon}} = \bar{\Upsilon}/(\bar{\mathcal{K}}\bar{F}^{3/2})$. As we argued, however, this effective coefficient need not be the physical one in the Einstein frame. This stems from the fact that, once the conformal transformation is performed, all matter fields couple non-trivially to the scalar field. These rescaled couplings contribute to the dissipation coefficient, so that a simple conformal transformation of the defining-frame quantities may not capture the full field dependence. Despite these issues of interpretation, the Einstein frame remains a useful mathematical tool for understanding physical phenomena. 

The regime in which warm inflation is realised depends on the ratio of the dissipation coefficient to the Hubble parameter, $Q = \bar{\Upsilon}/(3H)$, in the defining frame. We found that the corresponding effective ratio in the Einstein frame is $\tilde{Q} \simeq Q/(\bar{\mathcal{K}}\bar{F})$. This result implies that, given $\bar{\mathcal{K}}\bar{F}>1$, the system can be in a strong dissipative regime in the defining frame while being in an effective weak dissipative regime in the Einstein frame. In other words, the effects of modified gravity suppress dissipative effects during slow-roll inflation. Having obtained the slow-roll equations and the corresponding slow-roll conditions in each frame, we computed the inflationary observables. The presence of thermal effects during inflation leaves distinctive imprints in the curvature power spectrum, and consequently in the scalar spectral index and the tensor-to-scalar ratio. We found that the curvature power spectrum retains a form similar to that of General Relativity when expressed in terms of effective Einstein-frame quantities, and that the effective dynamics is constrained in a similar manner. The parameter $\tilde{Q}$ then dictates whether the curvature power spectrum is quantum- or thermally dominated, the crossover occurring at $\tilde{Q}\sim 1$. From the point of view of the defining frame, the crossover instead takes place at $Q\sim \bar{\mathcal{K}}\bar{F}$ and therefore depends on the modified-gravity model chosen. Warm inflation may thus be realised in the strong dissipative regime of the defining frame, while quantum perturbations still dominate the curvature power spectrum. From an observational standpoint, even if measurements were to favour cold over warm inflation, sufficiently strong modified-gravity effects could mask the presence of dissipation during inflation. 

To illustrate the analysis explicitly, we adopted a non-minimal coupling of the type considered in Higgs inflation, $F(\phi)=1+\xi(\phi/\m)^2$, with a quartic potential $V(\phi)=\lambda \phi^4/4$, and confronted two different dissipation coefficients with observations. For constant $\bar{\Upsilon}=M C_{\Upsilon}$, we surveyed the parameter space over $\xi\in[10^{-3},10^4]$ and $\sigma_0\equiv MC_{\Upsilon}/(\m \sqrt{3\lambda})\in[10^{-4},10]$, for $\hat{N}_\star = 50$ and $60$. At both values of $\hat{N}_\star$, no point is excluded by the observationally allowed range of the scalar spectral index. The constraints instead come from a band at $\sigma_0>4$, in which slow-roll inflation never ends, together with two wedges arising from the tensor-to-scalar ratio at small $\xi$ and from the requirement $T>H$ at small $\sigma_0$ and large $\xi$. Throughout the viable region, warm inflation proceeds in the weak dissipative regime of the Einstein frame, with $\tilde{Q}<1$ everywhere, while retaining the possibility of strong dissipation in the defining frame, and satisfying all observational constraints. For $\Upsilon(\phi)= C_{\Upsilon} \phi^2/M$, an analogous analysis was performed over $\xi\in[10^{-3},10^4]$ and $\sigma_{-2}\equiv \m C_{\Upsilon}/(M\sqrt{3\lambda})\in[10^{-4},10^6]$. Here the tensor-to-scalar ratio does not constrain the parameter space, and slow-roll inflation always ends, since $\tilde{Q}$ remains finite as $\phi \to 0$ while $\epsilon_U$ diverges. Two excluded regions nevertheless remain, one from the scalar spectral index and another one from the requirement $T>H$. In contrast to the case of constant $\bar{\Upsilon}$, warm inflation may here be realised in the strong-dissipation regime of the Einstein frame, which is attained over a substantial fraction of the viable parameter space.

A further condition involves the exit from inflation. Ending the accelerated expansion does not by itself guarantee that the universe is left close to radiation domination. Combining the slow-roll equations gives $\bar{\rho}_r/\bar{V}=\frac{1}{2}\tilde{Q}\epsilon_U/(1+\tilde{Q})^2$, so that $\left.\bar{\rho}_r/\bar{V}\right|_{\rm end}\simeq \tilde{Q}/(2(1+\tilde{Q}))$. We have indicated in the viability maps the region in which $\hat{\bar{\rho}}_r/\bar{U}=\bar{\rho}_r/\bar{V}<10^{-2}$ at the end of inflation. 

Finally, this work rests on some assumptions and leaves open a number of questions that we intend to pursue. One such assumption is local equilibrium in both frames. Although we give a brief argument for why this should hold during inflation, we will address the question in more generic settings in future work. We also intend to carry out a more detailed analysis of the cosmological perturbations, by going beyond zeroth order in the slow-roll expansion and by allowing for temperature-dependent dissipation coefficients. A numerical treatment of the full perturbation equations would allow the higher-order inflationary observables, such as the running of the spectral indices and the non-Gaussianity parameters, to be computed for both cold and warm inflation without recourse to the slow-roll expansion. A further avenue worth pursuing, which may help to elucidate the quantum equivalence between frames, is the explicit computation of the dissipation coefficient in different frames (that is, of $\bar{\Upsilon}$ and $\tilde{\bar{\Upsilon}}$), for a given particle-physics model.

\section*{Acknowledgements}
\noindent
A.~C.~T.~and M.~K.~acknowledge that this project has received funding from the Research Council of Lithuania (LMTLT), agreement No.~S-PD-24-135. A.~C.~T.~is also partially supported by grant PID2022-137003NB-I00 funded by MCIN/AEI/10.13039/501100011033/ and by ERDF `A way of making Europe.' P.~B.~F was supported by the FCT-Fundação para a Ciência e Tecnologia (FCT), I.P. fellowship SFRH/BD/151475/2021 with DOI identifier 10.54499/SFRH/BD/151475/2021. J.~J.~T.~D. acknowledges the financial support provided by FCT-Fundação para a Ciência e Tecnologia (FCT), I.P., through the Strategic Funding UID/04650/2025 and UID/04564/2025 and national funds with DOI identifiers 10.54499/2023.11681.PEX, 10.54499/2024.00249.CERN funded by measure RE-C06-i06.m02-``Reinforcement of funding for International Partnerships in Science, Technology and Innovation'' of the Recovery and Resilience Plan–RRP, within the framework of the financing contract signed between the Recover Portugal Mission Structure (EMRP) and the Foundation for Science and Technology I.P. (FCT), as an intermediate beneficiary, as well as the advanced computing projects 2024.00249.CERN.F1.

\appendix
\section{\label{app:dissipative-Minkowski-dyn}Dissipative Dynamics and Thermal Noise}
\subsection{Minkowski Spacetime}
\noindent
In thermal field theory, we are typically interested in the ensemble average of an observable evaluated at a given time, with the system specified by an initial state at $t=t_i$. The Schwinger-Keldysh formalism \cite{Schwinger:1960qe, Keldysh:1964ud} (also known as the Closed-Time-Path or `in-in' formalism), realises this by performing the time integration over a closed contour that runs first along the forward branch $\mathcal{C}_+$, from $t_{i}$ to $+\infty$, and then back on the backward branch $\mathcal{C}_{-}$, from $+\infty$ to $t_{i}$. The associated generating functional $W$ reads \cite{Berges:2004yj,Berera:2008ar}\footnote{The formalism presented in this appendix is formulated for a generic scalar field $\Phi$, which will later be identified with the non-minimally coupled one in Eq.~\eqref{eq:the_action_nonminimal}.}
\begin{equation}
e^{\iu W[J_+,J_-]} = \text{const.}\times\int_{\mathcal{C}_-} \mathcal{D}\Phi_-~\exp\left(-\iu S[\Phi_-]-\iu \int J_-\Phi_-\right)\int_{\mathcal{C}_+} \mathcal{D}\Phi_+~\exp\left(\iu S[\Phi_+]+\iu \int J_+\Phi_+\right),
\end{equation}
where `const.' is a (field-independent) normalisation constant that drops out of the connected Green functions,
\begin{equation}
G_{ab}(x,x') = \left.\frac{\delta^2 W[J_+,J_-]}{\delta J_a(x)\delta J_b(x')}\right|^{\Phi_-=\Phi_+=\Phi}_{J_-=J_+=0}~, 
\end{equation}
and $\iu = \sqrt{-1}$ is the imaginary unit. $J_{\pm}$ are auxiliary source terms, set to zero in the computation of the Green functions, and $\mathcal{D}\Phi_{\pm}$ denote the path-integral measures over field configurations on the two branches of the contour. The effective action $\Gamma$ is defined analogously \cite{Rivers_1987}:
\begin{equation}
e^{\iu \Gamma[\Phi_+,\Phi_-]} = \text{const.}\times\int_{\cal C_-} \mathcal{D}\eta_-~\exp\left(-\iu S[\Phi_-+\eta_-]\right)\int_{\cal C_+} \mathcal{D}\eta_+~\exp\left(\iu S[\Phi_++\eta_+]\right)~,
\end{equation}
with fluctuations $\eta_{\pm}$ around classical backgrounds $\Phi_{\pm}$. The loop expansion of $\Gamma$ contains only one-particle-irreducible diagrams. 

In the Keldysh representation, a linear transformation of the fields renders some components of the propagator matrix identically zero. The new fields are given by
\begin{equation}
\Phi_c = \frac{1}{2}(\Phi_{+}+\Phi_-)~, \qquad \Phi_{\Delta} = \Phi_+-\Phi_-~,
\end{equation}
and, in this basis ($a', b' = c, \Delta$), the matrix of connected two-point functions takes the form \cite{Calzetta:2008iqa}
\begin{equation}
G_{a'b'}=
\begin{pmatrix}
     F(x,x') & G_R(x,x')\\
     G_A(x,x') & 0
\end{pmatrix}~,
\end{equation}
where $G_{R}$ and $G_{A}$ are the retarded and advanced propagators, respectively, while $F$ is the Keldysh two-point function \cite{Berera:2008ar}. The self-energy matrix takes the complementary block structure 
\begin{equation}
\Sigma^{a'b'}=
\begin{pmatrix}
     0 & \Sigma_A(x,x')\\
     \Sigma_R(x,x') & -\iu \Sigma_F(x,x')
\end{pmatrix}~.
\end{equation}
Both $G_{ab}$ and $\Sigma_{ab}$ ($a, b = +, -$) can be computed perturbatively by solving the Schwinger-Dyson equations \cite{Schwartz:2014sze},
\begin{equation}
\Box G_{ab}(x,x') = \int \dif^4y~\Sigma_{a}^{\; c}(x,y)G_{cb}(y,x')-\iu c_{ab}\delta^{(4)}(x-x')~,
\end{equation}
where $\Box \equiv \eta^{\mu\nu}\partial_\mu\partial_\nu$ is the d'Alembertian in flat spacetime with Minkowski metric $\eta_{\mu\nu}=\textrm{diag}(-1,1,1,1)$, and $\delta(x)$ is Dirac's delta function. Also, the field has been taken to be massless for simplicity, and $c_{ab} = \text{diag}(1,-1)$ is a constant $2\times2$ source matrix.

A useful feature of the Keldysh representation is that the effective action admits the generic form \cite{Berera:2008ar}
\begin{equation}
\Gamma[\Phi_c,\Phi_\Delta] = -\int \dif^4x~\mathcal{F}(x)\Phi_\Delta(x)+\frac{1}{2}\int \dif^4x~\dif^4x'~\Phi_\Delta \iu \Sigma_F(x,x')\Phi_\Delta(x')+\mathcal{O}(\Phi_\Delta^3)~,
\end{equation}
truncated at second order in $\Phi_\Delta$. Variation with respect to $\Phi_{\Delta}$ followed by the identification $\Phi_\Delta = 0$ yields the effective equation of motion $\mathcal{F} = 0$. The second term, which depends only on the self-energy, acquires a transparent interpretation after a Hubbard-Stratonovich transformation. This trades the quantum fluctuations for statistical fluctuations of an auxiliary random field $\xi(x)$:
\begin{equation}
\exp\left(-\frac{1}{2}\int \dif^4x~\dif^4x'~\Phi_\Delta \iu \Sigma_F(x,x')\Phi_\Delta(x')\right) \propto \int \mathcal{D}\xi~\exp\left(-\frac{1}{2}\int \dif^4x~\dif^4x'~\xi(x) \iu \Sigma^{-1}_F(x,x')\xi(x')+\iu \int \dif^4x~\xi(x)\Phi_\Delta(x)\right).
\end{equation}
Hence $\xi(x)$ can be interpreted as a Gaussian random field of zero mean, $\langle \xi \rangle_\xi = 0$, and two-point correlation function
\begin{equation}
\langle \xi(x)\xi(x')\rangle_\xi = \Sigma_F(x,x')~,\label{eq: sigmaxi}
\end{equation}
where `$\langle \cdot \rangle_\xi$' denotes the statistical average over $\xi$ (we suppress the index `$\xi$' hereafter for simplicity). The effective equation of motion for the classical field then reads
\begin{equation}
\Box\Phi_c(x)+\int \dif^4x'~\Sigma_R(x,x')\Phi_c(x')+V_{,\Phi_c}(\Phi_c) = \xi(x)~,\label{eq: effeom1}
\end{equation}
which is a Langevin-type equation with memory kernel and stochastic source. 

For a homogeneous, slowly varying scalar field $\phi$, Eq.~\eqref{eq: effeom1} reduces to \cite{Berera:2008ar}
\begin{equation}
\ddot{\phi}+\Upsilon\dot{\phi}+V_{,\phi}(\phi)=0~,
\end{equation}
and, at high temperatures $T$, the correlation function in Eq.~\eqref{eq: sigmaxi} takes on the local form \cite{Berera:2007qm}
\begin{equation}
    \langle \xi(x)\xi(x')\rangle \simeq 2\Upsilon T\delta^{(4)}(x-x')~,
\end{equation}
with the dissipation coefficient $\Upsilon$ related to the retarded self-energy by
\begin{equation}
    \Upsilon \equiv -\int \dif^4x'~\Sigma_R(x-x')(t-t')~.\label{eq: dissipation_coef_1}
\end{equation}
The regime in which these approximations are valid is the so-called \emph{adiabatic regime} \cite{Berera:2008ar}.

\subsection{\label{app:flat-spacetime-subsection}Flat FLRW Spacetime}
\noindent
For application to warm inflation, in which the scalar field plays the role of the inflaton and the matter sector constitutes a thermal bath \cite{Bartrum:2013fia}, we take the geometric background to be FLRW. Within the scalar sector, the non-minimal term
$F(\Phi)R$ generates additional interactions, including couplings to metric perturbations. In principle, gravitons could propagate in loop diagrams contributing to the dissipation coefficient. In our semi-classical treatment, however, the metric is not quantised; moreover, such contributions would be Planck-suppressed and therefore negligible compared with the thermal contributions from matter fields that dominate in the regime of interest. From the Lagrangian (see Eq.~\eqref{eq:the_action_nonminimal})
\begin{equation}
\mathcal{L}_{\Phi} = \sqrt{-g}\left[-\frac{1}{2}\nabla_\mu\Phi\nabla^\mu\Phi - V(\Phi)+\frac{\m^2}{2}F(\Phi)R\right],
\end{equation}
the effective equation of motion follows \cite{Berera:2008ar}:
\begin{equation}
\left[\partial_t^2+3H\partial_t-\frac{\nabla^2}{a^2}\right]\Phi_c(x)+V_{,\Phi_c}-\frac{\m^2}{2}F_{,\Phi_c}R+\int \dif^4x'~\Sigma_R(x,x')\Phi_c(x')=\xi(x)~,
\end{equation}
where $\xi(x)$ again satisfies $\langle \xi(x)\rangle =0$ together with Eq.~\eqref{eq: sigmaxi}. The self-energy depends on propagators in curved spacetime, which in general are quite complicated functions. In the regime relevant for warm inflation, however, thermal effects dominate over curvature ($T \gg H$), and flat-space thermal propagators are a good approximation. Furthermore, when the inflaton evolves adiabatically on the timescale set by the thermal interactions, one may Taylor-expand $\phi(t') \simeq \phi(t)+\dot{\phi}(t)(t'-t)$ inside the convolution; the zeroth-order term is absorbed into the effective potential, while the first-order term yields a local friction $\Upsilon\dot\phi$ on the left-hand side, with $\Upsilon$ given by Eq.~\eqref{eq: dissipation_coef_1}. Restricting to the homogeneous mode, the equation of motion becomes
\begin{equation}
\ddot{\phi}+3H\dot{\phi}+V_\phi(\phi)-\frac{\m^2}{2}F_{,\phi}(\phi)R=-\Upsilon \dot{\phi}~.\label{eq: eomeff2}
\end{equation}
This equation generalises to an inhomogeneous field $\Phi(x)$ by writing $\Phi(x) = \phi(t)+\delta\phi(x)$, with the noise sourcing only the perturbation $\delta\phi(x)$:
\begin{equation}
    \ddot{\Phi}(x) -\frac{\nabla^2}{a^2}\Phi(x)+3H\dot{\Phi}(x)+V_{,\Phi}-\frac{\m^2}{2}F_{,\Phi}R=-\Upsilon\dot{\Phi}(x)+\xi(x)~,
\end{equation}
with the two-point correlation function for the noise
\begin{equation}
\langle\xi(t,\mathbf{x})\xi(t',\mathbf{x}')\rangle = \frac{2\Upsilon T}{a^3}\delta(t-t')\delta^{(3)}(\mathbf{x}-\mathbf{x}')~.\label{eq: corr_noise_1}
\end{equation}
For a general lapse $\mathcal{N}$, with $\dif \tau = \mathcal{N} \dif t$, the same correlator reads
\begin{equation}
\langle \xi(t,\mathbf{x})\xi(t',\mathbf{x}')\rangle =\frac{2\Upsilon T}{a^{3}\mathcal{N}}\delta(t-t')\delta^{(3)}(\mathbf{x}-\mathbf{x}')=\frac{2\Upsilon T}{a^{3}}\delta(\tau-\tau')\delta^{(3)}(\mathbf{x}-\mathbf{x}')~,\label{eq:correlator-noise-term-stilldontunderstandwhythiswasnotlabelled}
\end{equation}
where we have used the standard rescaling of the Dirac delta, $\delta(f(x)) = \delta(x-x_0)/|f'(x_0)|$ \cite{Ramos:2013nsa}. This recovers the form of $\mathcal{J}$ given in Eq.~\eqref{eq: source1}, with the unit-variance noise $\mathring{\xi}(x)$ defined by
\begin{equation}
    \langle \mathring{\xi}(t,\mathbf{x})\mathring{\xi}(t',\mathbf{x}')\rangle = \delta(t-t')\delta^{(3)}(\mathbf{x}-\mathbf{x}')~. 
\end{equation}

Since we work with several frames, it is useful to show explicitly how the noise transforms between them. As seen in Eqs.~\eqref{eq:sf-eq-rescaledandredefined}--\eqref{eq:Euler-eq-rescaledandredefined}, the noise is rescaled as
\begin{equation}
    \tilde{\xi}(t,\mathbf{x}) \equiv \frac{\xi(t,\mathbf{x})}{F^2(\phi)\sqrt{\mathcal{K}(\phi)}}=\sqrt{\frac{2\Upsilon T}{a^3 \mathcal{N} F^4(\phi) \mathcal{K}(\phi)}}\mathring{\xi}(t,\mathbf{x})~.\label{eq:rescaling-noise-term-whywasthisnotlabelled}
\end{equation}
The conformal rescaling of $\mathcal{N}$ and $a$ (Eqs.~\eqref{eq:transformation-background-lapse-conformal} and \eqref{eq:transformation-background-scale-factor}, respectively) brings this to the Einstein-frame form
\begin{equation}
\langle \tilde{\xi}(\hat{\tau},\mathbf{x})\tilde{\xi}(\hat{\tau}',\mathbf{x}')\rangle =\frac{2\tilde{\Upsilon}\hat{T}}{\hat{a}^3\hat{\mathcal{N}}}\delta(t-t')\delta^{(3)}(\mathbf{x}-\mathbf{x}')~,
\end{equation}
with the rescaled dissipation coefficient and temperature \cite{Faraoni:2023gqg,Karolinski:2024ukr}:
\begin{eqnarray}
    \tilde{\Upsilon} &\equiv& \frac{\Upsilon}{\mathcal{K}(\phi)F^{3/2}(\phi)}~, \\
    \hat{T} &\equiv& \frac{T}{\sqrt{F(\phi)}}~, \label{eq:hatted temperature2}
\end{eqnarray}
respectively. The tilde (rather than a hat) on the rescaled $\Upsilon$ is used for the reasons discussed in Sec.~\ref{sec:the_eqs_WI_effective}; the same logic applies to the noise term.

\section{\label{sec:app-linearperturbations}Linear Scalar and Tensor Perturbations in a Flat FLRW Universe with the Lapse Function}
\noindent
In this appendix we derive a number of linearly perturbed geometric quantities relevant to Sec.~\ref{sec:cosmoperturb-einsteinframe}: the Christoffel symbols, the components of the Ricci and Einstein tensors, and the Ricci scalar. Attention is restricted to scalar and tensor perturbations of the metric, and we retain an arbitrary lapse function $\mathcal{N}$ for the background.\footnote{Although the background lapse function is a Lagrange multiplier in GR \cite{Remmen:2013eja} and is conveniently set to unity, we leave it arbitrary in this appendix because of its relation to the non-minimally coupled scalar field in the Einstein frame. The freedom to choose $\mathcal{N}$ in the defining and Einstein frames is intertwined with the choice of $\bar{F}$; see Eq.~\eqref{eq:transformation-background-lapse-conformal} and Refs.~\cite{Karciauskas:2022jzd,Diaz:2023tma}.} The matter energy-momentum tensor (EMT) is also perturbed, although the focus is placed on its scalar perturbations. 

For ease of reference, we begin by rewriting the perturbed line element of Eq.~\eqref{eq:perturbed-line-element} in any frame:
\begin{equation}
    \textrm{d}s^2 = -\left(1+2A\right)\mathcal{N}^2 \textrm{d}t^2-2a\mathcal{N}\partial_i B \textrm{d}x^{i} \textrm{d}t +a^2\left[\left(1-2\psi\right) \delta_{ij}+2\partial_i \partial_j E+h_{ij}\right]\textrm{d}x^{i}\textrm{d}x^{j}~,
\end{equation}
in which $A(x)$ is the linear perturbation of the lapse function, $B(x)$ the scalar shift perturbation, $\psi(x)$ the intrinsic curvature perturbation, $E(x)$ the scalar shear perturbation, and $h_{ij}(x)$ the traceless-transverse tensor mode, satisfying $\partial_i \tensor{h}{^{i}_{j}} = \tensor{h}{^{i}_{i}}=0$ \cite{Baumann:2022mni}. The background metric components are\footnote{\label{raising-lowering-indices}Overbars are employed to distinguish background (spatially homogeneous and isotropic) quantities from their full counterparts, unless otherwise stated. We use $\delta$ for linear perturbations. As noted at the beginning of Sec.~\ref{sec:cosmoperturb-einsteinframe}, Latin indices are lowered and raised with the flat spatial metric $\delta_{ij}$ and its inverse $\delta^{ij}$, respectively.}
\begin{eqnarray}
    \bar{g}_{00} &=& -\mathcal{N}^2~, \ \ \ \ \ \ \ \ \ \ \bar{g}^{00} = -\frac{1}{\mathcal{N}^2}~,\\
    \bar{g}_{0i} &=& \bar{g}_{i0} = 0~, \ \ \ \ \ \ \ \ \bar{g}^{0i} = \bar{g}^{i0} = 0~,\\
    \bar{g}_{ij} &=& a^2 \delta_{ij}~, \ \ \ \ \ \ \ \ \ \ \ \bar{g}^{ij} = a^{-2} \delta^{ij}~,
\end{eqnarray}
while the linear perturbations are
\begin{eqnarray}
    \label{eq:deltag00}\delta g_{00} &=& -2\mathcal{N}^2A~, \ \ \ \ \ \ \ \ \ \ \ \ \ \ \ \ \ \ \ \ \ \ \ \ \ \ \ \ \ \ \ \ \ \ \ \delta g^{00} = 2\frac{A}{\mathcal{N}^2}~,\\
    \delta g_{0i} &=& \delta g_{i0} = -a\mathcal{N}\partial_i B~, \ \ \ \ \ \ \ \ \ \ \ \ \ \ \ \ \ \ \ \ \ \ \ \ \ \delta g^{0i} = \delta g^{i0} = -\frac{\partial^i B}{a\mathcal{N}}~,\\
    \label{eq:deltagij}\delta g_{ij} &=& -2a^2\left(\psi \delta_{ij}-\partial_i \partial_j E-\frac{1}{2}h_{ij}\right), \ \ \ \ \ \ \ \ \delta g^{ij} = \frac{2}{a^2}\left(\psi \delta^{ij}-\partial^i \partial^j E-\frac{1}{2}h^{ij}\right).
\end{eqnarray}
Using Jacobi's formula \cite{Malik:2008im}, we find that the square root of the determinant of the metric at linear order is
\begin{equation}
    \delta(\sqrt{-g}) = -\frac{1}{2}\sqrt{-\bar{g}}~\bar{g}_{\mu\nu}\delta g^{\mu\nu} = \mathcal{N}a^3 \left(A-3\psi +\partial_i \partial^{i} E\right),
\end{equation}
while $\sqrt{-\bar{g}} = \mathcal{N}a^3$.

The non-vanishing background Christoffel symbols read 
\begin{eqnarray}
    \bar{\Gamma}^{0}_{00} &=&\frac{\dot{\mathcal{N}}}{\mathcal{N}}~,\\
    \bar{\Gamma}^{i}_{0j} = \bar{\Gamma}^{i}_{j0} &=& \mathcal{N}H \delta^{i}_j~,\\
    \bar{\Gamma}^{0}_{ij} &=& a^2 \frac{H}{\mathcal{N}}\delta_{ij}~.
\end{eqnarray}
Recall from Sec.~\ref{sec:the_eqs_WI_effective} that $H\equiv \dot a/(\mathcal{N}a)$, although here we make no reference to any specific frame \cite{Karciauskas:2022jzd}, where overdots denote derivatives with respect to the coordinate time $t$. The linear perturbations take the form
\begin{eqnarray}
    \delta \Gamma^{0}_{00} &=& \dot A~,\\
    \delta \Gamma^{0}_{i0} = \delta \Gamma^{0}_{0i} &=& \partial_i\left(A -aHB\right),\\
    \delta \Gamma^{i}_{00} &=& \frac{\mathcal{N}^2}{a^2}\partial^{i}\left(A-aHB -\frac{a}{\mathcal{N}}\dot B\right),\\
    \delta \Gamma^{0}_{ij} &=& -\frac{a^2}{\mathcal{N}^2}\left\{\left[\dot \psi +2\mathcal{N}H\left(\psi+A\right)\right]\delta_{ij}-2\mathcal{N}H\partial_i \partial_j E-\mathcal{N}Hh_{ij}-\frac{1}{2}\dot h_{ij}\right\}+\partial_i \partial_j \chi~,\\
    \delta \Gamma^{i}_{0j} = \delta \Gamma^{i}_{j0} &=& -\dot \psi \delta^{i}_j +\partial^{i}\partial_j \dot E+\frac{1}{2}\tensor{\dot{h}}{^{i}_{j}}~,\\
    \delta \Gamma^{k}_{ij} &=& -\left(\delta_i^{k} \partial_j \psi +\delta_j^{k} \partial_i \psi -\delta_{ij} \partial^{k} \psi \right) +aH\delta_{ij}\partial^{k} B+\partial_{i}\partial_j \partial^{k} E+\frac{1}{2}\left(\partial_i\tensor{h}{^{k}_{j}}+\partial_j \tensor{h}{^{k}_i}-\partial^{k} h_{ij}\right).
\end{eqnarray}
Furthermore,
\begin{equation}
    \label{eq:shear-potential}\chi \equiv \frac{a}{\mathcal{N}}\left(B+\frac{a}{\mathcal{N}}\dot E\right)
\end{equation} 
is the shear potential. 

The non-vanishing background Ricci tensor components are
\begin{eqnarray}
    \bar{R}_{00} &=& -3\mathcal{N}^2\left(\frac{\dot H}{\mathcal{N}}+H^2\right),\\
    \bar{R}_{ij} &=& a^2\left(\frac{\dot H}{\mathcal{N}}+3H^2\right)\delta_{ij}~,
\end{eqnarray}
and the mixed components of the linear perturbation read
\begin{align}
    &\delta \tensor{R}{^{0}_{0}} = -\frac{3}{\mathcal{N}^2}\left[\ddot \psi -\frac{\dot{\mathcal{N}}}{\mathcal{N}}\dot \psi +\mathcal{N}H\left(2\dot \psi +\dot A\right)+2\mathcal{N}^2\left(\frac{\dot H}{\mathcal{N}}+H^2\right)A\right]+a^{-2}\partial_i \partial^{i}\left(\dot \chi +\frac{\dot{\mathcal{N}}}{\mathcal{N}}\chi -A\right),\\
    &\delta \tensor{R}{^{0}_{i}} = -\frac{2}{\mathcal{N}^2}\partial_i \left(\dot \psi +\mathcal{N}HA\right),\\
    &\delta \tensor{R}{^{i}_{0}} = \frac{2}{a^2}\partial^{i}\left(\dot \psi +\mathcal{N}H A+a\dot H B\right),\\
    \nonumber&\delta \tensor{R}{^{i}_{j}} = -\frac{1}{\mathcal{N}^2}\left[\ddot \psi-\frac{\dot{\mathcal{N}}}{\mathcal{N}}\dot \psi +\mathcal{N}H\left(6\dot \psi +\dot A\right)+2\mathcal{N}^2\left(\frac{\dot H}{\mathcal{N}}+3H^2\right)A-\frac{\mathcal{N}^2}{a^2}\partial_k \partial^{k}\left(\psi+\mathcal{N}H\chi\right)\right]\delta^{i}_{j}\\
    &+a^{-2}\partial_j \partial^i\left(\dot \chi +\frac{\dot{\mathcal{N}}}{\mathcal{N}}\chi+\mathcal{N}H\chi+\psi - A \right)+\frac{1}{2\mathcal{N}^2}\left(\tensor{\ddot{h}}{^{i}_{j}}-\frac{\dot{\mathcal{N}}}{\mathcal{N}}\tensor{\dot{h}}{^{i}_{j}}+3\mathcal{N}H\tensor{\dot{h}}{^{i}_{j}}\right)-\frac{1}{2}a^{-2}\partial_k \partial^{k} \tensor{h}{^{i}_{j}}~.
\end{align}
The background Ricci scalar is 
\begin{equation}
    \label{eq:background-Ricci}\bar{R} = 6\left(\frac{\dot H}{\mathcal{N}}+2H^2\right),
\end{equation}
and its linear perturbation reads
\begin{equation}
    \label{eq:Ricci-scalar-pert}\delta R = 2\left[\frac{\dot \kappa}{\mathcal{N}}+4H\kappa-\left(a^{-2}\partial_i \partial^{i}A+3\frac{\dot H}{\mathcal{N}} A\right)+2a^{-2}\partial_i \partial^{i}\psi\right],
\end{equation}
where
\begin{equation}
    \label{eq:kappa-expansion}\kappa \equiv -3\left(HA+\frac{\dot \psi}{\mathcal{N}}\right)+a^{-2}\mathcal{N}\partial_i \partial^{i}\chi
\end{equation}
is the scalar expansion rate perturbation at linear order in the comoving frame, \textit{i.e.} the rest frame of the matter fluid \cite{Hwang:1996xh,Hwang:2001qk}.  

Using the Ricci scalar and Ricci tensor, we obtain the Einstein tensor components, which at background level are
\begin{eqnarray}
    \bar{G}_{00} &=& 3\mathcal{N}^2 H^2~,\\
    \bar{G}_{ij} &=& -a^2\left(2\frac{\dot H}{\mathcal{N}}+3H^2\right)\delta_{ij}~,
\end{eqnarray}
while the corresponding linear perturbations read
\begin{align}
    &\delta \tensor{G}{^{0}_{0}} = -2\left(H\kappa +a^{-2} \partial_i \partial^{i}\psi\right),\\
    &\delta \tensor{G}{^{0}_{i}} = \delta \tensor{R}{^{0}_{i}}~,\\
    &\delta \tensor{G}{^{i}_{0}} = \delta \tensor{R}{^{i}_{0}}~,\\
    \nonumber&\delta \tensor{G}{^{i}_{j}} = -\frac{1}{\mathcal{N}^2} \left[\ddot \psi -\frac{\dot{\mathcal{N}}}{\mathcal{N}}\dot \psi+\mathcal{N}\dot\kappa +3\mathcal{N}^2 H \kappa+\mathcal{N}H\left(3\dot \psi +\dot A\right)-\mathcal{N}^2\left(\frac{\dot H}{\mathcal{N}}-3H^2\right)A+\frac{\mathcal{N}^2}{a^2}\partial_k \partial^{k}\left(\psi-A\right)\right]\delta^{i}_{j}\\
    &+a^{-2}\partial_j \partial^i \left(\dot \chi +\frac{\dot{\mathcal{N}}}{\mathcal{N}}\chi +\mathcal{N}H\chi+\psi-A\right)+\frac{1}{2\mathcal{N}^2}\left(\tensor{\ddot{h}}{^{i}_{j}}-\frac{\dot{\mathcal{N}}}{\mathcal{N}}\tensor{\dot{h}}{^{i}_{j}}+3\mathcal{N}H\tensor{\dot{h}}{^{i}_{j}}\right)-\frac{1}{2}a^{-2}\partial_k \partial^{k} \tensor{h}{^{i}_{j}}~.
\end{align}

For the four-velocity of a fluid element, $u^{\mu}$, we parametrise the components as
\begin{equation}
    u^{\mu} = \mathcal{N}^{-1}\delta^{\mu}_0 +\delta u^{\mu}~,
\end{equation}
with $\delta u^{\mu}$ the corresponding linear perturbation. The (spatially) homogeneous and isotropic components are then $\bar{u}^{0} = \mathcal{N}^{-1}$ and $\bar{u}^{i} = 0$. The normalisation condition $g_{\mu\nu}u^{\mu}u^{\nu} = -1$, combined with the perturbed metric components in Eqs.~\eqref{eq:deltag00}--\eqref{eq:deltagij}, yields
\begin{eqnarray}
    \delta u^{0} &=& -\mathcal{N}^{-1}A~,\\
    \label{eq:deltaui-intermsof-comoving}\delta u^{i} &=& a^{-1}\partial^{i}v~,
\end{eqnarray}
where $\delta u^{i}$ has been parametrised in terms of the comoving vector $\partial^{i}v$ (since only scalar perturbations are retained, $\delta u^{i}$ is the gradient of a single scalar). The covariant components are
\begin{eqnarray}
    \delta u_0 &=& -\mathcal{N}A~,\\
    \delta u_i &=& a\partial_i\left(v-B\right),
\end{eqnarray}
with background values $\bar{u}_0 = -\mathcal{N}$ and $\bar{u}_i = 0$. The linear perturbation of the scalar expansion rate, $\Theta \equiv \nabla_{\mu} u^{\mu}$, is then given by
\begin{equation}
    \label{eq:pert-scalar-exp-rate}\delta \Theta = \kappa +a^{-1}\partial_i \partial^{i}\left(v-B\right),
\end{equation}
the background value being $\bar{\Theta}=3H$.

To understand why $\kappa$ has been identified as the scalar expansion rate in the `comoving' frame, it is useful to compute the perturbed components of the matter EMT, $\tensor{\mathcal{T}}{^{\mu}_{\nu}}$, defined in Eq.~\eqref{eq:perfect-fluid-form-emtensor}:
\begin{eqnarray}
    \label{eq:comp-00-pert-matter}\delta \tensor{\mathcal{T}}{^{0}_{0}} &=& -\delta \rho_{\textrm{m}}~,\\
    \delta \tensor{\mathcal{T}}{^{0}_{i}} &=& \frac{a\left(\bar{\rho}_{\textrm{m}}+\bar{P}_{\textrm{m}}\right)}{\mathcal{N}}\partial_i \left(v-B\right),\\
    \delta \tensor{\mathcal{T}}{^{i}_{0}} &=& -\frac{\mathcal{N}\left(\bar{\rho}_{\textrm{m}}+\bar{P}_{\textrm{m}}\right)}{a}\partial^{i}v~,\\
    \label{eq:comp-ij-pert-matter}\delta \tensor{\mathcal{T}}{^{i}_{j}} &=& \delta P_{\textrm{m}}\delta^{i}_j~,
\end{eqnarray}
where $\bar{\rho}_{\textrm{m}}$ and $\bar{P}_{\textrm{m}}$ stand for the homogeneous energy and isotropic pressure densities, respectively. Setting $v=B$ yields $\delta \tensor{\mathcal{T}}{^{0}_i}=0$, \textit{i.e.} the spatial flow of energy or momentum density of matter,
\begin{equation}
    \label{eq:momentum-matter}\Psi_{\textrm{m}}\equiv a\left(\bar{\rho}_{\textrm{m}}+\bar{P}_{\textrm{m}}\right)\left(v-B\right),
\end{equation}
vanishes; the observer is then comoving with the matter fluid, hence the name `comoving frame'.

The four-acceleration, defined below Eq.~\eqref{eq:conservation-eq-mattersector} as $a^{\mu} \equiv u^{\nu}\nabla_{\nu}u^{\mu}$, has vanishing background components, $\bar{a}^{0} = 0$ and $\bar{a}^{i} = 0$. Its linear perturbations are 
\begin{eqnarray}
    \label{eq:contravariant-a0}\delta a^{0} &=& 0~, \\
    \label{eq:contravariant-ai}\delta a^{i} &=& \frac{1}{a\mathcal{N}}\partial^{i}\left[\dot v -\dot B +\mathcal{N}H\left(v-B\right)+\frac{\mathcal{N}}{a}A\right]. 
\end{eqnarray}
The covariant components similarly vanish at the background level, $\bar{a}_0 =0$ and $\bar{a}_i = 0$, while their linear perturbations read
\begin{eqnarray}
    \delta a_0 &=& 0~,\\
    \label{eq:covariant-ai}\delta a_i &=& \frac{a}{\mathcal{N}} \partial_i \left[\dot v -\dot B +\mathcal{N}H\left(v-B\right)+\frac{\mathcal{N}}{a}A\right].
\end{eqnarray}
The four-acceleration does not vanish in the comoving frame, where it reduces to the gradient of the lapse perturbation; this is to be expected, as comoving observers do not in general follow geodesics.  

Finally, the background covariant components of the EMT read
\begin{eqnarray}
    \label{eq:cov-back-matter-00}\bar{\mathcal{T}}_{00} &=& \mathcal{N}^2 \bar{\rho}_{\textrm{m}}~,\\
    \bar{\mathcal{T}}_{0i} = \bar{\mathcal{T}}_{i0} &=& 0~,\\
    \label{eq:cov-back-matter-ij}\bar{\mathcal{T}}_{ij} &=& a^2 \bar{P}_{\textrm{m}}\delta_{ij}~.
\end{eqnarray}

\section{\label{sec:app-lineargaugeconf}Linearised Gauge and Conformal Transformations of Linear Scalar and Tensor Perturbations}
\subsection{Gauge Transformations}
\noindent
In relativistic cosmological perturbation theory, the split between background and perturbed spacetimes is not unique, and perturbations are therefore \emph{gauge-dependent} \cite{Bardeen:1980kt,Kodama:1984ziu,Hwang:2001fb,Malik:2008im}. For a gauge-transformation vector parametrised, in the scalar sector, as
\begin{equation}
    \label{eq:gauge-trans-vector-xi}\xi^{\mu} = \left(\alpha,\partial^{i}\beta\right),
\end{equation}
with $\partial^{i}\beta = \delta^{ij}\partial_j \beta$ (see footnote~\ref{raising-lowering-indices}), and $\alpha = \alpha(x)$, $\beta = \beta(x)$, the scalar metric perturbations in Eq.~\eqref{eq:perturbed-line-element} transform under the infinitesimal coordinate change $x^{\mu} \rightarrow x^{\mu}+\xi^{\mu}$ (a first-order gauge transformation) as\footnote{\label{footnote:explanation-no-one-thought-of}Spatially constant integration terms have been omitted from the transformations of $B$ and $E$ in Eqs.~\eqref{eq:gauge-trans-B} and \eqref{eq:gauge-trans-E}, respectively; they correspond to global boosts (equivalently, a background redefinition), and are conventionally set to zero.} 
\begin{eqnarray}
    \label{eq:gauge-trans-A}A &\rightarrow& A -\dot \alpha -\frac{\dot{\mathcal{N}}}{\mathcal{N}}\alpha~,\\
    \label{eq:gauge-trans-B}B &\rightarrow& B + a\frac{\dot\beta}{\mathcal{N}}-\frac{\mathcal{N}}{a}\alpha~,\\
    \label{eq:gauge-trans-E}E &\rightarrow& E - \beta~,\\
    \label{eq:gauge-trans-psi}\psi &\rightarrow& \psi+\mathcal{N}H\alpha~.
\end{eqnarray}
These follow from the standard transformation rules for tensor perturbations under infinitesimal diffeomorphisms. The shear potential of Eq.~\eqref{eq:shear-potential} consequently transforms as
\begin{equation}
    \label{eq:gauge-trans-chi}\chi \rightarrow \chi - \alpha~,
\end{equation}
which depends on $\alpha$ alone: $\chi$ is therefore invariant under a change of threading, as are $A$ and $\psi$. By contrast, $E$ is invariant under a change of slicing. The following two combinations are fully gauge-invariant at first order (see footnote~\ref{footnote:superscript-perturbation} regarding the superscript notation):
\begin{eqnarray}
    \label{eq:BardeenA}A^{\chi} &\equiv& A-\frac{\dot{\left(\mathcal{N}\chi\right)}}{\mathcal{N}}~,\\ 
    \label{eq:BardeenPsi}\psi^{\chi} &\equiv& \psi+\mathcal{N}H \chi~.
\end{eqnarray}
These are the \emph{Bardeen potentials} \cite{Bardeen:1980kt,Malik:2008im,Baumann:2009ds}, that is, the lapse and curvature perturbations on slicings of zero shear. The longitudinal (or Newtonian) gauge is defined by $B=E=0$, which together imply $\chi =0$; this fixes both the slicing and the threading, and constitutes a complete gauge fixing, with the resulting metric fully diagonal. By contrast, the synchronous gauge ($A=B=0$) is an example of incomplete gauge fixing \cite{Ma:1995ey,Weinberg:2003sw}, given that the imposed conditions leave residual coordinate freedom in the form of time-independent spatial reparametrisations. 

The four-velocity perturbation transforms as
\begin{equation}
    \label{eq:gauge-trans-v}v \rightarrow v+a\frac{\dot \beta}{\mathcal{N}}~,
\end{equation}
whereas the four-acceleration is gauge-invariant at linear order (see Eqs.~\eqref{eq:contravariant-a0}--\eqref{eq:covariant-ai}). This is consistent with the fact that a gauge transformation cannot alter the physical worldline, and follows from the Stewart-Walker lemma \cite{Stewart:1974uz} as the background worldline has zero acceleration. The scalar expansion rate perturbation in Eq.~\eqref{eq:pert-scalar-exp-rate} obeys the transformation rule
\begin{equation}
    \label{eq:gauge-expansionrate}\delta \Theta \rightarrow \delta \Theta -3\dot H \alpha~,
\end{equation}
and the matter momentum density $\Psi_{\textrm{m}}$, defined in Eq.~\eqref{eq:momentum-matter}, transforms as
\begin{equation}
    \label{eq:gauge-moment-matter}\Psi_{\textrm{m}}\rightarrow \Psi_{\textrm{m}} +\mathcal{N}\left(\bar{\rho}_{\textrm{m}}+\bar{P}_{\textrm{m}}\right)\alpha~.
\end{equation}
From the gauge transformations of the mixed components of the matter EMT (Eqs.~\eqref{eq:comp-00-pert-matter}--\eqref{eq:comp-ij-pert-matter}) one obtains\footnote{\label{footnote:gauge-trans-scalar}Transformations of the type below are the standard linear gauge-transformation law for a scalar quantity $s$ with time-dependent background value $\bar{s}$, namely $\delta s\rightarrow \delta s-\dot{\bar{s}}\alpha$ \cite{Lyth:2009zz}. For Eq.~\eqref{eq:gauge-expansionrate}, the background value of the scalar expansion rate is $3H$, so that $\frac{1}{3}\delta \Theta \rightarrow \frac{1}{3}\delta \Theta-\dot H \alpha$. This rule does not apply to the matter momentum density in Eq.~\eqref{eq:gauge-moment-matter}, since it has no background counterpart and is not a genuine scalar under diffeomorphisms.}
\begin{eqnarray}
    \delta \rho_{\textrm{m}} &\rightarrow& \delta \rho_{\textrm{m}}-\dot{\bar{\rho}}_{\textrm{m}}\alpha~,\\
    \label{eq:gauge-trans-deltapm}\delta P_{\textrm{m}} &\rightarrow& \delta P_{\textrm{m}}-\dot{\bar{P}}_{\textrm{m}}\alpha~.
\end{eqnarray}
A gauge-invariant quantity that follows from these transformations is the non-adiabatic pressure perturbation (equivalently the pressure perturbation on slicings of uniform energy density):
\begin{equation}
    \delta P_i^{\textrm{nad}} \equiv \delta P_i-\frac{\dot{\bar{P}}_i}{\dot{\bar{\rho}}_i} \delta \rho_i~,
\end{equation}
where $i=\textrm{m},\Phi$,~... runs over the matter sector, the scalar field, or any other field included in the effective fluid description. 

Other relevant gauge-invariant quantities are the curvature perturbation on slicings of uniform energy density,
\begin{equation}
    -\zeta_i \equiv \psi +\frac{\mathcal{N}H}{\dot{\bar{\rho}}_i}\delta \rho_i~, 
\end{equation}
and the comoving curvature perturbation \cite{Weinberg:2003sw,Malik:2008im} (the curvature perturbation on slicings of zero momentum),
\begin{equation}
    \mathcal{R}_i\equiv \psi -\frac{H}{\bar{\rho}_{i}+\bar{P}_{i}}\Psi_i~,
\end{equation}
where the index $i$ spans the same set as above. Using the gauge transformations of the total energy density and momentum perturbations, 
\begin{eqnarray}
    \delta \rho &\rightarrow& \delta \rho -\dot{\bar{\rho}} \alpha~,\\
    \Psi &\rightarrow& \Psi+\mathcal{N}\left(\bar{\rho}+\bar{P}\right)\alpha~,
\end{eqnarray}
the corresponding total curvature perturbation are obtained:
\begin{eqnarray}
    \label{eq:total-zeta-definition-allincluded}-\zeta &\equiv& \psi +\frac{\mathcal{N}H}{\dot{\bar{\rho}}}\delta \rho~,\\
    \label{eq:total-R-definition-allincluded}\mathcal{R} &\equiv& \psi-\frac{H}{\bar{\rho}+\bar{P}}\Psi~.
\end{eqnarray}
Note that, in general, $\zeta \neq \sum_i \zeta_i$ and $\mathcal{R}\neq \sum_i \mathcal{R}_i$.

Finally, the linear tensor perturbations $h_{ij}$ are gauge-invariant \cite{Malik:2008im} by virtue of the aforementioned Stewart-Walker lemma, since FLRW spacetimes admit no background tensor counterpart. 

\subsection{Conformal Transformations}
\noindent
We are also interested in conformal transformations \cite{Dabrowski:2008kx}. As already noted, these \emph{do not} correspond to a change of coordinates (see footnote~\ref{footnote:clarification-conformalrescaling}), in contrast to gauge transformations. The gauge transformation vector $\xi^{\mu}$ in Eq.~\eqref{eq:gauge-trans-vector-xi} is therefore unchanged by a conformal rescaling. Since the background metric components do change, however, the gauge transformations of $A$ and $\psi$ must be modified accordingly in the conformally related frame \cite{Fakir:1992cg,Hwang:1996xh,Hwang:2001qk,Gong:2011qe}. 

To see this, we first determine how the scalar metric perturbations change under a conformal rescaling of the form
\begin{equation}
    \hat{g}_{\mu\nu}(x) = \Omega^2(x) g_{\mu\nu}(x)~,
\end{equation}
where $\Omega^2$ is an arbitrary conformal factor (not necessarily $F$, as in Eq.~\eqref{eq:conformal-Einstein}). Linearising \cite{Fakir:1992cg}:
\begin{equation}
    \delta \hat{g}_{\mu\nu}(x) = \bar{\Omega}^2(t) \delta g_{\mu\nu}(x)+2\bar{\Omega}(t) \bar{g}_{\mu\nu}(t) \delta \Omega(x)~,
\end{equation}
from which\footnote{Note that $\mathcal{N}$ and $a$ are themselves rescaled by the conformal transformation. For instance, $\delta g_{00}$ in Eq.~\eqref{eq:deltag00} yields $\delta \hat{g}_{00} = -2\hat{\mathcal{N}}^2 \hat{A}$ and not $\delta \hat{g}_{00} = -2\mathcal{N}^2 \hat{A}$.}
\begin{eqnarray}
    A &\rightarrow& A+\frac{\delta \Omega}{\bar{\Omega}}~,\\
    B &\rightarrow& B~,\\
    E &\rightarrow& E~,\\
    \psi &\rightarrow& \psi -\frac{\delta \Omega}{\bar{\Omega}}~.
\end{eqnarray}
The invariance of $B$ and $E$ is consistent with that of their respective gauge-transformation rules in Eqs.~\eqref{eq:gauge-trans-B} and \eqref{eq:gauge-trans-E} (we notice that $\hat{\mathcal{N}} = \bar{\Omega} \mathcal{N}$ and $\hat{a}= \bar{\Omega} a$). The traceless-transverse tensor mode $h_{ij}$ is likewise invariant \cite{Gong:2011qe}. Combining these results with the rescalings of $\mathcal{N}$ and $a$, the shear potential $\chi$ in Eq.~\eqref{eq:shear-potential} is also unchanged:
\begin{equation}
    \label{eq:frame-transformation-shear-potential}\hat{\chi} = \frac{\hat{a}}{\hat{\mathcal{N}}}\left(\hat{B}+\frac{\hat{a}}{\hat{\mathcal{N}}}\dot{\hat{E}}\right) = \frac{a}{\mathcal{N}}\left(B+\frac{a}{\mathcal{N}}\dot E\right)=\chi~.
\end{equation}
Consequently the longitudinal gauge ($\chi=0$) is preserved by any conformal rescaling of the metric, in any frame \cite{Brown:2011eh}. This is not true of the uniform-curvature ($\psi=0$) or synchronous ($A=0$) gauges, which involve quantities that are altered by the conformal transformation. The corresponding gauge-transformation rules for $A$ and $\psi$ must therefore be readjusted. As a result, the gauge-invariant Bardeen potentials in Eqs.~\eqref{eq:BardeenA} and \eqref{eq:BardeenPsi} are \emph{not} conformally invariant:
\begin{eqnarray}
    \hat{A}^{\hat{\chi}} &=& A^{\chi}+\bar{\Omega}^{-1}\left(\delta \Omega - \dot{\bar{\Omega}} \chi\right),\\
    \hat{\psi}^{\hat{\chi}} &=& \psi^{\chi} -\bar{\Omega}^{-1}\left(\delta \Omega -\dot{\bar{\Omega}} \chi\right).
\end{eqnarray}
The combination $A^{\chi}+\psi^{\chi}$ is, however, conformally invariant, as is the gauge-dependent combination $A+\psi$.

As explained in footnote~\ref{footnote:fourveloc}, the four-velocity $u^{\mu}$ transforms as $u^{\mu}(x) \rightarrow \Omega^{-1}(x)u^{\mu}(x)$ under the conformal rescaling (see Eq.~\eqref{eq:rescale-velocity} for the special case $\Omega^2(x)=F[\Phi(x)]$). The corresponding scalar velocity perturbation is invariant: $\hat{v}=v$ (see Eq.~\eqref{eq:deltaui-intermsof-comoving}), as expected, because the rescaling is absorbed into the redefinition of the scale factor $a$. This is consistent with the invariance of the gauge-transformation rule for $v$ in Eq.~\eqref{eq:gauge-trans-v}. However, the four-acceleration components are not invariant, owing to the lapse perturbation $A$ (see Eq.~\eqref{eq:contravariant-ai}). Although a gauge transformation merely relabels spacetime points and does not alter the worldline, a conformal transformation modifies the affine structure and induces a genuine acceleration of the worldline in the rescaled spacetime (see Eq.~\eqref{eq:rescale-acceleration}): $a^{\mu}(x)\rightarrow \Omega^{-2}(x)\left[a^{\mu}(x)+h^{\mu\nu}(x)\nabla_{\nu}\ln \Omega(x)\right]$, where $h_{\mu\nu}(x) \equiv g_{\mu\nu}(x) + u_{\mu}(x)u_{\nu}(x)$ is the projection tensor introduced below Eq.~\eqref{eq:perfect-fluid-form-emtensor}.

Since $v-B$ is conformally invariant (and, as shown below in Eqs.~\eqref{eq:trans-conf-Omega-rhomback} and \eqref{eq:trans-conf-Omega-pmback}, $\bar{\rho}_{\textrm{m}}$ and $\bar{P}_{\textrm{m}}$ are rescaled by a multiplicative factor only), the comoving frame with respect to the matter fluid (defined by $\Psi_{\textrm{m}}=0$) is conformally invariant. The scalar expansion rate perturbation $\kappa$ in Eq.~\eqref{eq:kappa-expansion}, by contrast, transforms in a more involved way:
\begin{equation}
    \kappa \rightarrow \bar{\Omega}^{-1}\left\{\kappa-3\left[\frac{\dot{\bar{\Omega}}}{\mathcal{N}\bar{\Omega}}A+\left(H+2\frac{\dot{\bar{\Omega}}}{\mathcal{N}\bar{\Omega}}\right)\frac{\delta \Omega}{\bar{\Omega}}-\frac{\dot{\delta \Omega}}{\mathcal{N}\bar{\Omega}}\right]\right\}, 
\end{equation}
and the same transformation law applies to $\delta \Theta$ in Eq.~\eqref{eq:pert-scalar-exp-rate}:
\begin{equation}
    \label{eq:pert-scalar-exp-rate-conformal-rescaling-kindofdifficulttounderstand}\delta \Theta \rightarrow\bar{\Omega}^{-1}\left\{\delta\Theta-3\left[\frac{\dot{\bar{\Omega}}}{\mathcal{N}\bar{\Omega}}A+\left(H+2\frac{\dot{\bar{\Omega}}}{\mathcal{N}\bar{\Omega}}\right)\frac{\delta \Omega}{\bar{\Omega}}-\frac{\dot{\delta \Omega}}{\mathcal{N}\bar{\Omega}}\right]\right\}. 
\end{equation} 

As with the four-velocity, the energy and pressure densities of the matter fluid are rescaled by the conformal transformation. From the definition of the EMT in Eq.~\eqref{eq:energy-momentum-tensor}, and assuming that $S_{\textrm{m}}$ remains unchanged after the conformal rescaling and accompanying field redefinitions (see footnote~\ref{footnote:clarification-conformalrescaling}), one has
\begin{equation}
    \hat{\mathcal{T}}_{\mu\nu} = \frac{1}{\Omega^2}\mathcal{T}_{\mu\nu}~,
\end{equation}
and hence $\tensor{\hat{\mathcal{T}}}{^{\hat{\mu}}_{\nu}} = \hat{g}^{\mu\alpha}\hat{\mathcal{T}}_{\alpha\nu} = \Omega^{-4} \tensor{\mathcal{T}}{^{\mu}_{\nu}}$. At background level:
\begin{eqnarray}
    \label{eq:trans-conf-Omega-rhomback}\bar{\rho}_{\textrm{m}} &\rightarrow& \bar{\Omega}^{-4} \bar{\rho}_{\textrm{m}}~,\\
    \label{eq:trans-conf-Omega-pmback}\bar{P}_{\textrm{m}} &\rightarrow& \bar{\Omega}^{-4} \bar{P}_{\textrm{m}}~,
\end{eqnarray}
because $\bar{\rho}_{\textrm{m}}\equiv -\tensor{\bar{\mathcal{T}}}{^{0}_{0}}$ and $\bar{P}_{\textrm{m}} \equiv \frac{1}{3} \tensor{\bar{\mathcal{T}}}{^{i}_{i}}$. The equation-of-state (EoS) parameter, $w_{\textrm{m}} \equiv \bar{P}_{\textrm{m}}/\bar{\rho}_{\textrm{m}}$, is therefore conformally invariant, while the adiabatic sound speed, $c_s^2 \equiv \dot{\bar{P}}_{\textrm{m}}/\dot{\bar{\rho}}_{\textrm{m}}$, transforms as
\begin{equation}
    \hat{c}_s^2 = \frac{\dot{\hat{\bar{P}}}_{\textrm{m}}}{\dot{\hat{\bar{\rho}}}_{\textrm{m}}} = \frac{\dot{\bar{\rho}}_{\textrm{m}}c_s^2-4\frac{\dot{\bar{\Omega}}}{\bar{\Omega}}\bar{\rho}_{\textrm{m}}w_{\textrm{m}}}{\dot{\bar{\rho}}_{\textrm{m}}-4\frac{\dot{\bar{\Omega}}}{\bar{\Omega}}\bar{\rho}_{\textrm{m}}}~.
\end{equation}
The two coincide, $\hat{c}_s^2 = c_s^2$, only when $c_s^2 = w_{\textrm{m}}$ (\textit{i.e.} for constant EoS parameter). The transformations above agree with Eqs.~\eqref{eq:rescale-rhom} and \eqref{eq:rescale-pm}, respectively, when $\Omega^{2}(x) = F[\Phi(x)]$. At the level of perturbations:
\begin{eqnarray}
    \delta \rho_{\textrm{m}} &\rightarrow& \bar{\Omega}^{-4}\left(\delta \rho_{\textrm{m}}-4\frac{\delta \Omega}{\bar{\Omega}}\bar{\rho}_{\textrm{m}}\right),\\
    \delta P_{\textrm{m}} &\rightarrow& \bar{\Omega}^{-4}\left(\delta P_{\textrm{m}}-4\frac{\delta \Omega}{\bar{\Omega}}\bar{P}_{\textrm{m}}\right).
\end{eqnarray}

The non-adiabatic pressure perturbation associated with the matter sector, shown earlier to be gauge-invariant, transforms as 
\begin{equation}
    \delta \hat{P}_{\textrm{m}}^{\textrm{nad}} = \delta \hat{P}_{\textrm{m}}-\hat{c}_s^2 \delta \hat{\rho}_{\textrm{m}} = \frac{\bar{\Omega}^{-4}}{\dot{\bar{\rho}}_{\textrm{m}}-4\frac{\dot{\bar{\Omega}}}{\bar{\Omega}}\bar{\rho}_{\textrm{m}}}\left[\dot{\bar{\rho}}_{\textrm{m}}\delta P^{\textrm{nad}}_{\textrm{m}}-4\frac{\dot{\bar{\Omega}}}{\bar{\Omega}}\bar{\rho}_{\textrm{m}}\left(\delta P_{\textrm{m}}-w_{\textrm{m}}\delta \rho_{\textrm{m}}\right)-4\frac{\delta \Omega}{\bar{\Omega}}\dot{\bar{\rho}}_{\textrm{m}}\left(\bar{P}_{\textrm{m}}-c_s^2\bar{\rho}_{\textrm{m}}\right)\right].
\end{equation}
For constant $w_{\textrm{m}}$, the $\delta \Omega$ contribution vanishes, and, although the non-adiabatic pressure perturbation is not yet conformally invariant, it rescales by the same overall factor as the background energy and pressure densities,
\begin{equation}
    \left.\delta \hat{P}^{\textrm{nad}}_{\textrm{m}}\right|_{w_{\textrm{m}}=\textrm{const.}} = \bar{\Omega}^{-4}\left.\delta P^{\textrm{nad}}_{\textrm{m}}\right|_{w_{\textrm{m}}=\textrm{const.}}.
\end{equation}
In the absence of entropy perturbations ($\delta P_{\textrm{m}} = c_s^2 \delta \rho_{\textrm{m}}$), $\delta P^{\textrm{nad}}_{\textrm{m}} = 0$ in all frames, provided $w_{\textrm{m}}$ is constant. If $w_{\textrm{m}}$ is time-dependent, however, conformal transformations can induce a non-adiabatic pressure perturbation in the new frame:
\begin{equation}
    \delta \hat{P}_{\textrm{m}}^{\textrm{nad}} = \frac{4\bar{\Omega}^{-5}\bar{\rho}_{\textrm{m}}(w_{\textrm{m}}-c_s^2)}{\dot{\bar{\rho}}_{\textrm{m}}-4\frac{\dot{\bar{\Omega}}}{\bar{\Omega}}\bar{\rho}_{\textrm{m}}}\left(\dot{\bar{\Omega}}\delta \rho_{\textrm{m}}-\dot{\bar{\rho}}_{\textrm{m}}\delta \Omega\right).
\end{equation}
If $\Omega^2(x)$ is chosen such that its perturbation is adiabatic with respect to $\rho_{\textrm{m}}(x)$ (that is, proportional to the energy density perturbation along the background trajectory),
\begin{equation}
    \delta \Omega(x) = \frac{\dot{\bar{\Omega}}(t)}{\dot{\bar{\rho}}_{\textrm{m}}(t)}\delta\rho_{\textrm{m}}(x)~,
\end{equation}
then $\delta \hat{P}^{\textrm{nad}}_{\textrm{m}} = 0$, and adiabaticity in the matter sector is preserved after the conformal rescaling, as expected \cite{White:2012ya}. 

The gauge-invariant curvature perturbation on slicings of uniform-$\Phi$,
\begin{equation}
    \psi^{\delta \phi}\equiv \psi+\frac{\mathcal{N}H}{\dot \phi}\delta\phi~,
\end{equation}
where $\Phi(x) = \phi(t) +\delta \phi(x)$, is conformally invariant when the conformal factor, $\Omega^2(x)$, satisfies $\Omega =\Omega(\Phi)$ \cite{Chiba:2008ia}. Indeed,
\begin{equation}
    \hat{\psi}^{\delta \phi} = \hat{\psi}+\frac{\dot{\hat{a}}}{\hat{a}}\frac{\delta \phi}{\dot \phi} = \psi^{\delta \phi}-\bar{\Omega}^{-1}\left(\delta \Omega -\frac{\dot{\bar{\Omega}}}{\dot \phi}\delta \phi\right),
\end{equation}
which is consistent because a uniform-$\Phi$ slicing implies $\delta \phi=0$ and hence $\delta \Omega = 0$, restoring the conformal invariance of $\psi$ on this slicing (although not of all perturbations, \textit{e.g.} $\delta \Theta$ in Eq.~\eqref{eq:pert-scalar-exp-rate-conformal-rescaling-kindofdifficulttounderstand} \cite{Gong:2011qe}). The uniform-$\Phi$ slicing coincides with the rest frame of the scalar field $\Phi$ regarded as a fluid with canonical EMT shown in Eq.~\eqref{eq:Defining-frame-field} (\textit{i.e.} $\Psi_{\Phi}\equiv -\dot \phi \delta \phi = 0$). It does not, however, coincide with the comoving slicing of the effective fluid whose EMT is covariantly conserved by virtue of the contracted Bianchi identity \cite{Diaz:2023tma} (see below Eqs.~\eqref{eq:Eofg}--\eqref{eq:source-term}). The effective-fluid comoving slicing is preferable, as the covariant conservation of the associated EMT ensures the continuous matching of the curvature perturbation to its post-inflationary counterpart. This follows from the continuity equation satisfied by the effective energy and pressure densities at background level \cite{TerenteDiaz:2024uxb}. On sufficiently large scales, where gradient terms remain negligible despite a possibly non-standard time evolution of $H$, the curvature perturbations on the uniform-effective-density and zero-total-momentum slicings coincide, and both pass unchanged through reheating into the radiation-dominated era; this is the curvature perturbation relevant for late-time cosmology \cite{Mukhanov:1990me,Hwang:1991aj,Hwang:2001qk,Weinberg:2003sw,Lyth:2009zz,Allahverdi:2010xz}.\footnote{Some of the authors of the present work (M.~K. and J.~J.~T.~D.) have analysed the conformal invariance of the curvature perturbation on slicings of zero total momentum in the Jordan frame during `cold' inflation for a $F(\Phi)R$ modified gravity (see Ref.~\cite{Diaz:2023tma}). Conformal invariance is guaranteed on superhorizon scales but is contingent on the relative variation of the two time-dependent homogeneous scales of the theory: the Hubble parameter $H$, and the non-minimal coupling function (or equivalently, the varying Newton's gravitational constant, $\sqrt{F}$, in Planck units).}

\section{Quantisation of Perturbations} \label{app:quantisation}
\noindent
In this appendix, we explain why the quantisation schemes based on $\delta\varphi$ and $\delta\phi$ are equivalent. To this end, we first consider their Fourier expansions in terms of the corresponding complete sets $\{\delta\varphi_k\}$ and $\{\delta\phi_{k}\}$ of mode solutions to their respective classical equations of motion:
\begin{eqnarray}
    \delta\varphi &=& \int\frac{\dif^3\mathbf{k}}{(2\pi)^{3/2}}\left(e^{-\iu k_ix^i}\delta\varphi_k \mathtt{a}_k+\textrm{c.c.}\right), \label{eq:delta varphi Fourier expansion} \\
    \delta\phi &=& \int\frac{\dif^3\mathbf{k}}{(2\pi)^{3/2}}\left(e^{-\iu k_ix^i}\delta\phi_{k} \tilde{\mathtt{a}}_{k}+\textrm{c.c.}\right), \label{eq:delta phi Fourier expansion}
\end{eqnarray}
where $\{\mathtt{a}_k\}$ and $\{\tilde{\mathtt{a}}_k\}$ are expansion coefficients, to be promoted to annihilation operators upon quantisation. Notice that, in the expressions above, we have used the same label $k$ for the modes $\{\delta\varphi_k\}$ and $\{\delta\phi_k\}$, the reason for this being that, if $\delta\phi_k$ is a mode solution of the equation for $\delta\phi$, then
\begin{equation} \label{eq:delta varphi delta phi modes}
    \delta\varphi_k=\sqrt{\bar{\mathcal{K}}}\delta\phi_{k}
\end{equation}
is a mode solution of the equation for $\delta\varphi$ with the same $k$, by virtue of Eq.~\eqref{eq:delta varphi delta phi}. Therefore, we can multiply Eq.~\eqref{eq:delta phi Fourier expansion} by $\sqrt{\bar{\mathcal{K}}}$, use the fact that $\sqrt{\bar{\mathcal{K}}}$ is a classical background quantity independent of momenta or the quantum fields $\delta\varphi$ and $\delta\phi$ to introduce it inside the integral on the right-hand side, and then substitute Eq.~\eqref{eq:delta varphi delta phi modes} to find the exact same mode decomposition as in Eq.~\eqref{eq:delta varphi Fourier expansion}, upon the identification
\begin{equation}
    \tilde{\mathtt{a}}_k=\mathtt{a}_k~.
\end{equation}
This establishes the equality of the quantisation schemes based on the two mode expansions for $\delta\varphi$. A similar argument can be provided for $\delta\phi$.

\section{Conformal Transformation of Momenta} \label{app:comoving momenta}
\noindent
Our discussion on cosmological perturbations and thermal equilibrium states as seen from the defining and hatted frames relies on the transformation properties of comoving momenta under conformal transformations. As we see here, comoving momenta turn out to be conformally invariant.

As is well-known, the defining property of the comoving momentum $k_i$ is that its magnitude remains constant in coordinate time $t$, in contrast with the magnitude of the (three-dimensional) physical momentum $p_i$, which scales as
\begin{equation} \label{eq:momentum magnitude}
    g^{ij}p_ip_j=\dfrac{\const}{a^2(t)}\equiv\dfrac{\delta^{ij}k_ik_j}{a^2(t)}\equiv\dfrac{k^2}{a^2(t)}~.
\end{equation}
The physical momentum one-form $p_\mu$ is invariant under conformal transformations (that is to say, $\hat{p}_\mu=p_\mu$), since its definition does not rely on the concept of the metric tensor. In consequence, the physical momentum vector $p^\mu$ transforms as
\begin{equation}
    p^{\hat{\mu}}=\hat{g}^{\mu\nu}p_\nu=\Omega^{-2}p^\mu~,
\end{equation}
and thus
\begin{equation}
    \hat{g}^{ij}p_ip_j=\Omega^{-2}p_ip^i~.
\end{equation}
Because the scale factor on the right-hand side of Eq.~\eqref{eq:momentum magnitude} transforms as $\hat{a}=\Omega a$, we have that the magnitude of $k_i$ is insensitive to frame changes:
\begin{equation}
    k^2=\delta^{ij}k_ik_j=\delta^{ij}\hat{k}_i\hat{k}_j=\hat{k}^2~.
\end{equation}
This implies that
\begin{equation} \label{eq:k and p}
    k_i=p_i~, \qquad k^i=a^2(t)p^i~.
\end{equation}
From the transformation rules for $p_i$, $p^i$, and $a$, it is then clear that $k_i$ and $k^i$ are both conformally invariant:
\begin{equation} \label{eq:k transformation}
    \hat{k}_i=k_i~, \qquad \hat{k}^{\hat{i}}=k^i~.
\end{equation}

\bibliography{bibfile}
\end{document}